\documentclass[%
 longbibliography,
 reprint,onecolumn,12pt,
 amsmath,amssymb,
 aps,
]{revtex4-1}

\usepackage{graphicx}
\usepackage{dcolumn}
\usepackage{bm}
\usepackage{color} 
\usepackage{bm} 

\begin{document}

\preprint{APS/123-QED}

\title{Wave Scattering in Spatially Inhomogeneous Currents}

\author{Semyon Churilov}
\affiliation{Institute of Solar-Terrestrial Physics of the
Siberian Branch of Russian Academy of Sciences, Irkutsk-33, PO Box
291, 664033, Russia.}
\author{Andrei Ermakov}
\affiliation{School of Agricultural, Computational and
Environmental Sciences,\\ University of Southern Queensland, QLD
4350, Australia.}%
\author{Yury Stepanyants}
\email{Corresponding author: Yury.Stepanyants@usq.edu.au}
\affiliation{Department of Applied Mathematics, Nizhny Novgorod
State Technical University, Nizhny Novgorod, 603950, Russia and \\
School of Agricultural, Computational and Environmental Sciences,
University of Southern Queensland, QLD 4350, Australia.}%

\date{\today}

\begin{abstract}
We analytically study a scattering of long linear surface waves on
stationary currents in a duct (canal) of constant depth and
variable width. It is assumed that the background velocity
linearly increases or decreases with the longitudinal coordinate
due to the gradual variation of duct width. Such a model admits
analytical solution of the problem in hand, and we calculate the
scattering coefficients as functions of incident wave frequency
for all possible cases of sub-, super, and trans-critical
currents. For completeness we study both co-current and
counter-current wave propagation in accelerating and decelerating
currents. The results obtained are analysed in application to
recent analog gravity experiments and shed light on the problem of
hydrodynamic modelling of Hawking radiation. The paper is published
in Physical Review D.%
\end{abstract}

\pacs{Valid PACS appear here}
\maketitle

\section{\label{sec:level1}Introduction}

Since 1981 when Unruh established the analogy in wave
transformation occurring at the horizon of a black hole and at a
critical point of a hydrodynamic flow \cite{Unruh-1981} there have
been many attempts to calculate the transformation coefficients
and find the analytical expression for the excitation coefficient
of a negative energy mode (see, for instance, \cite{CoutPar-2014,
RobMichPar-2016, Coutant-2016, Philbin-2016} and references
therein). In parallel with theoretical study there were several
attempts to model the wave scattering in spatially inhomogeneous
currents experimentally and determine this coefficient through the
measurement data \cite{Weinfurtner-2011, Euve-2016} (similar
experiments were performed or suggested in other media, for
example, in the atomic Bose–-Einstein condensate -- see
\cite{Steinhauer-2016} and numerous references therein). In
particular, the dependence of amplitude of a negative energy mode
on frequency of incident wave in a water tank was determined
experimentally \cite{Weinfurtner-2011}; however several aspects of
the results obtained in this paper were subject to criticism.

The problem of water wave transformation in spatially
inhomogeneous currents is of significant interest itself and there
is a vast number of publications devoted to theoretical and
experimental study of this problem. However in applying to the
modelling of Hawking's effect, the majority of these publications
suffer a drawback which is related to the parasitic effect of
dispersion, whereas the dispersion is absent in the pure
gravitational Hawking effect.

Below we consider a model which describes a propagation of
small-amplitude long surface water waves in a duct (canal) of
constant depth but variable width. The dispersion is absent, and
the model is relevant to the analytical study of the Hawking
effect. We show that the transformation coefficients can be found
in the exact analytical forms both for co-current and
counter-current wave propagation in gradually accelerating and
decelerating currents.

We believe that the results obtained can be of wider interest, not
only as a model of Hawking's effect, but in application to real
physical phenomena occurring in currents in non-homogenous ducts,
at least at relatively small Froude numbers. We consider all
possible configurations of the background current and incident
wave. The paper's contents is presented below. %

\newpage
\tableofcontents

\section{\label{sec:level2}Derivation of the Governing Equation}

Let us consider the set of equations for water waves on the
surface of a perfect fluid of a constant density $\rho$ and depth
$h$. Assume that the water moves along the $x$-axis with a
stationary velocity $U(x)$ which can be either an increasing or a
decreasing function of $x$. Physically such a current can be
thought as a model of water flow in a horizontal duct with a
properly varying width $b(x)$. We will bear in mind such a model,
although we do not pretend here to consider a current in a real
duct, but rather to investigate an idealized hydrodynamic model
which is described by the equation analogous to that appearing in
the context of black hole evaporation due to Hocking radiation
\cite{Unruh-1981, Jacobson-1991, Unruh-1995, Faccio-2013,
CoutPar-2014, RobMichPar-2016, Coutant-2016, Philbin-2016}.

In contrast to other papers also dealing with the surface waves on
a spatially varying current (see, e.g., \cite{CoutPar-2014,
RobMichPar-2016, Coutant-2016, Philbin-2016}), we consider here
the case of shallow-water waves when there is no dispersion,
assuming that the wavelengths $\lambda\gg h$.

In the hydrostatic approximation, which is relevant to long waves
in shallow water \cite{Landafshitz-1987}, the pressure can be
presented in the form $p = p_0 + \rho g (\eta - z)$, where $p_0$
is the atmospheric pressure, $g$ is the acceleration due to
gravity, $z$ is the vertical coordinate, and $\eta(x, t)$ is the
perturbation of free surface ($-h \le z \le \eta$). Then the
linearized Euler equation for small perturbations having also only
one velocity component $u(x, t)$ takes the form:
\begin{equation}
\label{LinEurEq} %
\frac{\partial u}{\partial t} + \frac{\partial (Uu)}{\partial x} =
-g\frac{\partial \eta}{\partial x}.
\end{equation}

The second equation is the continuity equation which is equivalent
to the mass conservation equation for shallow-water waves:
\begin{equation}
\label{MassCons} %
\frac{\partial S}{\partial t} + \frac{\partial }{\partial
x}\left[S\left(U + u\right)\right] = 0,
\end{equation}
where $S(x, t)$ is the portion of the cross-section of a duct
occupied by water, $S(x, t) = b(x)[h + \eta(x, t)]$, where $b(x)$
is the width of the duct.

For the background current Eq.~(\ref{MassCons}) gives the mass
flux conservation $Q \equiv \rho\, U(x)S(x) = \rho\, U(x)b(x)h = $
const. Inasmuch as $h = $ const, we have $U(x)b(x) = Q/\rho h = $
const, and Eq.~(\ref{MassCons}) in the linear approximation
reduces to:
\begin{equation}
\label{LinMassCons} %
b(x)\frac{\partial \eta}{\partial t} + \frac{\partial }{\partial
x}\left[b(x)\left(U\eta + uh\right)\right] = 0.
\end{equation}

Thus, the complete set of equations for shallow water waves in a
duct of a variable width consists of Eqs.~(\ref{LinEurEq}) and
(\ref{LinMassCons}). This set can be reduced to one equation of
the second order. To this end let us divide first
Eq.~(\ref{LinMassCons}) by $b(x)$ and rewrite it in the equivalent
form:
\begin{equation}
\label{RedMassCons1} %
\frac{\partial \eta}{\partial t} + U\frac{\partial \eta}{\partial
x} = -hU\frac{\partial}{\partial x}\frac{u}{U}.
\end{equation}

Expressing now the velocity component $u$ in terms of the velocity
potential $\varphi$, $u = \partial \varphi/\partial x$, and
combining Eqs.~(\ref{LinEurEq}) and (\ref{RedMassCons1}), we
derive
\begin{equation}
\label{ansatz} %
\left(\frac{\partial }{\partial t} + U\frac{\partial }{\partial x}
\right)\left(\frac{\partial \varphi}{\partial t} + U\frac{\partial
\varphi}{\partial x} \right) = c_0^2U\frac{\partial}{\partial
x}\left(\frac{1}{U}\frac{\partial \varphi}{\partial x}\right),
\end{equation}
where $c_0 = \sqrt{gh\vphantom{^2}}$ is the speed of linear long
waves in shallow water without a background current.

As this equation describes wave propagation on the stationary
moving current of perfect fluid, it provides the law of wave
energy conservation which can be presented in the form (its
derivation is given in Appendix \ref{appA}):
\begin{equation}
\label{Chur57} %
\frac{\partial{\cal E}}{\partial t} + \frac{\partial J}{\partial x} = 0, %
\end{equation}
where
\[
{\cal E} = \frac{{\rm
i}}{U}\left[\overline{\varphi}\left(\frac{\partial\varphi}{\partial
t} + U\frac{\partial\varphi}{\partial x}\right) - \varphi
\left(\frac{\partial\overline{\varphi}}{\partial t} +
U\frac{\partial\overline{\varphi}}{\partial x}\right)\right],
\quad
J = {\cal E}U - \frac{{\rm i}\,c_0^2}{U}
\left(\overline{\varphi}\frac{\partial\varphi}{\partial x} -
\varphi\frac{\partial\overline{\varphi}}{\partial x}\right),
\]
and the over-bar denotes complex conjugation.

Solution of the linear equation (\ref{ansatz}) can be sought in
the form $\varphi(x, t) = \Phi(x)\mbox{e}^{-{\rm i}\,\omega t}$,
then it reduces to the ODE for the function $\Phi(x)$:
\begin{equation}
\label{Spat} %
\left(-\mbox{i}\,\omega + U\frac{d}{dx}
\right)\left(-\mbox{i}\,\omega \Phi + U\frac{d\Phi}{dx} \right) =
c_0^2U\frac{d}{dx}\left(\frac{1}{U}\frac{d\Phi}{dx}\right).
\end{equation}

If we normalize the variables such that $U/c_0 = V$, $x/L = \xi$,
and $\omega L/c_0 = \hat\omega$, where $L$ is the characteristic
spatial scale of the basic current, then we can present the main
equation in the final form:
\begin{equation}
\label{NormEq} %
V\left(1 - V^2\right)\frac{d^2\Phi}{d\xi^2} - \left[\left(1 +
V^2\right)V' - 2\,\mbox{i}\,\hat\omega
V^2\right]\frac{d\Phi}{d\xi} + V\hat\omega^2\Phi = 0,
\end{equation}
where the prime stands for here and below differentiation with
respect to the entire function argument (in this particular case
with respect to $\xi$).

If the perturbations are monochromatic in time, as above, then the
wave energy $\cal E$ and energy flux $J$ do not depend on time,
therefore, as follows from Eq.~(\ref{Chur57}), the energy flux
does not depend on $x$ too, so $J = $ const.

\begin{figure}[h]
\includegraphics[width=90mm]{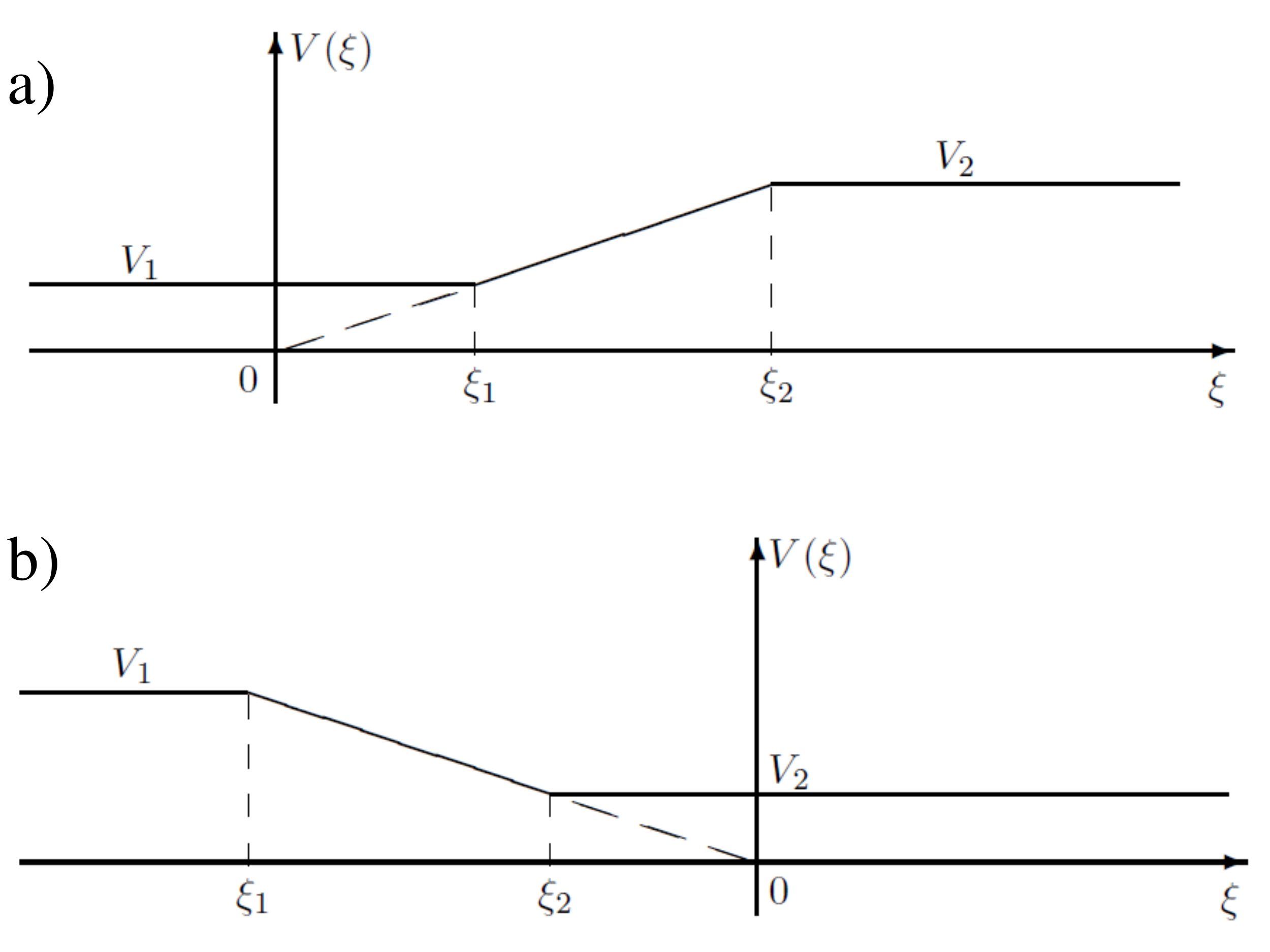}
\caption{Sketch of accelerating (a) and decelerating (b)
background currents.}%
\label{f01}%
\end{figure}

For the concrete calculations we chose the piece-linear velocity
profile, assuming that the current varies linearly within a finite
interval of $x$ and remains constant out of this interval (see
Fig. \ref{f01}):
\begin{equation}
\label{BasicFlow1} %
V_a(\xi) = \left\{
\begin{array}{rcl}%
V_1 \equiv \xi_1, \quad \phantom{0123456} &\xi& \le \xi_1, \\
\xi, \quad \phantom{0123} 0 < \xi_1 < &\xi& < \xi_2, \\
V_2 \equiv \xi_2, \quad \phantom{0123456} &\xi& \ge \xi_2;
\end{array}
\right.%
\quad \;\;\; V_d(\xi) = \left\{
\begin{array}{rcl}%
V_1 \equiv -\xi_1, \quad  \phantom{0123} &\xi& \le \xi_1, \\
-\xi, \quad \phantom{012345} \xi_1 < &\xi& < \xi_2 < 0, \\
V_2 \equiv -\xi_2, \quad \phantom{0123} &\xi& \ge \xi_2,
\end{array}
\right.
\end{equation}
where $V_a(\xi)$ pertains to the accelerating current, and
$V_d(\xi)$ -- to the decelerating current. To simplify further
calculations, we have chosen, without the loss of generality, the
origin of the coordinate frame such that the velocity profile is
directly proportional to $\pm\xi$ in the interval $\xi_1 \le \xi
\le \xi_2$ as shown in Fig. \ref{f01}). For such velocity
configurations it is convenient to set $L = (x_2 - x_1)c_0/(U_2 -
U_1) = (x_2 - x_1)/|V_2 - V_1|$.

\begin{figure}[h]
\includegraphics[width=90mm]{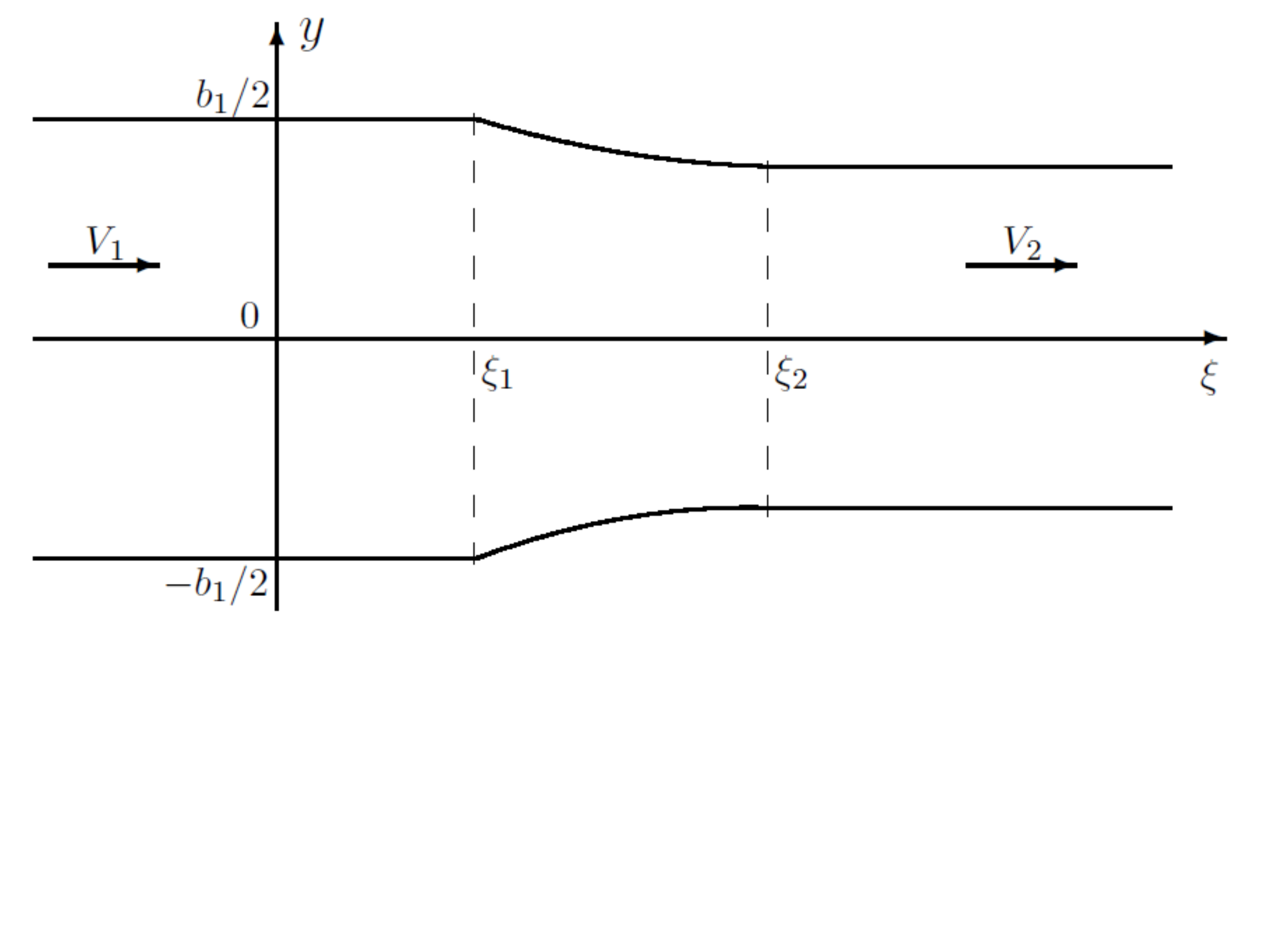}
\vspace*{-2.5cm}%
\caption{The sketch of a duct with the decreasing width that
provides spatially accelerating background current.}
\label{f02}%
\end{figure}

The choice of piece-linear velocity profile allows us to reduce
the governing equation (\ref{NormEq}) to the analytically solvable
equation and obtain exact solutions. The corresponding water flow
can be realized in a duct with a variable width, which is
constant, $b = b_1$, when $\xi \le \xi_1$, then gradually varies
along the $\xi$-axis as $b(\xi) = b_1\xi_1/\xi$ in the interval
$\xi_1 \le \xi \le \xi_2$, and after that remains constant again,
$b_2 = b_1\xi_1/\xi_2$ when $\xi \ge \xi_2$. Schematically the
sketch of a duct with gradually decreasing width that provides an
accelerating current is shown in Fig. \ref{f02}.

Equation (\ref{NormEq}) should be augmented by the boundary
conditions at $\xi \to \pm \infty$ which specify the scattering
problem, as well as by the matching conditions at $\xi=\xi_1$ and
$\xi=\xi_2$. The latter conditions reduce to the continuity of the
function $\Phi(\xi)$ and its derivative $\Phi'(\xi)$ (see Appendix
\ref{appB} for the derivation):
\begin{equation}
\label{Matching} %
\Phi(\xi_{1,2} + 0) = \Phi(\xi_{1,2} - 0),\ \ \ \
\Phi'(\xi_{1,2} + 0) = \Phi'(\xi_{1,2} - 0).
\end{equation}

On the basis of Eq.~(\ref{NormEq}) and matching conditions
(\ref{Matching}), we are able to study analytically all possible
cases of orientation of an incident wave and a current, assuming
that the current can be sub-critical ($V_{1,2} < 1$),
trans-critical ($V_{1} > 1$, $V_{2} < 1$ or vice versa $V_{1} <
1$, $V_{2} > 1$), or super-critical ($V_{1,2} > 1$).

\section{\label{sec:level3}Qualitative Analysis of the Problem Based
on the JWKB Approximation}

Before the construction of an exact solution for wave scattering
in currents with the piece-linear velocity profiles, it seems
reasonable to consider the problem qualitatively to reveal its
specific features which will help in the interpretation of results
obtained.

Consider first a long sinusoidal wave propagating on a current
with constant $U$. Assume, in accordance with the shallow-water
approximation, that the wavelength $\lambda \gg h$. The dispersion
relation for such waves is
\begin{equation}
\label{DispRel} %
\left(\omega - {\bf kU}\right)^2 = c_0^2k^2, %
\end{equation}
where ${\bf k} = (k, 0, 0)$ is a wave vector related with a
wavelength $\lambda = 2\pi/|{\bf k}|$.

\begin{figure}[h]
\centering
\includegraphics[width=90mm]{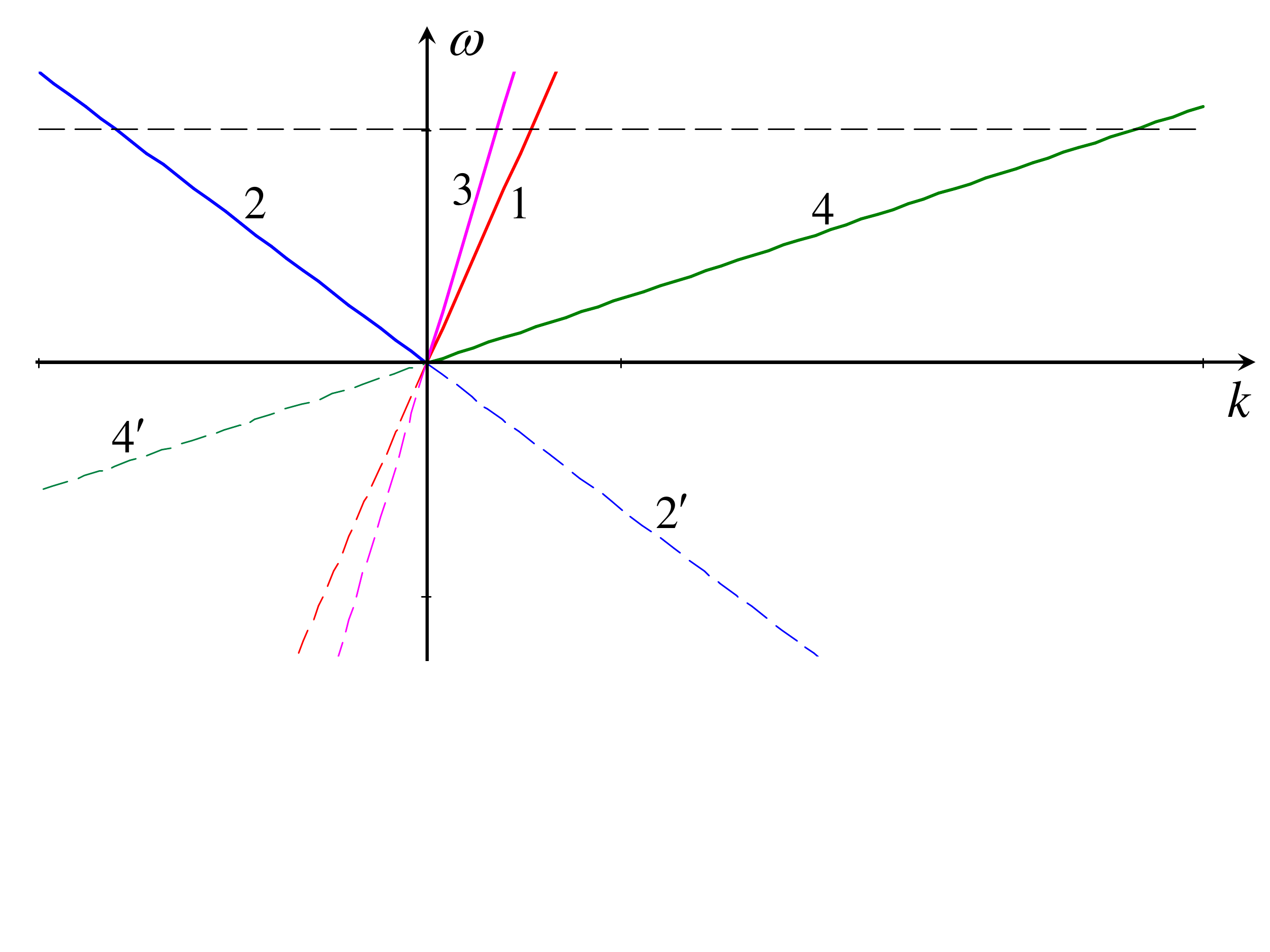}
\vspace*{-23mm}%
\caption{(color online) The dispersion dependences for surface
waves on uniformly moving shallow water. Lines 1 and 2 pertain to
co-current and counter-current propagating waves respectively in a
sub-critical current ($U < c_0$). Lines 3 and 4 pertain to
positive- and negative-energy waves respectively in a
super-critical current ($U
> c_0$) both propagating downstream.}
\label{f03}%
\end{figure}

A graphic of the dispersion relation is shown in Fig.~\ref{f03}
for two values of the current speed, sub-critical, $U < c_0$, and
super-critical, $U > c_0$. Since we consider dispersionless
shallow-water waves, graphics of the dependences $\omega(k)$ are
straight lines formally extending from minus to plus infinity. We
suppose, however, that the frequency $\omega$ is a non-negative
quantity which is inversely proportional to the wave period;
therefore, without loss of generality, we can ignore those
portions of dispersion lines which correspond to negative
frequencies (in Fig.~\ref{f03} they are shown by inclined dashed
lines). The dashed horizontal line in Fig.~\ref{f03} shows a
particular fixed frequency of all waves participating in the
scattering process.

For co-current propagating waves with ${\pmb k} \uparrow \uparrow
{\bf U}$ the dispersion relation (\ref{DispRel}) reduces to
$\omega = (U + c_0)|{\bf k}|$, whereas for counter-current
propagating waves with ${\pmb k} \downarrow \uparrow {\bf U}$ it
is $\omega = |U - c_0||{\bf k}|$. Thus, the dispersion lines for
surface waves on a current are not symmetrical with respect to the
vertical axis $k = 0$. When the current speed $U$ increases, the
right branch 1 turns toward the vertical axis (cf. lines 1 and 3
in Fig.~\ref{f03}). The left branch 2 in this case tilts toward
the negative half-axis $k$; coincides with it when $U = c_0$, and
then, when $U > c_0$, it goes to the lower half-plane and becomes
negative. However, its negative portion $2'$ goes up, passes
through the axis $k$ and appears in the upper half-plane as the
dispersion line 4. Thus, waves corresponding to lines 3 and 4 are
downstream propagating waves, whereas there are no upstream
propagating waves, if $U > c_0$. From the physical point of view
this means that the current is so strong that it pulls downstream
even counter-current propagating waves. As was shown, for
instance, in Refs. \cite{FabrStep-1998, MaiRusStep-2016,
ChurErRousStep-2017}, in a such strong current, waves on branch 3
have positive energy, whereas waves on branch 4 have negative
energy.

To consider wave propagation on a spatially variable current when
it accelerates or decelerates along $x$-axis, let us use the JWKB
method, which physically presumes that the wavelengths are much
less than the characteristic scale of inhomogeneity, $\lambda\ll
L$ (whereas still $\lambda\gg h$ and the shallow-water
approximation is valid). This condition can be presented in the
form $L/\lambda = L/(c_0T) = L\omega/(2\pi c_0) =
\hat{\omega}/2\pi \gg 1$ (where $T = 2\pi/\omega$ is the wave
period) and if it is fulfilled, the JWKB solution of
Eq.~(\ref{NormEq}) can be sought in the form (see, e.g.,
\cite{Dingle-1973, Olver-1974}):
\begin{equation}%
\label{JWKB-solution}%
\Phi(\xi) = \exp{\left[{\rm i}\,\hat\omega\!\int\! q(\xi)\,d
\xi\right]}, \quad q(\xi) = q_0(\xi) + \hat\omega^{-1}q_1(\xi) +
\hat\omega^{-2}q_2(\xi) + \ldots
\end{equation}

Substitution of these expressions into Eq.~(\ref{NormEq}) gives
two linearly independent solutions:
\begin{equation}%
\label{Chur59} %
\Phi^{(\pm)}(\xi) = \sqrt{V(\xi)}\exp{\left[{\rm
i}\,\hat\omega\!\int\!\frac{d\xi}{V(\xi)\pm 1} +
O(\hat\omega^{-1})\right]},
\end{equation}
and the general solution of Eq.~(\ref{NormEq}) is the linear
combination of these two particular solutions:%
\begin{equation}%
\label{Chur60} %
\Phi(x) = A_F\,\Phi^{(+)}(x) + A_B\,\Phi^{(-)}(x),
\end{equation}
where $A_F$ and $A_B$ are amplitudes of co-current propagating
F-wave and counter-current propagating B-wave, respectively.

In the current with a spatially varying velocity $V(\xi)$, wave
propagation and transformation has a regular character, if $V(\xi)
\ne 1$ (i.e., if $U(x) \ne c_0$); then Eq.~(\ref{NormEq}) does not
contain critical points.

In sub-critical currents, when $0 < V_{1,2} < 1$ everywhere, an
incident wave arriving from the left (F-wave) or from the right
(B-wave) partially transmits through the domain of inhomogeneity
and partially transforms into the reflected wave of B- or F-type
respectively. Notice that in this case waves of both types have
positive energy.

In super-critical currents, when $V_{1,2} > 1$ is everywhere, as
was mentioned above, both F-wave and B-wave can propagate only in
the direction of the current; however F-wave has positive energy
whereas B-wave has negative energy. An incident wave of any type
propagating from left to right partially transforms into the wave
of another type, so that at the infinity, $\xi\to\infty$, waves of
both types appear.

In contrast to these cases, in a trans-critical current there is a
critical point where $V(\xi) = 1$. The existence of such a point
has only a minor influence on the co-current propagating F-wave,
but exerts a crucial action on the B-wave, because its ``wave
number'' $q_0^{(-)} \to \infty$ when $V(\xi)\to 1$. Due to this,
an arbitrarily small but finite viscosity leads to dissipation of
a B-wave that attains a neighborhood of the critical point. As the
result of this the energy flux $J$ does not conserve, in general,
when waves pass through this critical point. However, as will be
shown below, the energy flux conserves in spatially accelerating
trans-critical currents, but does not conserve in decelerating
currents.

Indeed, in an accelerating current where $0<V_1<1<V_2$, an
incident wave can arrive only from the left as the F-type wave
only. In the sub-critical domain ($\xi < 1$) it transforms into
the B-wave that runs backwards, towards $\xi = -\infty$. After
passing the critical point, being in the supercritical domain
($\xi>1$) it transforms into the B-wave that runs forward towards
$\xi = +\infty$. As a result, there is no B-wave that attains the
critical point; hence, there is no dissipation, and energy flux
conserves. On the contrary, in decelerating currents (where
$V_1>1>V_2>0$) B-waves, no matter incident or ``reflected'', run
to the critical point and dissipate there; therefore the energy
flux does not conserve in this case.

A specific situation occurs when the incident B-wave propagates
from plus infinity in the sub-critical current towards the
critical point and generates an F-wave on the current
inhomogeneity. If the current is super-critical on the left of the
critical point, then no one wave can penetrate into that domain.
Thus, the wave energy of incident B-wave partially converts into a
reflected F-wave and partially absorbs in the vicinity of the
critical point due to vanishingly small viscosity. We will come to
the discussion of these issues in Section \ref{sec:level5} when we
construct exact solutions of scattering problem for
Eq.~(\ref{NormEq}) where it is possible.

A qualitative analysis presented above demonstrates that the most
interesting results can be obtained for the trans-critical
currents and that the critical points play a crucial role in such
currents. However, in the vicinity of a critical point the
velocity of arbitrary type $U(x)$ can be generally approximated by
a linear function, $U(x) \sim x$. This makes an additional
argument in favor of studying wave scattering in currents with
piece-linear velocity profiles.

\section{\label{sec:level4}Wave Scattering in Inhomogeneous Currents
With a Piece-Linear Velocity Profile}

Consider now exact solutions of the problem on surface wave
scattering in inhomogeneous currents with piece-linear velocity
profiles described by Eqs.~(\ref{BasicFlow1}) and shown in Fig.
\ref{f01}. The basic equation (\ref{NormEq}) has constant
coefficients out of the interval $\xi_1 < \xi < \xi_2$, where the
current velocity linearly varies with $\xi$ (either increasing or
decreasing). Therefore out of this interval, solutions to this
equation can be presented in terms of exponential functions with
the purely imaginary exponents describing sinusoidal travelling
waves.

Within the interval $\xi_1 < \xi < \xi_2$ Eq.~(\ref{NormEq}) with
the help of change of variable $\zeta = \xi^2$ reduces to one of
the hypergeometric equations:%
\begin{equation}%
\label{hyperAccel}%
\zeta(1 - \zeta)\frac{d^2\Phi}{d\zeta^2} - (1 \mp {\rm
i}\,\hat\omega)\zeta\,\frac{d\Phi}{d\zeta} +
\frac{\hat\omega^2}{4}\,\Phi = 0,
\end{equation}
where upper sign pertains to the case of accelerating current, and
lower sign -- to the case of decelerating current.

The matching conditions at $\xi = \xi_1$ and $\xi = \xi_2$ are
given by Eqs.~(\ref{Matching}).

\subsection{\label{sec:level4l}Wave transformation in sub-critical currents}

Assume first that an incident wave propagates from left to right
parallel to the main current which is sub-critical in all domains,
$V_1 < V_2 < 1$. As mentioned above, in the left ($\xi < \xi_1$)
and right ($\xi > \xi_2$) domains Eq.~(\ref{NormEq}) has constant
coefficients, and in the intermediate domain ($\xi_1 < \xi <
\xi_2$), where $V(\xi) = \xi$, this equation reduces to one of
hypergeometric equations (\ref{hyperAccel}). These equations are
regular in the sub-critical case, and their coefficients do not
turn to zero. Two linearly independent solutions can be expressed
in terms of Gauss hypergeometric function $_2F_1(a,b;c;\zeta)$
(see \S 6.4 in book \cite{Luke-1975}). Thus, the general solution
of Eq.~(\ref{NormEq}) for the accelerating current in three
different domains can be presented as follows:
\begin{eqnarray}
\Phi(\xi) &=& A_1\mbox{e}^{{\rm i}\kappa_1(\xi-\xi_1)} +
A_2\mbox{e}^{-{\rm i}\kappa_2(\xi-\xi_1)}, \hspace{17mm} \xi \le \xi_1, \phantom{www} \label{SolA1}\\
\Phi(\xi) &=& B_1w_2(\xi^2) + B_2w_3(\xi^2), \hspace{17mm} \xi_1 \le \xi \le \xi_2, \phantom{www} \label{SolA2}\\
\Phi(\xi) &=& C_1\mbox{e}^{{\rm i}\kappa_3(\xi-\xi_2)} +
C_2\mbox{e}^{-{\rm i}\kappa_4(\xi-\xi_2)}, \hspace{17mm} \xi \ge
\xi_2, \phantom{www} \label{SolA3}
\end{eqnarray}
where $\kappa_1 = \hat\omega/(1 + V_1)$, $\kappa_2 = \hat\omega/(1
- V_1)$, $\kappa_3 = \hat\omega/(1 + V_2)$, $\kappa_4 =
\hat\omega/(1 - V_2)$, $A_{1,2}$, $B_{1,2}$, $C_{1,2}$ are
arbitrary constants, and
\begin{equation}
\label{AccelSub1}%
w_2(\zeta) = \; \zeta\, {_2F_1}(1 - {\rm i}\,\hat\omega/2,\,1 -
{\rm i}\,\hat\omega/2;\,2;\zeta), \quad w_3(\zeta)
=\;{_2F_1}({-\rm i}\hat\omega/2,-{\rm i}\,\hat\omega/2;1 - {\rm
i}\,\hat\omega;1-\zeta).
\end{equation}

The Wronskian of these linearly independent functions is
\cite{Luke-1975}:
\begin{equation}%
\label{Wronsk1}%
W =
w'_2(\zeta)w_3(\zeta)-w_2(\zeta)w'_3(\zeta)=\frac{\Gamma(1-{\rm
i}\,\hat\omega)} {\Gamma^2(1-{\rm i}\,\hat\omega/2)}
(1-\zeta)^{{\rm i}\,\hat\omega-1}.
\end{equation}

Similarly the general solution of Eq.~(\ref{NormEq}) for the
decelerating current can be presented. In the domains $\xi <
\xi_1$ and $\xi > \xi_2$ solutions are the same as above, whereas
in the intermediate domain $\xi_1 < \xi < \xi_2$ the general
solution is:
\begin{equation}
\label{SolD2}%
\Phi(\xi) = B_1\tilde w_2(\xi^2) + B_2\tilde w_3(\xi^2),
\end{equation}
where the linearly independent functions are
\begin{equation}
\label{DecelSub1}%
\tilde w_2(\zeta) = \zeta\, {_2F_1}(1 + {\rm i}\,\hat\omega/2,\,1
+ {\rm i}\,\hat\omega/2;\,2;\zeta), \quad \tilde w_3(\zeta) =
{_2F_1}({\rm i}\hat\omega/2,{\rm i}\,\hat\omega/2;1 + {\rm
i}\,\hat\omega;1-\zeta).
\end{equation}
with the Wronskian:
\begin{equation}%
\label{Wronsk2}%
\tilde W = \tilde w'_2(\zeta)\tilde w_3(\zeta) - \tilde
w_2(\zeta)\tilde w'_3(\zeta) = \frac{\Gamma(1+{\rm
i}\,\hat\omega)} {\Gamma^2(1+{\rm i}\,\hat\omega/2)}
(1-\zeta)^{-{\rm i}\,\hat\omega-1}.
\end{equation}

\subsubsection{\label{sec:level4l1}Accelerating currents. Transformation
of downstream propagating incident wave}

Assume that the incident wave has a unit amplitude $A_1 = 1$ and
calculate the transformation coefficients, setting $C_2 = 0$ and
denoting the amplitudes of the reflected wave by $R \equiv A_2$
and the transmitted wave by $T \equiv C_1$ ($R$ and $T$ play a
role of transformation coefficients, as they are usually
determined in hydrodynamics -- see, e.g. \cite{Massel-1989,
Kurkin-2015} and references therein).

Using the matching conditions at the boundaries of domains (see
Appendix \ref{appB}), we find:
\begin{eqnarray}
B_1\,w_2(V_1^2) &+& B_2\,w_3(V_1^2) = R + 1, \label{BCond1}\\
B_1\,w'_2(V_1^2) &+& B_2\,w'_3(V_1^2) = \frac{{\rm
i}\,\hat\omega}{2V_1}\left(\frac{1}{1+V_1} -
\frac{R}{1-V_1}\right), \phantom{ww}
\label{BCond2}\\
B_1\,w_2(V_2^2) &+& B_2\,w_3(V_2^2) = T, \label{BCond3}\\
B_1\,w'_2(V_2^2) &+& B_2\,w'_3(V_2^2) = \frac{{\rm
i}\,\hat\omega}{2V_2}\frac{T}{1+V_2}. \label{BCond4}
\end{eqnarray}

From these equations we derive the transformation coefficients:
\begin{widetext}
\begin{eqnarray}
R &=&
\frac{1}{\Delta}\left\{\frac{\hat\omega^2\left[w_2(V_1^2)w_3(V_2^2)
- w_2(V_2^2)w_3(V_1^2)\right]}{4V_1V_2(1+V_1)(1+V_2)} -
w'_2(V_1^2)w'_3(V_2^2) + w'_2(V_2^2)w'_3(V_1^2) + {}\right.
\nonumber \\
{} && \left. \frac{{\rm
i}\,\hat\omega}{2}\left[\frac{w_2(V_1^2)w'_3(V_2^2) -
w'_2(V_2^2)w_3(V_1^2)}{V_1(1 + V_1)} - \frac{w_2(V_2^2)w'_3(V_1^2)
- w'_2(V_1^2)w_3(V_2^2)}{V_2(1+V_2)}\right]%
\vphantom{\frac{\hat\omega^2\left[w_2(V_1^2)w_3(V_2^2)
- w_2(V_2^2)w_3(V_1^2)\right]}{4V_1V_2(1+V_1)(1+V_2)}}%
\right\}, \label{Trans1} \\%
&& {} \nonumber \\
T &=& -\frac{{\rm i}\,\hat\omega}{\Delta}\frac{(1-V_2^2)^{{\rm
i}\, \hat\omega - 1}}{V_1(1-V_1^2)}\,\frac{\Gamma(1-{\rm i}\,
\hat\omega)}{\Gamma^2(1-{\rm i}\, \hat\omega/2)}\,, \label{Trans2}\\
&& {} \nonumber \\
B_1 &=& -\frac{{\rm
i}\,\hat\omega}{\Delta}\frac{1}{V_1(1-V_1^2)}\left[\frac{{\rm
i}\,\hat\omega}{2V_2(1+V_2)}\,w_3(V_2^2) - w'_3(V_2^2)\right], \phantom{www} \label{Trans3}\\
&& {} \nonumber \\
B_2 &=& \frac{{\rm
i}\,\hat\omega}{\Delta}\frac{1}{V_1(1-V_1^2)}\left[\frac{{\rm
i}\,\hat\omega}{2V_2(1+V_2)}\,w_2(V_2^2) - w'_2(V_2^2)\right],
\label{Trans4}
\end{eqnarray}
\end{widetext}

\noindent where %
\[
\Delta = w'_2(V_1^2)w'_3(V_2^2) - w'_2(V_2^2)w'_3(V_1^2) +
\frac{\hat\omega^2\left[w_2(V_1^2)w_3(V_2^2)-w_2(V_2^2)w_3(V_1^2)\right]}{4V_1V_2(1-V_1)(1+V_2)}
+ {}
\]
\begin{equation}
\label{Delta}%
\frac{{\rm i}\,\hat\omega}{2}\left[\frac{w_2(V_1^2)w'_3(V_2^2) -
w'_2(V_2^2)w_3(V_1^2)}{V_1(1-V_1)} + \frac{w_2(V_2^2)w'_3(V_1^2) -
w'_2(V_1^2)w_3(V_2^2)}{V_2(1+V_2)}\right].
\end{equation}

\bigskip

\begin{figure}[h]
\centering
\includegraphics[width=90mm]{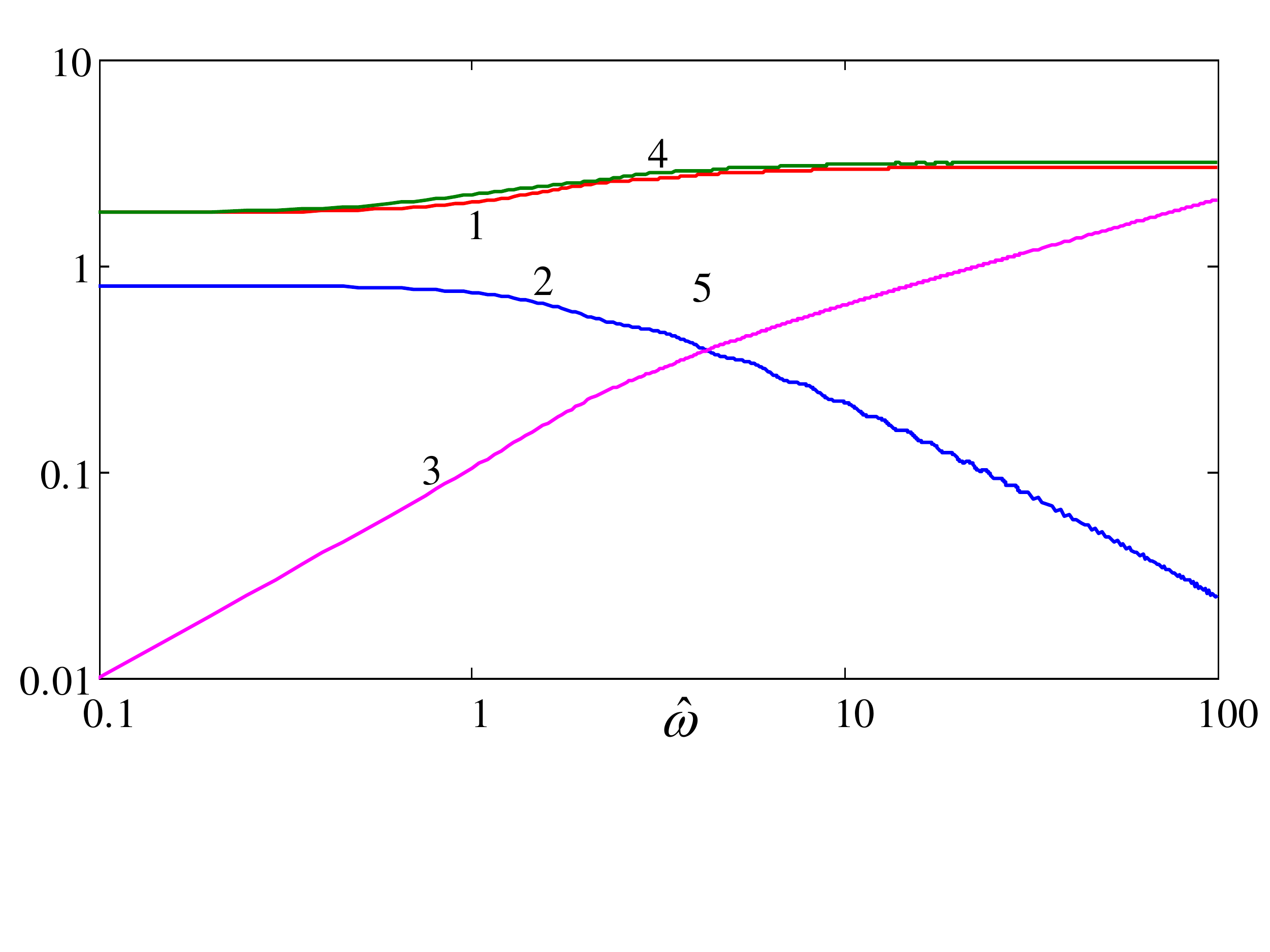}
\vspace*{-15mm}%
\caption{(color online) Modules of transformation coefficient as
functions of dimensionless frequency $\hat\omega$ for $V_1 = 0.1$,
$V_2 = 0.9$. Line 1 -- $|T|$, line 2 -- $|R|$, line 3 -- $|B_1|$,
line 4 -- $|B_2|$ Dashed line 5 represents the asymptotic for the
reflection coefficient $R \sim \hat\omega^{-1}$.}
\label{f04}%
\end{figure}

The modules of transformation coefficients $|T|$ and $|R|$, as
well as modules of intermediate coefficients of wave excitation in
the transient domain, $|B_1|$ and $|B_2|$, are shown in
Fig.~\ref{f04} as functions of dimensionless frequency
$\hat\omega$ for the particular values of $V_1 = 0.1$ and $V_2 =
0.9$. Qualitatively similar graphics were obtained for other
values of $V_1$ and $V_2$.

In the long-wave approximation, when $\hat\omega \to 0$, the
hypergeometric function $_2F_1(a,b;c;d)$ degenerates (see Appendix
\ref{appC}), then the transformation coefficients reduce to
\begin{equation}
\label{Reduct1}%
R = \frac{1 - V_1/V_2}{1 + V_1/V_2}, \quad T = 1 + R = \frac{2}{1
+ V_1/V_2}.
\end{equation}

These values are purely real and agree with the transformation
coefficients derived in Ref. \cite{ChurErRousStep-2017} for
surface waves in a duct with the stepwise change of cross-section
and velocity profile, and such an agreement takes place also for
other wave-current configurations considered below. Notice only
that here the transformation coefficients are presented in terms
of velocity potential $\varphi$, whereas in Ref.
\cite{ChurErRousStep-2017} they are presented in terms of free
surface elevation $\eta$. The relationship between these
quantities is given in the end of Appendix \ref{appA}).

In Fig.~\ref{f05}a) we present the graphic of $|\Phi(\xi)|$ (see
line 1) as per Eqs.~(\ref{SolA1})--(\ref{SolA3}) with $A_1 = 1$
and other determined transformation coefficients $A_2 = R$ as per
Eq.~(\ref{Trans1}), $C_1 = T$ as per Eq.~(\ref{Trans2}), and $C_2
= 0$. Coefficients $B_1$ and $B_2$ are given by
Eqs.~(\ref{Trans3}) and (\ref{Trans4}). The plot was generated for
the particular value of $\hat\omega = 1$; for other values of
$\hat\omega$ the graphics are qualitatively similar.

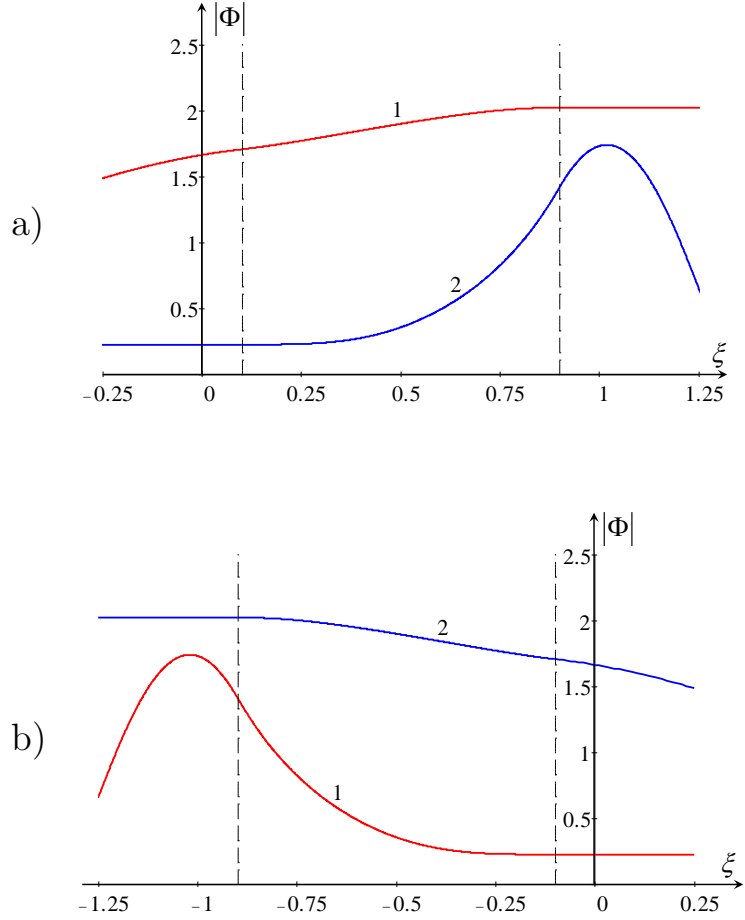
\begin{figure}[h]
\centering
\includegraphics[width=90mm]{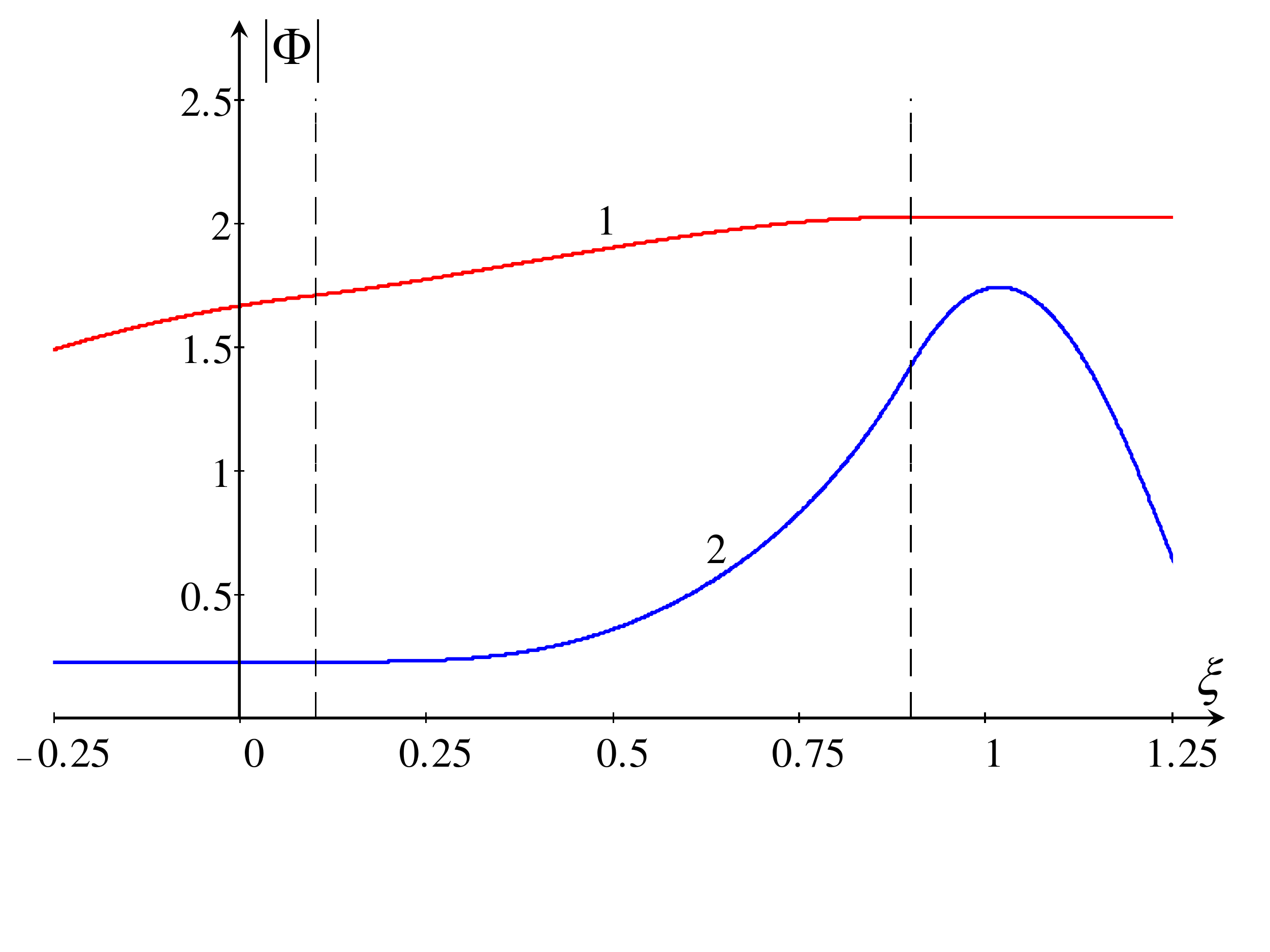}
\includegraphics[width=90mm]{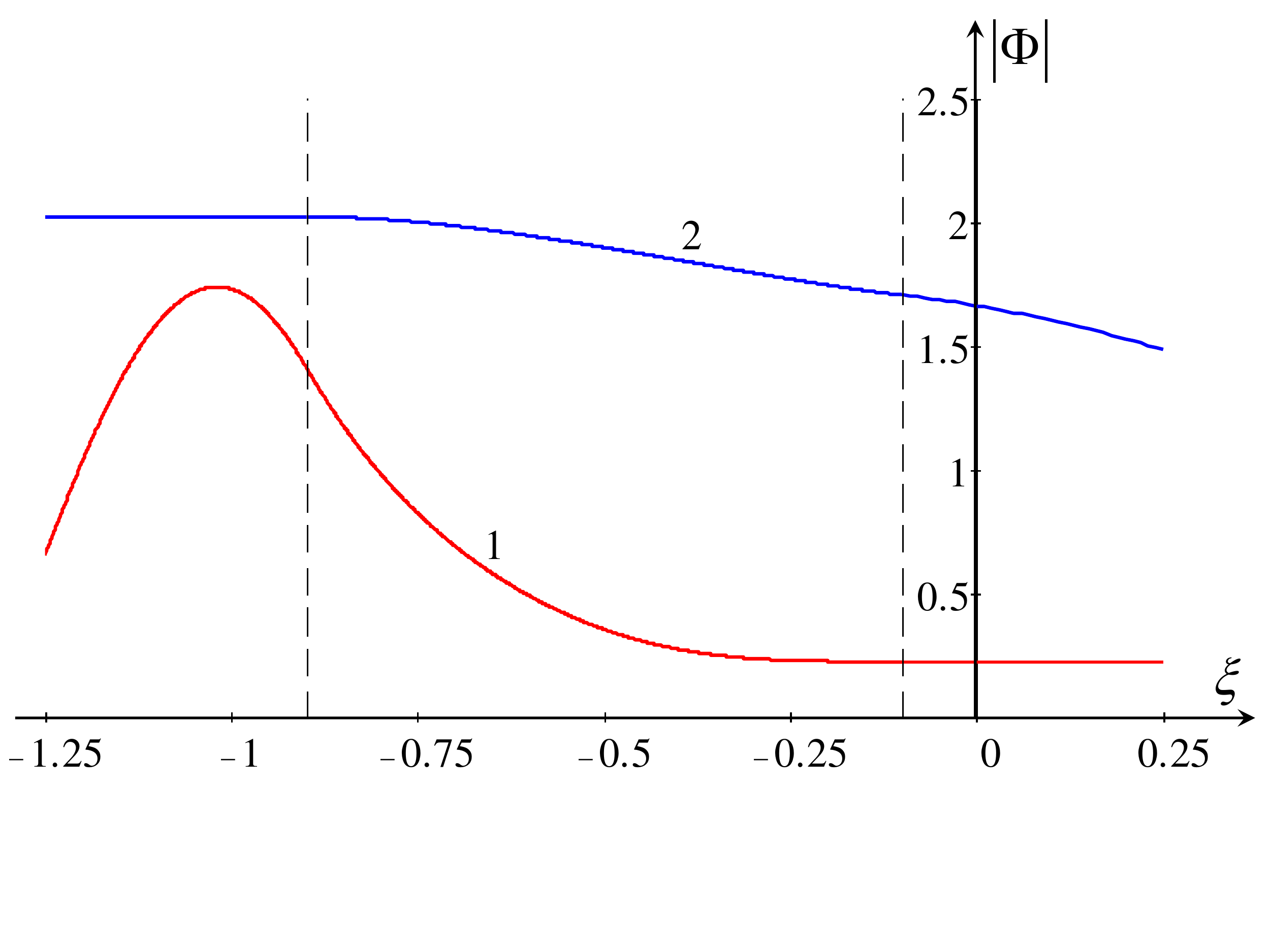}
\begin{picture}(300,6)%
\put(0,315){{\large a)}}%
\put(0,120){{\large b)}}%
\end{picture}
\vspace*{-15mm}%
\caption{(color online) Modules of function $\Phi(\xi)$ for wave
scattering in accelerating (a) and decelerating (b) sub-critical
currents with $V_1 = 0.1$ and $V_2 = 0.9$ in the former case and
$V_1 = 0.9$ and $V_2 = 0.1$ in the latter case. Line 1 in each
frame pertains to the co-current propagating incident wave, and
line 2 -- to the counter-current propagating incident wave. Dashed
vertical lines show the boundaries of the transient domain $\xi_1$
and $\xi_2$ where the speed of the background current linearly
changes.}
\label{f05}%
\end{figure}

The solution obtained should be in consistency with the energy
flux conservation \cite{MaiRusStep-2016, ChurErRousStep-2017},
which is derived in Appendix \ref{appA} in terms of the velocity
potential $\varphi$:
\begin{equation}
\label{FluxCons} %
V_2\left(1 - |R|^2\right) = V_1|T|^2.
\end{equation}

Substituting here the transformation coefficients $R$ and $T$ from
Eqs.~(\ref{Trans1}) and (\ref{Trans2}), we confirm that
Eq.~(\ref{FluxCons}) reduces to the identity.

To characterize the rate of energy flux transmission, one can
introduce the energy transmission factor
\begin{equation}
\label{Transfer} %
K_T = \frac{V_1}{V_2}\,|T|^2 \quad \stackrel{\hat\omega \; \to \;
0}{\longrightarrow} \quad \frac{4V_1/V_2}{\left(1 +
V_1/V_2\right)^2}.
\end{equation}

Then one can see that although the modulus of the transmission
coefficient is greater than one (see line 1 in Fig. \ref{f04}) the
total energy flux (\ref{FluxCons}) through the duct cross-section
conserves because the cross-section decreases in the transition
from the left to right domain, and the transmitted energy flux is
less than the incident one ($K_T < 1$).

\subsubsection{\label{sec:level4l2}Accelerating currents. Transformation
of upstream propagating incident wave}

If the incident wave arrives from plus infinity, then we set in
Eqs.~(\ref{SolA1}) and (\ref{SolA3}) its amplitude $C_2 = 1$, the
amplitude of reflected wave $C_1 = R$, and the amplitude of
transmitted wave $A_2 = T$, whereas $A_1 = 0$. Then from the
matching conditions we obtain
\begin{eqnarray}
B_1\,w_2(V_1^2) &+& B_2\,w_3(V_1^2) = T, \label{BCond11}\\
B_1\,w'_2(V_1^2) &+& B_2\,w'_3(V_1^2) = -\frac{{\rm
i}\,\hat\omega}{2V_1} \frac{T}{1-V_1}, \phantom{ww}
\label{BCond21}\\
B_1\,w_2(V_2^2) &+& B_2\,w_3(V_2^2) = 1 + R, \label{BCond31}\\
B_1\,w'_2(V_2^2) &+& B_2\,w'_3(V_2^2) = -\frac{{\rm
i}\,\hat\omega}{2V_2}\left(\frac{1}{1-V_2} -
\frac{R}{1+V_2}\right). \label{BCond41}
\end{eqnarray}

Solution of this set of equations is:
\begin{widetext}
\begin{eqnarray}
R &=&
\frac{1}{\Delta}\left\{\frac{\hat\omega^2\left[w_2(V_1^2)w_3(V_2^2)
- w_2(V_2^2)w_3(V_1^2)\right]}{4V_1V_2(1-V_1)(1-V_2)} -
w'_2(V_1^2)w'_3(V_2^2) + w'_2(V_2^2)w'_3(V_1^2) - {}\right.
\nonumber \\
{} && \left. \frac{{\rm
i}\,\hat\omega}{2}\left[\frac{w_2(V_1^2)w'_3(V_2^2) -
w'_2(V_2^2)w_3(V_1^2)}{V_1(1 - V_1)} - \frac{w_2(V_2^2)w'_3(V_1^2)
- w'_2(V_1^2)w_3(V_2^2)}{V_2(1-V_2)}\right]%
\vphantom{\frac{\hat\omega^2\left[w_2(V_1^2)w_3(V_2^2)
- w_2(V_2^2)w_3(V_1^2)\right]}{4V_1V_2(1+V_1)(1+V_2)}}%
\right\}, \label{Trans11} \\%
&& {} \nonumber \\
T &=& -\frac{{\rm i}\,\hat\omega}{\Delta}\frac{(1-V_1^2)^{{\rm
i}\, \hat\omega - 1}}{V_2(1-V_2^2)}\,\frac{\Gamma(1-{\rm i}\,
\hat\omega)}{\Gamma^2(1-{\rm i}\, \hat\omega/2)}\,, \label{Trans21}\\
&& {} \nonumber \\
B_1 &=& \frac{{\rm
i}\,\hat\omega}{\Delta}\frac{1}{V_2(1-V_2^2)}\left[\frac{{\rm
i}\,\hat\omega}{2V_1(1-V_1)}\,w_3(V_1^2) + w'_3(V_1^2)\right], \phantom{www} \label{Trans31}\\
&& {} \nonumber \\
B_2 &=& -\frac{{\rm
i}\,\hat\omega}{\Delta}\frac{1}{V_2(1-V_2^2)}\left[\frac{{\rm
i}\,\hat\omega}{2V_1(1-V_1)}\,w_2(V_1^2) + w'_2(V_1^2)\right],
\label{Trans41}
\end{eqnarray}
\end{widetext}
where $\Delta$ is the same as in Eq.~(\ref{Delta}).

In the long-wave approximation, $\hat\omega \to 0$, we obtain the
limiting values of transformation coefficients
\begin{equation}
\label{Reduct2}%
R = \frac{1 - V_2/V_1}{1 + V_2/V_1}, \quad T = 1 + R = \frac{2}{1
+ V_2/V_1}.
\end{equation}

These values again are purely real and agree with the
transformation coefficients derived in Ref.
\cite{ChurErRousStep-2017} for surface waves in a duct with the
stepwise change of cross-section and velocity profile.

This solution is also in consistency with the energy flux
conservation, which now takes the form:
\begin{equation}
\label{FluxCons2} %
V_1\left(1 - |R|^2\right) = V_2|T|^2.
\end{equation}

Substituting here the expressions for the transformation
coefficients, (\ref{Trans11}) and (\ref{Trans21}), we confirm that
Eq.~(\ref{FluxCons2}) reduces to the identity. The energy
transmission factor $K_T$ in the limit $\hat \omega \to 0$ remains
the same as in Eq.~(\ref{Transfer}).

The graphic of $|\Phi(\xi)|$ is presented in Fig.~\ref{f05}a) by
line 2. The plot was generated on the basis of solution
(\ref{SolA1})--(\ref{SolA3}) with $C_2 = 1$, $A_1 = 0$ and other
determined transformation coefficients $C_1 = R$ as per
Eq.~(\ref{Trans11}) and $A_2 = T$ as per Eq.~(\ref{Trans21}).
Coefficients $B_1$ and $B_2$ are given by Eqs.~(\ref{Trans31}) and
(\ref{Trans41}).

\subsubsection{\label{sec:level4l3}Wave transformation in a decelerating
sub-critical current}

The decelerating current can occur, for example, in a widening
duct. To calculate the transformation coefficients of waves in a
decelerating current with a piece-linear profile it is convenient
to choose the origin of coordinate frame such as shown in Fig.
\ref{f01}b).

The general solutions of the basic equation (\ref{NormEq}) in the
left and right domains beyond the interval $\xi_1 < \xi < \xi_2$
are the same as in Eqs. (\ref{SolA1}) and (\ref{SolA3}), whereas
in the transient domain the solution is given by
Eq.~(\ref{SolD2}).

To calculate the transformation coefficients one can repeat the
simple, but tedious calculations similar to the presented above.
The result shows that the expressions for the transformation
coefficients remain the same as in Eqs.
(\ref{Trans1})--(\ref{Delta}) for the co-current propagating
incident wave and Eqs. (\ref{Trans11})--(\ref{Trans41}), and
(\ref{Delta}) for the counter-current propagating incident wave,
but in both these cases $\hat\omega$ should be replaced by
$-\hat\omega$ and $w_i$ by $\tilde w_i$. The energy flux
Eq.~(\ref{FluxCons}) for the co-current propagating incident wave
or Eq.~(\ref{FluxCons2}) for the counter-current propagating
incident wave conserves in these cases too.

The graphics of $|\Phi(\xi)|$ are presented in Fig.~\ref{f05}b) by
line 1 for co-current propagating incident wave, and by line 2 for
counter-current propagating incident wave.

\subsection{\label{sec:level42}Wave transformation in a super-critical
current}

Assume now that the main current is super-critical everywhere,
$V_2 > V_1 > 1$. In this case, there are no upstream propagating
waves. Indeed in such strong current even waves propagating with
the speed $-c_0$ in the frame moving with the water are pulled
downstream by the current whose speed $U > c_0$, therefore in the
immovable laboratory frame the speed of such ``counter-current''
propagating waves is $U - c_0 > 0$. Such waves possess a negative
energy (see, for instance, \cite{FabrStep-1998, MaiRusStep-2016,
ChurErRousStep-2017}). Thus, the problem statement can contain an
incident sinusoidal wave propagating only downstream from the $\xi
< \xi_1$ domain; the wave can be of either positive energy with
$\hat\omega = (V_1 + 1)\kappa_1$ or negative energy with
$\hat\omega = (V_1 - 1)\kappa_2$. After transformation on the
inhomogeneous current in the interval $\xi_1 < \xi < \xi_2$ these
waves produce two transmitted waves in the right domain, $\xi >
\xi_2$ one of positive energy and another of negative energy.
Below we consider such transformation in detail.

In the super-critical case the basic equation (\ref{NormEq}) is
also regular and its coefficients do not turn to zero. To
construct its solutions in the intermediate domain $\xi_1\le \xi
\le \xi_2$ it is convenient to re-write the equation in slightly
different form:
\begin{equation}%
\label{hypergeom1}%
\eta(1-\eta)\frac{d^2\Psi}{d\eta^2} + \left[1-(2 \mp {\rm
i}\,\hat\omega)\eta\right]\frac{d\Psi}{d\eta} \pm \frac{{\rm
i}\,\hat\omega}{2}\left(1 \mp \frac{{\rm
i}\,\hat\omega}{2}\right)\Psi = 0,
\end{equation}
where $\eta = 1/\zeta$, $\Psi(\eta) = \eta^{\pm{\rm
i}\,\hat\omega/2} \Phi$, upper signs pertain to the accelerating
current, and lower signs -- to the decelerating currents.

Solutions of Eq.~(\ref{NormEq}) in the domains where the current
speed is constant are
\begin{eqnarray}
\Phi(\xi) &=& A_1\mbox{e}^{{\rm i}\kappa_1(\xi-\xi_1)} +
A_2\mbox{e}^{{\rm i}\kappa_2(\xi-\xi_1)}, \hspace{30mm} \xi \le \xi_1, \label{Sol11}\\
\Phi(\xi) &=& C_1\mbox{e}^{{\rm i}\kappa_3(\xi-\xi_2)} +
C_2\mbox{e}^{{\rm i}\kappa_4(\xi-\xi_2)}, \hspace{30mm} \xi \ge
\xi_2,
\label{Sol31} %
\end{eqnarray}
were $\kappa_1 = \hat\omega/(V_1 + 1)$, $\kappa_2 =
\hat\omega/(V_1 - 1)$, $\kappa_3 = \hat\omega/(V_2 + 1)$,
$\kappa_4 = \hat\omega/(V_2 - 1)$.

In the intermediate domain $\xi_1\le \xi \le \xi_2$ the solution
of hypergeometric Eq.~(\ref{hypergeom1}) in the case of
accelerating current is
\begin{equation}
\label{Sol21Ac}%
\Phi(\xi) = \xi^{{\rm i}\,\hat\omega}\left[B_1\breve
w_1\left(\xi^{-2}\right) + B_2\breve w_3
\left(\xi^{-2}\right)\right],
\end{equation}
where two linearly independent solutions of Eq.~(\ref{hypergeom1})
can be chosen in the form (see \S 6.4 in the book
\cite{Luke-1975}):
\begin{equation}
\label{Gauss11Ac}%
\breve w_1(\eta) =\; {_2F_1}(-{\rm i}\,\hat\omega/2,\,1 - {\rm
i}\,\hat\omega/2;\,1;\eta), \quad \breve w_3(\eta) =\;
{_2F_1}({-\rm i}\hat\omega/2, 1-{\rm i}\,\hat\omega/2; 1 - {\rm
i}\,\hat\omega; 1-\eta)
\end{equation}
with the Wronskian %
\begin{equation}
\label{Wronsk2Ac} %
\breve W = \breve w'_1(\eta)\breve w_3(\eta) - \breve w_1(\eta)
\breve w'_3(\eta) = \frac{(1-\eta)^{{\rm i}\,\hat\omega-1}}{\eta}
\frac{\Gamma(1-{\rm i}\,\hat\omega)}{\Gamma(-{\rm
i}\,\hat\omega/2)\Gamma(1-{\rm i}\,\hat\omega/2)}.
\end{equation}

In the case of decelerating current the solution of hypergeometric
Eq.~(\ref{hypergeom1}) is
\begin{equation}
\label{Sol21De}%
\Phi(\xi) = (-\xi)^{-{\rm i}\,\hat\omega}\left[B_1\hat
w_1\left(\xi^{-2}\right) + B_2\hat w_3
\left(\xi^{-2}\right)\right],
\end{equation}
and linearly independent solutions can be chosen in the form:
\begin{equation}
\label{Gauss11De}%
\hat w_1(\eta) =\; {_2F_1}({\rm i}\,\hat\omega/2,\,1 + {\rm
i}\,\hat\omega/2;\,1;\eta), \quad \hat w_3(\eta) =\; {_2F_1}({\rm
i}\hat\omega/2, 1+{\rm i}\,\hat\omega/2; 1 + {\rm i}\,\hat\omega;
1-\eta)
\end{equation}
with the Wronskian %
\begin{equation}
\label{Wronsk2De} %
\hat W = \hat w'_1(\eta)\hat w_3(\eta) - \hat w_1(\eta) \hat
w'_3(\eta) = \frac{(1-\eta)^{-{\rm i}\,\hat\omega-1}}{\eta}
\frac{\Gamma(1+{\rm i}\,\hat\omega)}{\Gamma({\rm
i}\,\hat\omega/2)\Gamma(1+{\rm i}\,\hat\omega/2)}.
\end{equation}

\subsubsection{\label{sec:level421}Transformation of a positive-energy
wave in an accelerating current}

Consider first transformation of a positive energy incident wave
(see line 3 in Fig.~\ref{f03}) with the unit amplitude ($A_1 = 1$,
$A_2 = 0$). Matching the solutions in different current domains
and using the chain rule $d/d\xi = -2\xi^{-3}d/d\eta$, we obtain
at $\xi = \xi_1$:
\begin{eqnarray}
B_1\,\breve w_1(V_1^{-2}) + B_2\,\breve w_3(V_1^{-2}) &=& V_1^{-{\rm i}\,\hat\omega}, \label{Match11}\\
B_1\,\breve w'_1(V_1^{-2}) + B_2\,\breve w'_3(V_1^{-2}) &=&
\frac{{\rm i}\,\hat\omega}{2}\frac{V_1^{2-{\rm i}\,\hat\omega}}{V_1 + 1}, \label{Match12}%
\end{eqnarray}
where prime stands for a derivative of a corresponding function
with respect to its entire argument.

Similarly from the matching conditions at $\xi = \xi_2$ we obtain:
\begin{widetext}
\begin{eqnarray}
C_1 + C_2 &=& V_2^{{\rm i}\,\hat\omega}\left[B_1\,\breve
w_1(V_2^{-2}) + B_2\,\breve w_3(V_2^{-2})\right], \label{Match13}\\
(V_2 - 1)C_1 - (V_2 + 1)C_2 &=& -\frac{2{\rm
i}}{\hat\omega}\,V_2^{{\rm i}\,\hat\omega -
2}(V_2^2-1)\left[B_1\,\breve w'_1(V_2^{-2})
\right. + \left. B_2\,\breve w'_3(V_2^{-2})\right]. \phantom{www}\label{Match14}%
\end{eqnarray}\\
\end{widetext}

From Eqs.~(\ref{Match11}) and (\ref{Match12}) we find
\begin{widetext}
\begin{eqnarray}
B_1 &=& -\frac{\Gamma(-{\rm i}\,\hat\omega/2)\, \Gamma(1-{\rm
i}\,\hat\omega/2)}{\Gamma(1-{\rm i}\,\hat\omega)}\, V_1^{{\rm
i}\,\hat\omega-2}\left(V_1^2 - 1\right)^{1-{\rm i}\,\hat\omega}
\left[\frac{\breve w'_3(V_1^{-2})}{V_1^2} - \frac{{\rm
i}\,\hat\omega\,\breve w_3(V_1^{-2})}{2(V_1+1)}\right],
\label{CoefB1p}\\
B_2 &=& \frac{\Gamma(-{\rm i}\,\hat\omega/2)\, \Gamma(1-{\rm
i}\,\hat\omega/2)}{\Gamma(1-{\rm i}\,\hat\omega)}\, V_1^{{\rm
i}\,\hat\omega - 2}\left(V_1^2 - 1\right)^{1-{\rm i}\,\hat\omega}
\left[\frac{\breve w'_1(V_1^{-2})}{V_1^2} - \frac{{\rm
i}\,\hat\omega\,\breve w_1(V_1^{-2})}
{2(V_1+1)}\right]. \label{CoefB2p}%
\end{eqnarray}
\end{widetext}

Substituting these in Eqs.~(\ref{Match13}) and (\ref{Match14}), we
find the transmission coefficients for the positive energy mode
$T_p \equiv C_1$ and negative energy mode $T_n \equiv C_2$:
\begin{widetext}
$$%
T_p = -\frac{\Gamma^2(-{\rm i}\,\hat\omega/2)} {2\Gamma(1-{\rm
i}\,\hat\omega)}\,V_1^{{\rm i}\,\hat\omega - 2}V_2^{{\rm
i}\,\hat\omega - 1}\left(V_1^2 - 1\right)^{1-{\rm
i}\,\hat\omega}\!\left(V_2^2 - 1\right) \times {} \nonumber %
$$%
$$%
\left\{\frac{\breve w'_1(V_1^{-2})\breve w'_3(V_2^{-2})\! -
\!\breve w'_1(V_2^{-2})\breve w'_3(V_1^{-2})}{V_1^2V_2^2} \right.
+ \frac{\hat\omega^2}{4}\,\frac{\breve w_1(V_1^{-2})\breve
w_3(V_2^{-2})\! - \!\breve w_1(V_2^{-2})\breve
w_3(V_1^{-2})}{(V_1+1)(V_2-1)} + {} \nonumber
$$%
\begin{equation}%
\label{CoefTp}%
\left.\frac{{\rm i}\,\hat\omega}{2}\left[ \frac{\breve
w'_1(V_1^{-2})\breve w_3(V_2^{-2})\! - \!\breve
w_1(V_2^{-2})\breve w'_3(V_1^{-2})}{V_1^2(V_2-1)} - \frac{\breve
w_1(V_1^{-2})\breve w'_3(V_2^{-2})\! - \!\breve
w'_1(V_2^{-2})\breve w_3(V_1^{-2})}{V_2^2(V_1+1)} \right]\right\},
\end{equation}

$$%
T_n = \frac{\Gamma^2(-{\rm i}\,\hat\omega/2)} {2\Gamma(1-{\rm
i}\,\hat\omega)}\,V_1^{{\rm i}\,\hat\omega - 2}V_2^{{\rm
i}\,\hat\omega - 1}\left(V_1^2 - 1\right)^{1-{\rm
i}\,\hat\omega}\!\left(V_2^2 - 1\right) \times {}%
$$%
$$%
\left\{\frac{\breve w'_1(V_1^{-2})\breve w'_3(V_2^{-2})\! -
\!\breve w'_1(V_2^{-2})\breve w'_3(V_1^{-2})}{V_1^2V_2^2} \right.
- \frac{\hat\omega^2}{4}\,\frac{\breve w_1(V_1^{-2})\breve
w_3(V_2^{-2})\! - \!\breve w_1(V_2^{-2})\breve
w_3(V_1^{-2})}{(V_1+1)(V_2+1)} - {} \nonumber
$$%
\begin{equation}%
\label{CoefTn}%
\left.\frac{{\rm i}\,\hat\omega}{2}\left[ \frac{\breve
w'_1(V_1^{-2})\breve w_3(V_2^{-2})\! - \!\breve
w_1(V_2^{-2})\breve w'_3(V_1^{-2})}{V_1^2(V_2+1)} + \frac{\breve
w_1(V_1^{-2})\breve w'_3(V_2^{-2})\! - \!\breve
w'_1(V_2^{-2})\breve w_3(V_1^{-2})}{V_2^2(V_1+1)} \right]\right\}.
\end{equation}%
\end{widetext}

\bigskip

The modules of transformation coefficients $|T_p|$ and $|T_n|$
together with the intermediate coefficients of wave excitation in
the transient zone $|B_1|$ and $|B_2|$ are shown below in
Fig.~\ref{f06}a) as functions of dimensionless frequency
$\hat\omega$ for the particular values of $V_1 = 1.1$ and $V_2 =
1.9$. Qualitatively similar graphics were obtained for other
values of $V_1$ and $V_2$.

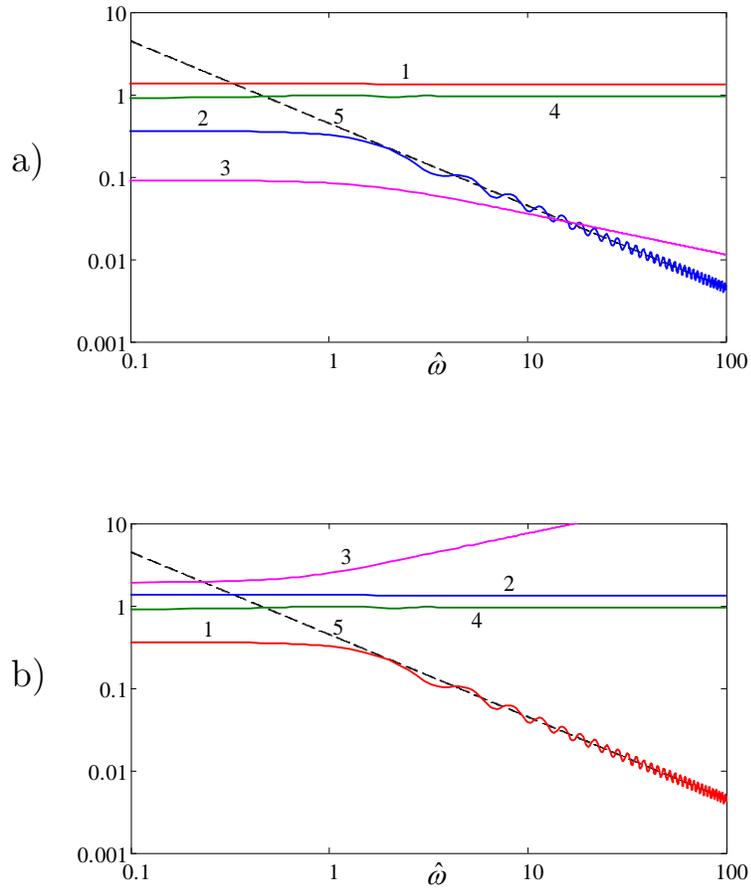
\begin{figure}[h]
\centering
\includegraphics[width=90mm]{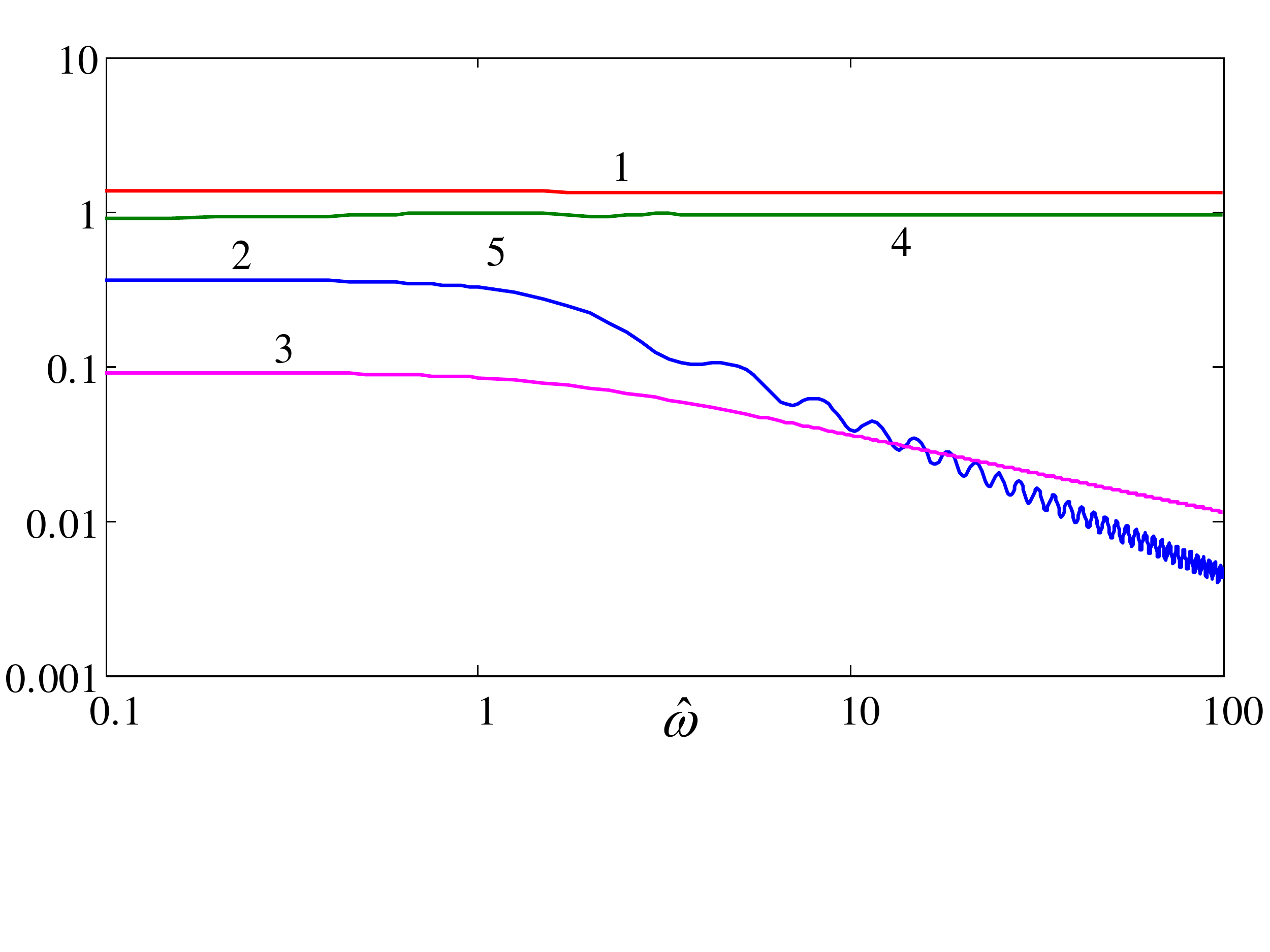}
\includegraphics[width=90mm]{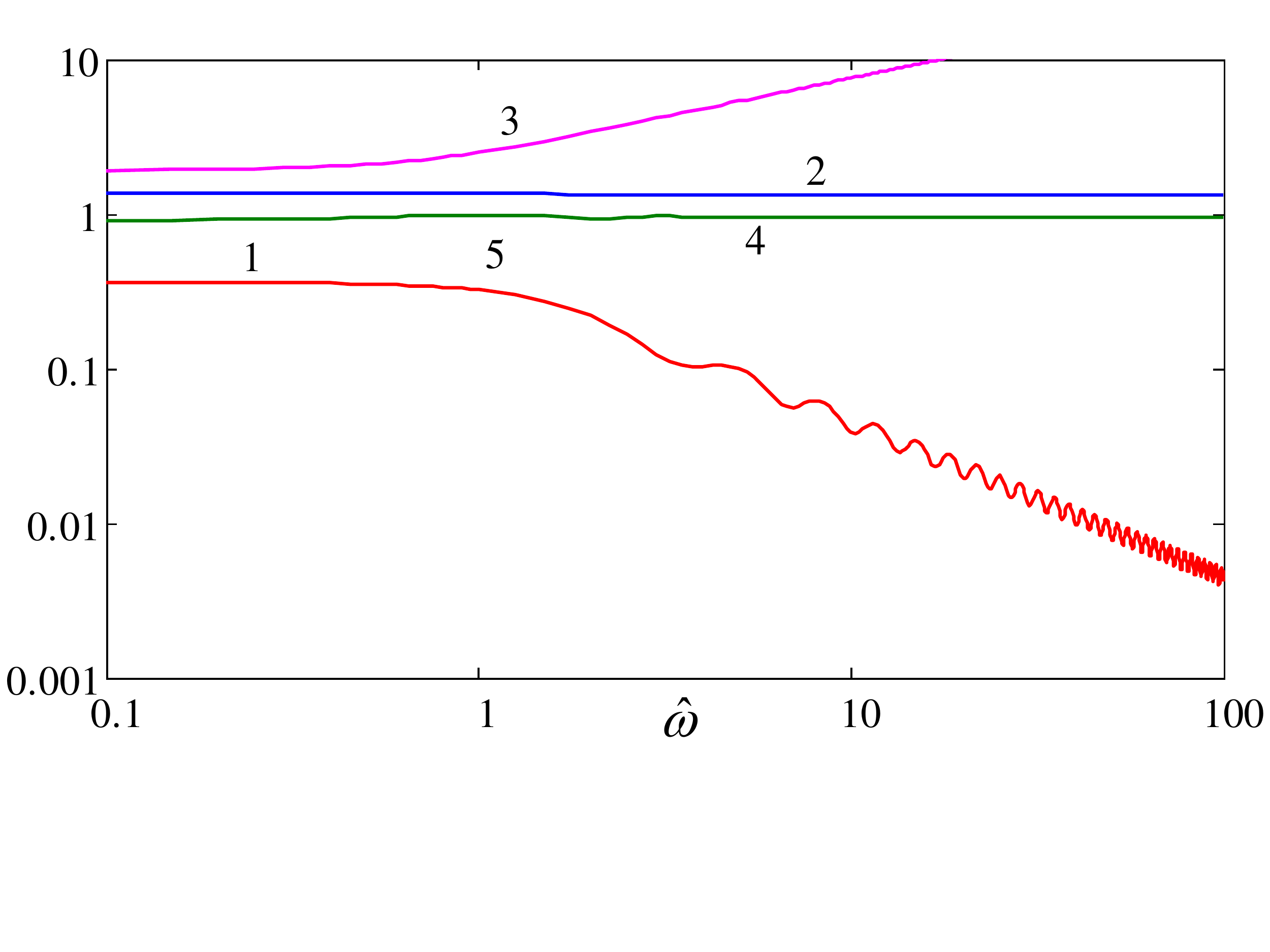}
\begin{picture}(300,6)%
\put(0,335){{\large a)}}%
\put(0,140){{\large b)}}%
\end{picture}
\vspace*{-23mm}%
\caption{(color online) Modules of transformation coefficient as
functions of dimensionless frequency $\hat\omega$ when a positive
energy wave scatters (panel a) and negative energy wave scatters
(panel b) in the current with $V_1 = 1.1$, $V_2 = 1.9$. Line 1 --
$|T_p|$, line 2 -- $|T_n|$, line 3 -- $|B_1|$, line 4 -- $|B_2|$.
Dashed lines 5 represent the asymptotics for $|T_n| \sim
\hat\omega^{-1}$ in panel a) and for $|T_p|
\sim \hat\omega^{-1}$ in panel b).} %
\label{f06}%
\end{figure}

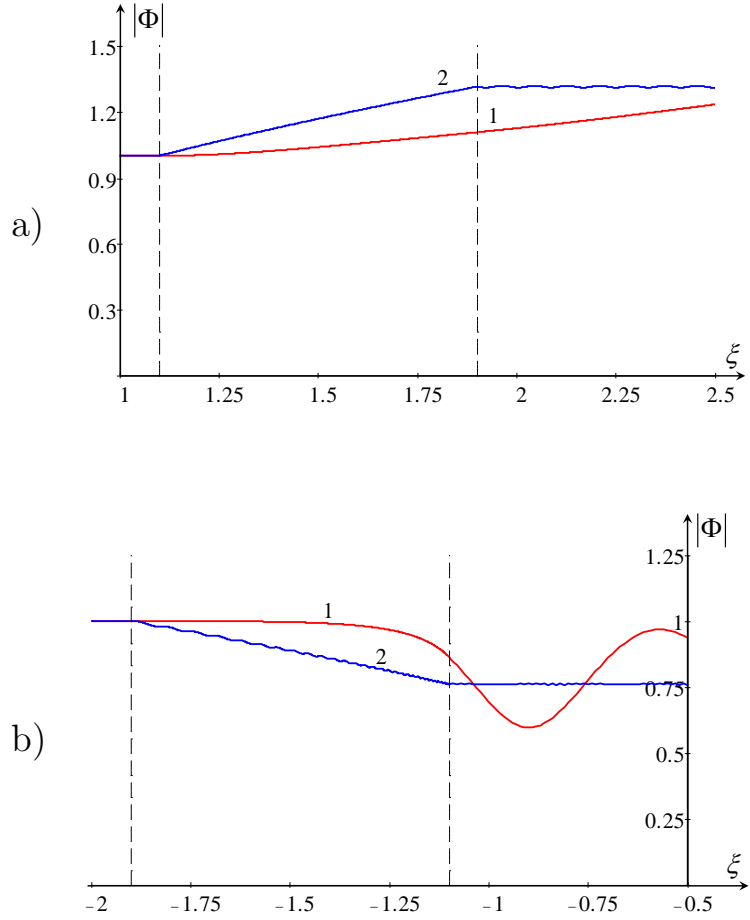
\begin{figure}[h]
\centering
\includegraphics[width=90mm]{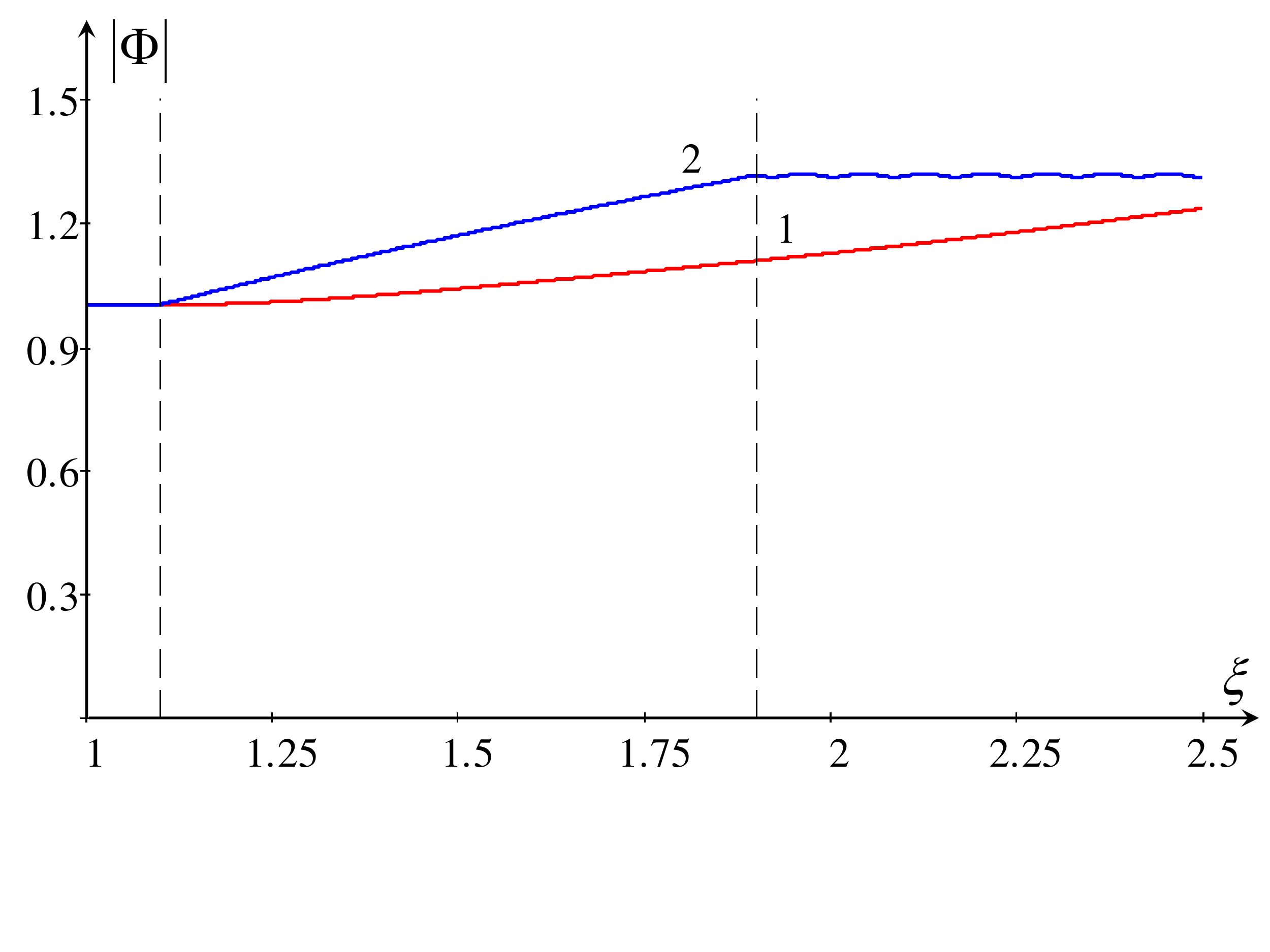}
\includegraphics[width=90mm]{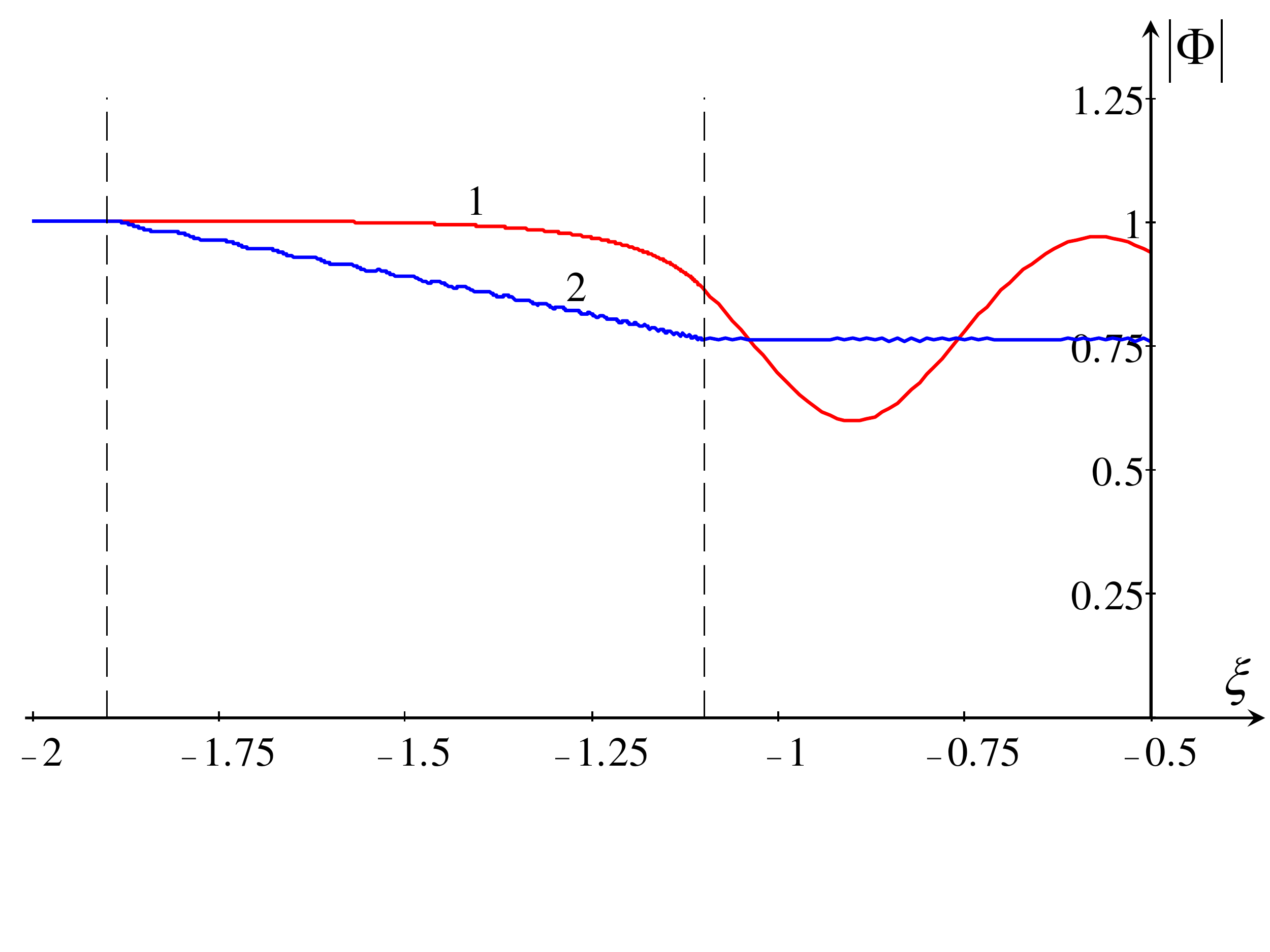}
\begin{picture}(300,6)%
\put(0,315){{\large a)}}%
\put(0,120){{\large b)}}%
\end{picture}
\vspace*{-20mm}%
\caption{(color online) Module of function $\Phi(\xi)$ for the
scattering of positive- and negative-energy waves in accelerating
with $V_1 = 1.1$ and $V_2 = 1.9$ (frame a) and decelerating with
$V_1 = 1.9$ and $V_2 = 1.1$ (frame b) super-critical currents for
two particular values of frequency, $\hat\omega = 1$ (line 1), and
$\hat\omega = 100$ (line 2).}
\label{f07}%
\end{figure}

In Fig.~\ref{f07}a) we present graphics of $|\Phi(\xi)|$ as per
Eqs.~(\ref{Sol11})--(\ref{Sol21Ac}) for $A_1 = 1$,  $A_2 = 0$,
$C_1 = T_p$ as per Eq.~(\ref{CoefTp}), and $C_2 = T_n$  as per
Eq.~(\ref{CoefTn}). Coefficients $B_1$ and $B_2$ are given by
Eqs.~(\ref{CoefB1p}) and (\ref{CoefB2p}). The plot was generated
for two particular values of frequency, $\hat\omega = 1$ (line 1),
and $\hat\omega = 100$ (line 2).

The transmission coefficients are in consistency with the energy
flux conservation law which has the following form:
\begin{equation}
\label{EnFluxCons2}%
J = \frac{2\hat\omega}{V_1} = \frac{2\hat\omega}{V_2}\left(|T_p|^2
-
|T_n|^2\right) \quad \mbox{or} \quad |T_p|^2 - |T_n|^2 = \frac{V_2}{V_1}.%
\end{equation}

\begin{figure}[h]
\centering
\includegraphics[width=90mm]{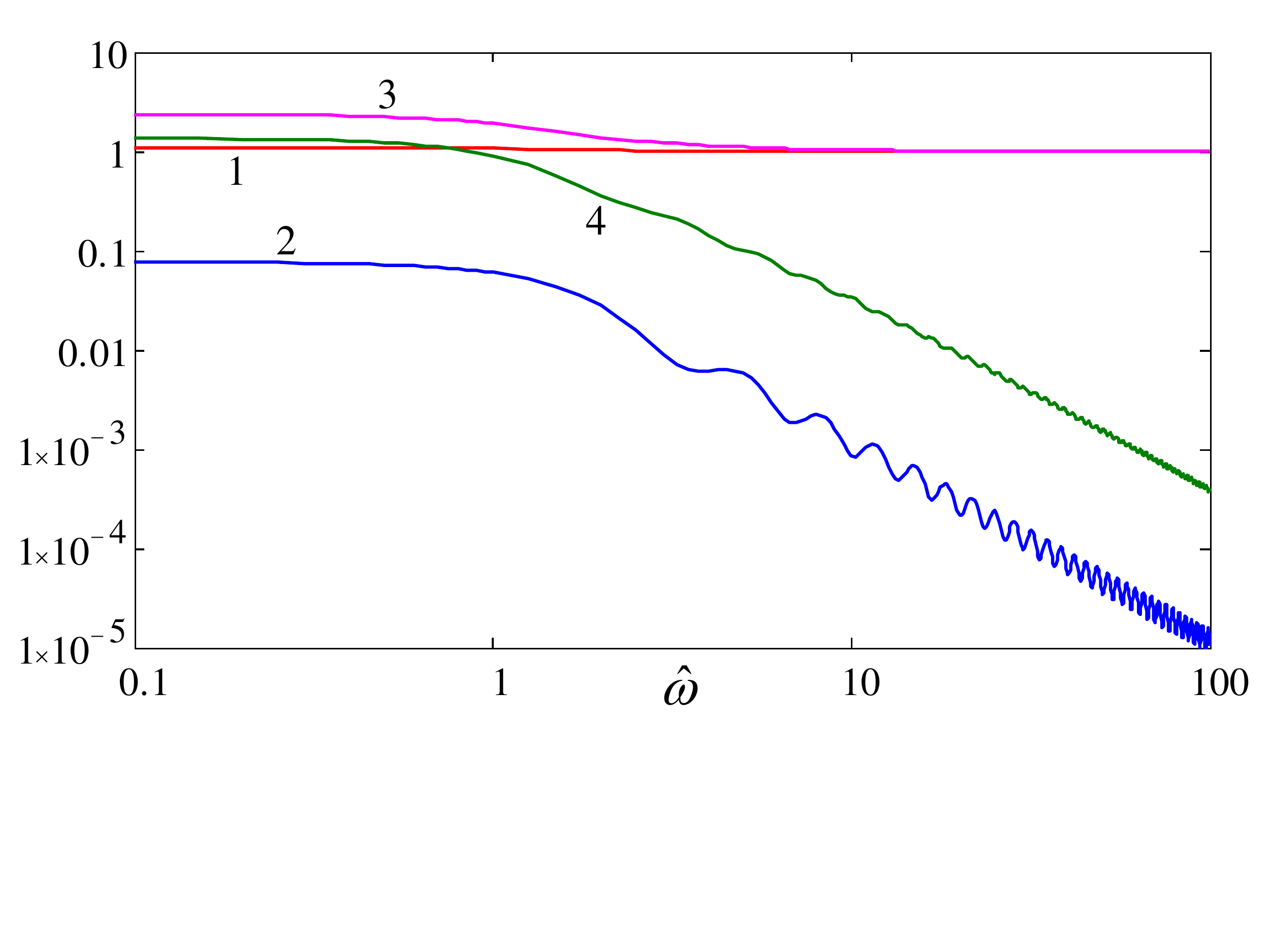}
\vspace*{-20mm}%
\caption{(color online) The dependences of energy transmission
factors $K_{Tp}$ and $K_{Tn}$ on the frequency for a relatively
small increase of current speed ($V_1 = 1.1$, $V_2 = 1.9$), lines
1 and 2 respectively, and a large increase of current speed ($V_1
= 1.1$, $V_2 = 8.0$), lines 3 and 4 respectively. Inclined dashed
lines show the asymptotic dependences $K_{Tn} \sim
\hat\omega^{-2}$.}
\label{f08}%
\end{figure}

If we introduce two energy transmission factors, for positive- and
negative-energy waves,
\begin{equation}
\label{Transfer1} %
K_{Tp} = \frac{V_1}{V_2}\,|T_p|^2 \ \ \ {\rm and} \ \ \
K_{Tn} = \frac{V_1}{V_2}\,|T_n|^2,
\end{equation}
then we can see that both waves grow in such a manner that $K_{Tp}
- K_{Tn} = 1$. This means that the positive-energy wave not only
dominates in the right domain (cf. lines 1 and 2 in
Fig.~\ref{f06}a), but it also carries a greater energy flux than
the incident one. Moreover, with a proper choice of $V_1$ and
$V_2$ even the energy flux of negative-energy wave can become
greater by modulus than that of incident wave, $K_{Tn} > 1$. Then
we have $K_{Tp} > K_{Tn} > 1$. Figure \ref{f08} illustrates the
dependences of energy transmission factors on the frequency for
relatively small increase of current speed ($V_1 = 1.1$, $V_2 =
1.9$) and big increase of current speed ($V_1 = 1.1$, $V_2 =
8.0$). In the latter case both $K_{Tp}$ and $K_{Tn}$ are greater
than 1 in a certain range of frequencies $\hat\omega <
\hat\omega_c$.

In the long-wave approximation, $\hat\omega \to 0$ (see Appendix
\ref{appC}) we obtain (cf. \cite{ChurErRousStep-2017}):
\begin{equation}
\label{Reduct3}%
T_p = \frac{1 + V_1/V_2}{2V_1/V_2}, \quad T_n = -\frac{1 -
V_1/V_2}{2V_1/V_2}, \quad K_{Tp} = \frac{(1 +
V_1/V_2)^2}{4V_1/V_2}\,, \quad K_{Tn} = \frac{(1 -
V_1/V_2)^2}{4V_1/V_2}.
\end{equation}

\subsubsection{\label{sec:level422}Transformation of negative-energy
wave in an accelerating current}

Consider now transformation of a negative energy incident wave
(see line 4 in Fig.~\ref{f03}) with unit amplitude ($A_1 = 0$,
$A_2 = 1$). From the matching conditions at $\xi = \xi_1$ we
obtain:
\begin{eqnarray}
B_1\,\breve w_1(V_1^{-2}) + B_2\,\breve
w_3(V_1^{-2}) &=& V_1^{-{\rm i}\,\hat\omega}, \label{Match21}\\
B_1\,\breve w'_1(V_1^{-2}) + B_2\,\breve w'_3(V_1^{-2}) &=&
-\frac{{\rm i}\,\hat\omega}{2}
\frac{V_1^{2-{\rm i}\,\hat\omega}}{V_1 - 1}. \label{Match22}%
\end{eqnarray}
The matching conditions at $\xi = \xi_2$ remain the same as in
Eqs. (\ref{Match13}) and (\ref{Match14}).

From Eqs.~(\ref{Match21}) and (\ref{Match22}) we find
\begin{widetext}%
\begin{eqnarray}
B_1 &=& -\frac{\Gamma(-{\rm i}\,\hat\omega/2)\, \Gamma(1-{\rm
i}\,\hat\omega/2)}{\Gamma(1-{\rm i}\,\hat\omega)}\, V_1^{{\rm
i}\,\hat\omega-2}\left(V_1^2 - 1\right)^{1-{\rm i}\,\hat\omega}
\left[\frac{\breve w'_3(V_1^{-2})}{V_1^2} + \frac{{\rm
i}\,\hat\omega\,
\breve w_3(V_1^{-2})}{2(V_1-1)}\right], \label{CoefB1n}\\
B_2 &=& \frac{\Gamma(-{\rm i}\,\hat\omega/2)\, \Gamma(1-{\rm
i}\,\hat\omega/2)}{\Gamma(1-{\rm i}\,\hat\omega)}\, V_1^{{\rm
i}\,\hat\omega - 2}\left(V_1^2 - 1\right)^{1-{\rm i}\,\hat\omega}
\left[\frac{\breve w'_1(V_1^{-2})}{V_1^2} + \frac{{\rm
i}\,\hat\omega\,\breve w_1(V_1^{-2})}{2(V_1-1)}\right]. \label{CoefB2n}%
\end{eqnarray}
\end{widetext}%

\bigskip

Substituting these in Eqs.~(\ref{Match13}) and (\ref{Match14}), we
find the transmission coefficients for the positive energy mode
$T_p \equiv C_1$ and negative energy mode $T_n \equiv C_2$:
\begin{widetext}%
$$%
T_p = -\frac{\Gamma^2(-{\rm i}\,\hat\omega/2)} {2\Gamma(1-{\rm
i}\,\hat\omega)}\,V_1^{{\rm i}\,\hat\omega - 2}V_2^{{\rm
i}\,\hat\omega - 1}\left(V_1^2 - 1\right)^{1-{\rm
i}\,\hat\omega}\!\left(V_2^2 - 1\right) \times {} \nonumber %
$$%
$$%
\left\{\frac{\breve w'_1(V_1^{-2})\breve w'_3(V_2^{-2})\! -
\!\breve w'_1(V_2^{-2})\breve w'_3(V_1^{-2})}{V_1^2V_2^2} \right.
- \frac{\hat\omega^2}{4}\,\frac{\breve w_1(V_1^{-2})\breve
w_3(V_2^{-2})\! - \!\breve w_1(V_2^{-2})\breve
w_3(V_1^{-2})}{(V_1-1)(V_2-1)} + {} \nonumber
$$%
\begin{equation}%
\label{CoefTpn}%
\left.\frac{{\rm i}\,\hat\omega}{2}\left[ \frac{\breve
w'_1(V_1^{-2})\breve w_3(V_2^{-2})\! - \!\breve
w_1(V_2^{-2})\breve w'_3(V_1^{-2})}{V_1^2(V_2-1)} + \frac{\breve
w_1(V_1^{-2})\breve w'_3(V_2^{-2})\! - \!\breve
w'_1(V_2^{-2})\breve w_3(V_1^{-2})}{V_2^2(V_1-1)} \right]\right\},
\end{equation}

$$%
T_n = \frac{\Gamma^2(-{\rm i}\,\hat\omega/2)} {2\Gamma(1-{\rm
i}\,\hat\omega)}\,V_1^{{\rm i}\,\hat\omega - 2}V_2^{{\rm
i}\,\hat\omega - 1}\left(V_1^2 - 1\right)^{1-{\rm
i}\,\hat\omega}\!\left(V_2^2 - 1\right) \times {}%
$$%
$$%
\left\{\frac{\breve w'_1(V_1^{-2})\breve w'_3(V_2^{-2})\! -
\!\breve w'_1(V_2^{-2})\breve w'_3(V_1^{-2})}{V_1^2V_2^2} \right.
+ \frac{\hat\omega^2}{4}\,\frac{\breve w_1(V_1^{-2})\breve
w_3(V_2^{-2})\! - \!\breve w_1(V_2^{-2})\breve
w_3(V_1^{-2})}{(V_1-1)(V_2+1)} - {}
$$%
\begin{equation}%
\label{CoefTnn}%
\left.\frac{{\rm i}\,\hat\omega}{2}\left[ \frac{\breve
w'_1(V_1^{-2})\breve w_3(V_2^{-2})\! - \!\breve
w_1(V_2^{-2})\breve w'_3(V_1^{-2})}{V_1^2(V_2+1)} - \frac{\breve
w_1(V_1^{-2})\breve w'_3(V_2^{-2})\! - \!\breve
w'_1(V_2^{-2})\breve w_3(V_1^{-2})}{V_2^2(V_1-1)} \right]\right\}.
\end{equation}
\end{widetext}%

\bigskip

The modules of transformation coefficients $|T_p|$ and $|T_n|$
together with the intermediate coefficients of wave excitation in
the transient zone $|B_1|$ and $|B_2|$ are shown in
Fig.~\ref{f06}b) as functions of dimensionless frequency
$\hat\omega$ for the particular values of $V_1 = 1.1$ and $V_2 =
1.9$. Qualitatively similar graphics were obtained for other
values of $V_1$ and $V_2$. The graphic of $|\Phi(\xi)|$ is the
same as the graphic shown in Fig.~\ref{f07}a) for the case of
scattering of positive-energy incident wave.

The transmission coefficients are again in consistency with the
energy flux conservation law which now has the following form:
\begin{equation}
\label{EnFluxCons3} %
J = -\frac{2\hat\omega}{V_1} =
-\frac{2\hat\omega}{V_2}\left(|T_n|^2 -
|T_p|^2\right) \quad \mbox{or} \quad |T_n|^2 - |T_p|^2 = \frac{V_2}{V_1}.%
\end{equation}

As follows from this equation, the energy flux $J$ is negative
everywhere, and the negative energy wave dominates in the right
domain (cf. lines 1 and 2 in Fig.~\ref{f06}b). Both transmitted
waves grow in a such manner that the energy transmission factors
(see Eq.~(\ref{Transfer1})) obey the equality $K_{Tn} - K_{Tp} =
1$. Thus, the negative-energy wave not only dominates in the right
domain, but also carries a greater energy flux than the incident
wave. At a certain relationship between $V_1$ and $V_2$ the energy
fluxes of positive- and negative-energy waves can be greater on
absolute value than that of incident wave, then we have $K_{Tn} >
K_{Tp} > 1$.

In the long-wave approximation, $\hat\omega \to 0$, we obtain (see
Appendix \ref{appC}):
\begin{equation}
\label{Reduct4}%
T_p = -\frac{1 - V_1/V_2}{2V_1/V_2}, \quad T_n = \frac{1 +
V_1/V_2}{2V_1/V_2}, \quad K_{Tp} = \frac{(1 -
V_1/V_2)^2}{4V_1/V_2}\,,
 \quad K_{Tn} = \frac{(1 + V_1/V_2)^2}{4V_1/V_2},
\end{equation}
i.e., in comparison with Eqs.~(\ref{Reduct3}), the energy
transmission factors are interchanged. The values of transmission
coefficients are purely real, but now $T_p < 0$ and $T_n > 0$;
they are in agreement with results derived in Ref.
\cite{ChurErRousStep-2017}.

\subsubsection{\label{sec:level423}Wave transformation in a
decelerating super-critical current}

In the case of decelerating super-critical current ($V_1 > V_2 >
1$) the configuration of the incident wave and current is the same
as above in this subsection. Again there is no reflected wave in
the left domain $\xi < \xi_1$ and there are two transmitted waves
in the right domain $\xi > \xi_2$.

The main equation describing wave propagation is the same as Eq.
(\ref{hypergeom1}) with only formal replacement of $\hat\omega$ by
$-\hat\omega$. The general solutions of the basic equation
(\ref{NormEq}) in the left and right domains beyond the interval
$\xi_1 < \xi < \xi_2$ are the same as in Eqs. (\ref{Sol11}) and
(\ref{Sol31}), whereas in the transient domain the solution is
given by Eq.~(\ref{Sol21De}).

To calculate the transformation coefficients one can repeat the
simple, but tedious calculations similar to those presented above.
The result shows that the expressions for the transformation
coefficients remain the same as in Eqs. (\ref{CoefTp}) and
(\ref{CoefTn}) for the incident wave of positive energy and Eqs.
(\ref{CoefTpn}) and (\ref{CoefTnn}) for the incident wave of
negative energy, but in both these cases $\hat\omega$ should be
replaced by $-\hat\omega$ and $\breve w_i$ by $\hat w_i$. The
corresponding energy fluxes for the incident waves of positive and
negative energies conserve, and Eqs.~(\ref{EnFluxCons2}) and
Eq.~(\ref{EnFluxCons3}) remain the same in these cases too.

The graphics of $|\Phi(\xi)|$ for the scattering of positive- and
negative-energy waves are also the same in the decelerating
currents. They are shown in Fig.~\ref{f07}b) in the subsubsection
\ref{sec:level421} for two particular values of frequency,
$\hat\omega = 1$ (line 1), and $\hat\omega = 100$ (line 2).

\subsection{\label{sec:level43}Wave transformation in trans-critical
accelerating currents $0 < V_1 < 1 < V_2$}

The specific feature of a trans-critical current is the transition
of the background current speed $U(x)$ through the critical wave
speed $c_0$. In this case the basic equation (\ref{NormEq})
contains a singular point where $V = 1$, therefore the behavior of
solutions in the vicinity of this point should be thoroughly
investigated.

The general solution of Eq.~(\ref{NormEq}) in different intervals
of $\xi$-axis can be presented in the form: %
\begin{eqnarray}
\Phi(\xi) &=& A_1e^{{\rm i}\,\kappa_1(\xi-\xi_1)} + A_2e^{-{\rm
i}\,\kappa_2(\xi-\xi_1)}, \hspace{4mm} \xi < \xi_1, \label{SD1} \\
\Phi(\xi) &=& B_1w_2(\xi^2) + B_2w_3(\xi^2), \hspace{15mm} \xi_1 < \xi < 1, \label{SD2} \\
\Phi(\xi) &=& \xi^{{\rm i}\,\hat\omega}\left[\breve B_1\,\breve
w_1(\xi^{-2}) + \breve B_2\,\breve w_3(\xi^{-2})\right],
\hspace{1mm} 1 < \xi < \xi_2, \phantom{ww} \label{SD3} \\
\Phi(\xi) &=& C_1\,e^{{\rm i}\,\kappa_3(\xi-\xi_2)} +
C_2\,e^{-{\rm i}\,\kappa_4(\xi-\xi_2)}, \hspace{5mm} \xi > \xi_2,
\label{SD4}
\end{eqnarray}
where $\kappa_1 = \hat\omega/(1 + V_1)$, $\kappa_2 = \hat\omega/(1
- V_1)$, $\kappa_3 = \hat\omega/(V_2 + 1)$, and $\kappa_4 =
\hat\omega/(V_2 - 1)$.

To pass through the singular point where $V(\xi) = 1$, let us
consider asymptotic behavior of solution $\Phi(\xi)$ in the
vicinity of the point $\xi = 1$. To
this end we use the formula valid for $|\arg(1-x)|<\pi$ (see
\citep{GR-2007}, formula 9.131.2.):
\begin{widetext}%
\begin{eqnarray}
{_2F_1}(a, b; c; x) &=&
\frac{\Gamma(c)\,\Gamma(c-a-b)}{\Gamma(c-a)\, \Gamma(c-b)}\,
{_2F_1}(a, b; a+b-c+1; 1-x) + {} \nonumber \\%
\phantom{_2F_1(a, b; c; x)} && \frac{\Gamma(c)\, \Gamma(a + b -
c)}{\Gamma(a)\, \Gamma(b)}\, (1 - x)^{c-a-b}\, {_2F_1}(c-a, c-b;
c-a-b+1; 1-x). \phantom{123} \label{HyperFuncEq}%
\end{eqnarray}
\end{widetext}%

With the help of this formula let us present the asymptotic
expansion of functions (\ref{SD2}) and (\ref{SD3}), keeping only
the leading terms:
\begin{widetext}%
\begin{eqnarray}
\hspace{-8mm} \Phi(\xi) &=& B_2 + \frac{\Gamma({\rm
i}\,\hat\omega)\, B_1}{\Gamma^2(1+{\rm i}\,\hat\omega/2)} +
\frac{\Gamma(-{\rm i}\,\hat\omega)\, B_1}{\Gamma^2(1-{\rm
i}\,\hat\omega/2)}\,(1-\xi^2)^{{\rm i}\,\hat\omega} +
O(1-\xi^2), \quad \xi^2 \to 1_{-0}, \label{Phi-m} \\%
{}&& \nonumber \\%
\hspace{-8mm} \Phi(\xi) &=& \breve B_2 +
\frac{\Gamma(1+{\rm i}\,\hat\omega)\,\breve B_1}{2\Gamma^2(1+{\rm
i}\,\hat\omega/2)} + \frac{\Gamma(1-{\rm i}\,\hat\omega)\,\breve
B_1}{2\Gamma^2(1-{\rm i}\,\hat\omega/2)}\,(\xi^2-1)^{{\rm
i}\,\hat\omega} + O(\xi^2-1), \quad \xi^2 \to  1_{+0}.
\label{Phi-p}
\end{eqnarray}
\end{widetext}%

\bigskip

As one can see from these formulae, for real $\hat\omega$
solutions contain fast oscillating functions from both sides of a
singular point $\xi^2 = 1$, which correspond to B-waves,
propagating against the current; these functions, however, remain
finite. To match the solutions across the singular point let us
take into consideration a small viscosity in Eq.~(\ref{LinEurEq}):
\begin{equation}
\label{ViscLinEurEq} %
\frac{\partial u}{\partial t} + \frac{\partial (Uu)}{\partial x} =
-g\frac{\partial \eta}{\partial x} + \nu\frac{\partial^2
u}{\partial x^2},
\end{equation}
where $\nu$ is the coefficient of kinematic viscosity.

Due to this correction to Eq.~(\ref{LinEurEq}) we obtain the
modified Eq.~(\ref{NormEq}) for $\Phi(\xi)$:
\begin{widetext}%
\begin{equation}
\label{NormViscEq} %
\nu V^2\frac{d^3\Phi}{d\xi^3} + V\left(1 - V^2 - {\rm
i}\,\nu\hat\omega\right)\frac{d^2\Phi}{d\xi^2} - \left[\left(1 +
V^2\right)V' - 2\,\mbox{i}\,\hat\omega
V^2\right]\frac{d\Phi}{d\xi} + V\hat\omega^2\Phi = 0.
\end{equation}
\end{widetext}%

Introducing a new variable $\zeta = \xi^2$ and bearing in mind
that $V(\xi) = \xi$ for the accelerating current, we re-write
Eq.~(\ref{NormViscEq}):
\begin{widetext}%
\begin{equation}
\label{NormViscEq1} %
2\nu \zeta^2\frac{d^3\Phi}{d\zeta^3} + \zeta\left[1 - \zeta +
\left(3 - {\rm
i}\,\hat\omega\right)\nu\right]\frac{d^2\Phi}{d\zeta^2} -
\left[\frac{{\rm i}\,\nu\hat\omega}{2} + \left(1 - {\rm
i}\,\hat\omega\right)\zeta\right]\frac{d\Phi}{d\zeta} +
\frac{\hat\omega^2}{4}\Phi = 0.
\end{equation}
\end{widetext}%

From this equation one can see that in the vicinity of the
critical point, where $|\zeta - 1| \sim \varepsilon \ll 1$, the
viscosity plays an important role, if $\nu \sim \varepsilon^2$.
Setting $\nu = \varepsilon^2/2$ and $\zeta = 1 + \varepsilon z$,
we obtain an equation containing the terms up to $\varepsilon^2$:
\begin{widetext}%
\begin{equation}
\label{NormViscEq2} %
\left(1 + \varepsilon z\right)^2\frac{d^3\Phi}{dz^3} - (1 +
\varepsilon z)\left(z - \frac{3 - {\rm
i}\,\hat\omega}{2}\varepsilon\right)\frac{d^2\Phi}{dz^2} -
\left[(1 - {\rm i}\,\hat\omega)(1 + \varepsilon z) + \frac{{\rm
i}\,\hat\omega}{4}\varepsilon^2\right]\frac{d\Phi}{dz} +
\frac{\varepsilon\hat\omega^2}{4}\Phi = 0.
\end{equation}
\end{widetext}%

Looking for a solution to this equation in the form of asymptotic
series with respect to parameter $\varepsilon$, $\Phi(z) =
\Phi_0(z) + \varepsilon\Phi_1(z) + \ldots$, we obtain in the
leading order
\begin{equation}
\label{LeadOrderEq1} %
\frac{d}{dz}\left(\frac{d^2\Phi_0}{dz^2} - z\frac{d\Phi_0}{dz} +
{\rm i}\,\hat\omega\Phi_0\right) = 0.
\end{equation}

Integration of this equation gives the second order equation
\begin{equation}
\label{2OrderEq} %
\frac{d^2\Phi_0}{dz^2} - z\frac{d\Phi_0}{dz} + {\rm
i}\,\hat\omega\left(\Phi_0 - D_0\right) = 0,
\end{equation}
where $D_0$ is a constant of integration.

This equation reduces to the equation of a parabolic cylinder with
the help of ansatz $\Phi_0(z) = e^{z^2/4}G(z) + D_0$:
\begin{equation}
\label{ParabCyl} %
\frac{d^2G}{dz^2} + \left({\rm i}\,\hat\omega + \frac 12 -
\frac{z^2}{4}\right)G = 0.
\end{equation}

Two linearly independent solutions of this equation can be
constructed from the following four functions ${\cal D}_{{\rm
i}\,\hat\omega}(\pm z)$ and ${\cal D}_{-{\rm i}\,\hat\omega-1}(\pm
{\rm i}z)$ (see \cite{GR-2007}, 9.255.1). Thus, in the vicinity
of the critical point $\xi = 1$ the solution can be presented in the form %
\begin{equation}
\label{ParCylSol} %
\Phi_0(z) = D_0 + e^{z^2/4}\left[D_1{\cal D}_{{\rm
i}\,\hat\omega}(z) + D_2{\cal D}_{{\rm i}\,\hat\omega}(-z)\right],
\end{equation}
where $D_0$, $D_1$, and $D_2$ are arbitrary constants.

This solution should be matched with the asymptotic expansions
(\ref{Phi-m}) and (\ref{Phi-p}) using the following asymptotics of
functions of the parabolic cylinder when $|s|\gg 1$ (see
\cite{GR-2007}, 9.246):
\begin{widetext}%
\begin{eqnarray}
{\cal D}_p(s) &\sim &
s^p\,e^{-s^2/4}{}_2F_0\left(-\frac{p}{2}\,,\frac{1-p}{2}\,;-\frac{2}{s^2}\right),
\quad |\arg s| < \frac{3\pi}{4}, \label{AssParCyl1} \\
{\cal D}_p(s) &\sim & s^p\,e^{-s^2/4}{}_2F_0\left(-\frac{p}{2}\,,
\frac{1-p}{2}\,;-\frac{2}{s^2}\right)- \frac{\sqrt{2\pi}\,e^{{\rm
i}\,\pi p}}{\Gamma(-p)}s^{-p-1}e^{s^2/4}
{}_2F_0\left(\frac{p}{2}\,,\frac{1+p}{2}\,;\frac{2}{s^2}\right),
\label{AssParCyl2} \\%
{\cal D}_p(s) &\sim & s^p\,e^{-s^2/4}{}_2F_0\left(-\frac{p}{2}\,,
\frac{1-p}{2}\,;-\frac{2}{s^2}\right)- \frac{\sqrt{2\pi}\,e^{-{\rm
i}\,\pi p}}{\Gamma(-p)}s^{-p-1}e^{s^2/4}{}_2F_0
\left(\frac{p}{2}\,,\frac{1+p}{2}\,;\frac{2}{s^2}\right),
\phantom{WW} \label{AssParCyl3}
\end{eqnarray}
\end{widetext}%
where Eq. (\ref{AssParCyl2}) is valid for $\pi/4 < \arg s <
5\pi/4$, and Eq. (\ref{AssParCyl3}) is valid for $-5\pi/4 < \arg s
< -\pi/4$.

With the help of these formulae it is easy to see that the
oscillating terms in expansions (\ref{Phi-m}) and (\ref{Phi-p})
should be matched with the last two terms in Eq.~(\ref{ParCylSol})
which, however, grow infinitely (the former grows, when $z \to
-\infty$, and the latter, when $z \to +\infty$). To remove
infinitely growing terms from the solution, we need to set $D_1 =
D_2 = 0$ in Eq.~(\ref{ParCylSol}), then after the matching, we
obtain in Eqs.~(\ref{Phi-m}), (\ref{Phi-p}) and (\ref{SD2}), (\ref{SD3})%
\begin{equation}
\label{ConstantsB}%
B_1 = \breve B_1 = 0, \quad \mbox{and} \quad B_2 = \breve B_2 = D_0. %
\end{equation}

Notice that from the physical point of view the former equality,
$B_1 = \breve B_1 = 0$, is just a consequence of the fact
mentioned in Sec.~\ref{sec:level3} that in the trans-critical
accelerating current the B-waves (i.e., counter-current
propagating waves on the left of critical point and
negative-energy waves on the right of it) cannot reach the
critical point.

After that assuming that the incident wave arriving from minus
infinity has a unit amplitude $A_1 = 1$, using matching conditions
(\ref{Matching}) and putting $T_p\equiv C_1$, $T_n\equiv C_2$, we
obtain
\begin{widetext}%
\begin{eqnarray}
B_2\,w_3\left(V_1^2\right) &=& R + 1, \label{MtCond1} \\
B_2\,w'_3(V_1^2) &=& \frac{-{\rm i}\,\hat\omega R}{2V_1(1 - V_1)}
+ \frac{{\rm i}\,\hat\omega}{2V_1(1 + V_1)}, \label{MtCond2} \\
T_n + T_p &=& V_2^{{\rm i}\,\hat\omega}\breve
w_3\left(V_2^{-2}\right)\breve B_2, \label{MtCond3} \\
(V_2 + 1)T_n - (V_2 - 1)T_p &=& \frac{2{\rm
i}}{\hat\omega}\,V_2^{{\rm i}\,\hat\omega - 2}\left(V_2^2 -
1\right)\breve w'_3\left(V_2^{-2}\right)\breve B_2.
\label{MtCond4}
\end{eqnarray}
\end{widetext}%

This set can be readily solved yielding the following
transformation coefficients:
\begin{widetext}%
\begin{eqnarray}
R &=& -\frac{w'_3\left(V_1^2\right) - \displaystyle\frac{{\rm
i}\,\hat\omega\,w_3\left(V_1^2\right)} {2V_1(1+V_1)}}
{w'_3\left(V_1^2\right) + \displaystyle\frac{{\rm
i}\,\hat\omega\,w_3\left(V_1^2\right)}{2V_1(1-V_1)}}, \label{TrCoef1} \\
B_2 &=& \breve B_2 = \frac{R + 1}{w_3\left(V_1^2\right)},
\label{TrCoef2} \\
T_n &=& \frac{{\rm i}}{\hat\omega}\,V_2^{{\rm i}\,\hat\omega -
1}\left(V_2^2 - 1\right)\left[\frac{\breve
w'_3\left(V_2^{-2}\right)}{V_2^2} - \frac{{\rm
i}\,\hat\omega}{2}\frac{\breve w_3\left(V_2^{-2}\right)}{V_2 + 1}\right]B_2, \label{TrCoef3} \\
T_p &=& -\frac{{\rm i}}{\hat\omega}\, V_2^{{\rm i}\,\hat\omega -
1}\left(V_2^2 - 1\right)\left[\frac{\breve
w'_3\left(V_2^{-2}\right)}{V_2^2} + \frac{{\rm
i}\,\hat\omega}{2}\frac{\breve w_3\left(V_2^{-2}\right)}{V_2 -
1}\right]B_2. \label{TrCoef4}
\end{eqnarray}
\end{widetext}%

In the long-wave approximation, $\hat\omega \to 0$, we obtain (see
Appendix \ref{appC}):
\begin{equation}
\label{Reduct5}%
R = \frac{1 - V_1}{1 + V_1}, \quad T_p = \frac{V_2 + 1}{V_1 + 1},
\quad T_n = -\frac{V_2 - 1}{V_1 + 1}, \quad K_{Tp,\,n} =
\frac{V_1}{V_2}\left(\frac{V_2 \pm 1}{V_1 + 1}\right)^2,
\end{equation}
where in the last formula sign plus pertains to the positive- and
sign minus -- to the negative-energy transmitted wave.

These values are purely real, $R > 0$ and $T_p > 0$, whereas $T_n
< 0$. The problem of surface wave transformation in a duct with
the stepwise change of cross-section and velocity profile is
undetermined for such current, therefore in Ref.
\cite{ChurErRousStep-2017} one of the parameters, $R_\eta$ -- the
reflection coefficient in terms of free surface perturbation, was
undefined. Now from Eq.~(\ref{Reduct5}) it follows that the
transformation coefficients in terms of free surface perturbation
in Ref. \cite{ChurErRousStep-2017} are $R_\eta = T_{p\eta} =
-T_{n\eta} = 1$ (for the relationships between the transformation
coefficients in terms of velocity potential and free surface
perturbation see Appendix \ref{appA}).

Because of the relationships between the coefficients
(\ref{ConstantsB}), the solution in the domain $\xi_1 < \xi <
\xi_2$ is described by the same analytical function
$w_3\left(\xi^2\right) \equiv \xi^{{\rm i}\,\hat\omega}\breve
w_3\left(\xi^{-2}\right)$ (see Eqs.~(9) and (11) in \S 6.4 of the
book \citep{Luke-1975}). In the result, the energy flux is still
conserved despite a small viscosity in the vicinity of the
critical
point $\xi = 1$: %
\begin{equation}%
\label{FluxCons023} %
J = \frac{2\hat\omega}{V_1}\left(1 - |R|^2\right) =
\frac{2\hat\omega}{V_2}\left(|T_p|^2 - |T_n|^2\right) > 0 \quad
\mbox{or} \quad V_2\left(1 - |R|^2\right) = V_1\left(|T_p|^2 -
|T_n|^2\right).
\end{equation}

As one can see from these expressions, the energy flux in the
reflected wave by modulus is always less than in the incident
wave, therefore over-reflection here is not possible. In the
meantime the energy transmission factors $K_{Tp,n}$ can be greater
than 1; this implies that the over-transmission can occur with
respect to both positive- and negative-energy waves.

\begin{figure}[h]
\centering
\includegraphics[width=90mm]{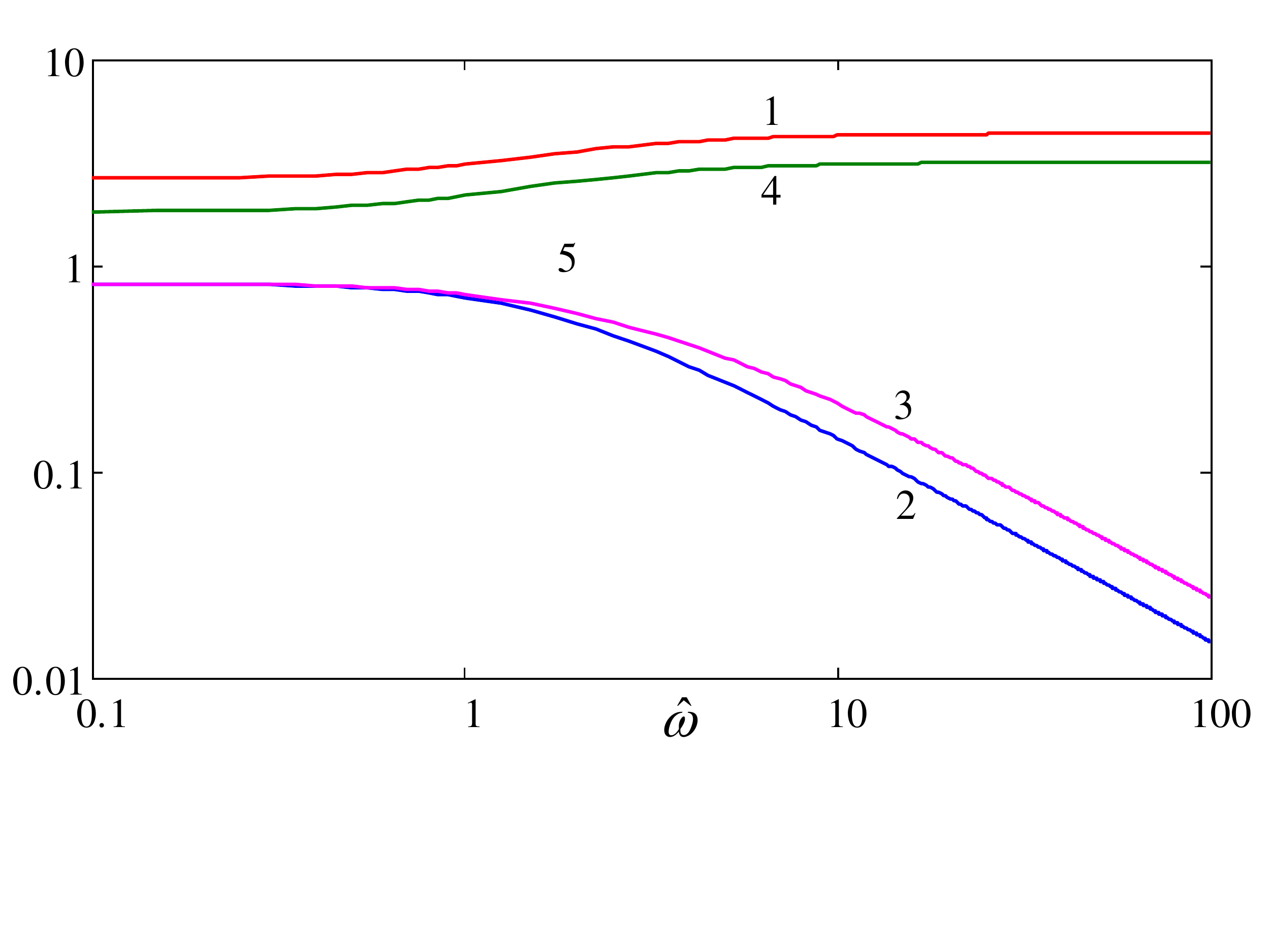}
\vspace*{-18mm}%
\caption{(color online) Modules of the transformation coefficient
as functions of dimensionless frequency $\hat\omega$ for $V_1 =
0.1$, $V_2 = 1.9$. Line 1 -- $|T_p|$, line 2 -- $|T_n|$, line 3 --
$|R|$, line 4 -- $|B_2| = |\breve B_2|$. Dashed line 5 represents
the asymptotic for $|T_n| \sim \hat\omega^{-1}$.}
\label{f09}%
\end{figure}

The transformation coefficients $|R|$, $|T_p|$ and $|T_n|$
together with the intermediate coefficients of wave excitation in
the transient zone, $|B_2| = |\breve B_2|$, are presented in
Fig.~\ref{f09} as functions of dimensionless frequency
$\hat\omega$ for the particular values of speed, $V_1 = 0.1$ and
$V_2 = 1.9$. Qualitatively similar graphics were obtained for
other values of $V_1$ and $V_2$.

Notice that both the transmission coefficient of negative energy
wave $|T_n|$ and reflection coefficient of positive energy wave
$|R|$ decay asymptotically with the same rate $\sim
\hat\omega^{-1}$.

\begin{figure}[h]
\centering
\includegraphics[width=90mm]{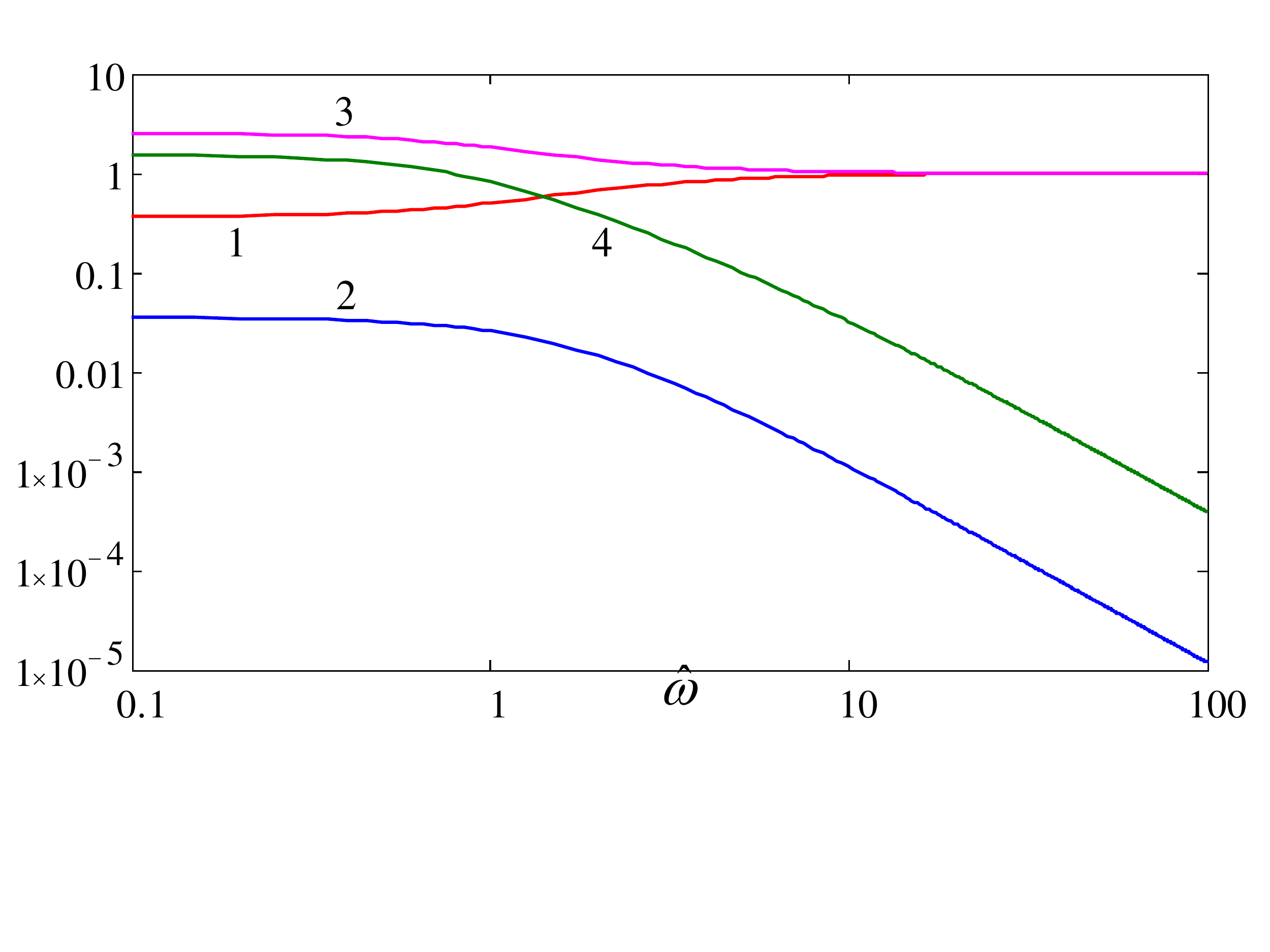}
\vspace*{-20mm}%
\caption{(color online) The dependences of energy transmission
factors on the frequency (i) when both $K_{Tp} < 1$ (line 1) and
$K_{Tn} < 1$ (line 2) (here $V_1 = 0.1$, $V_2 = 1.9$); and (ii)
when both $K_{Tp} > 1$ (line 3) and $K_{Tn} > 1$ (line 4) in a
certain range of frequencies $\hat\omega < \hat\omega_c$ (here
$V_1 = 0.9$, $V_2 = 8.0$). Inclined dashed lines show the
asymptotic dependences $K_{Tn} \sim \hat\omega^{-2}$.}
\label{f10}%
\end{figure}

Figure \ref{f10} illustrates the dependences of energy
transmission factors on the frequency for two cases: (i) when both
$K_{Tp,n} < 1$ ($V_1 = 0.1$, $V_2 = 1.9$) and (ii) when both
$K_{Tp,n} > 1$ in a certain range of frequencies $\hat\omega <
\hat\omega_c$ ($V_1 = 0.9$, $V_2 = 8.0$).

\begin{figure}[h]
\centering
\includegraphics[width=90mm]{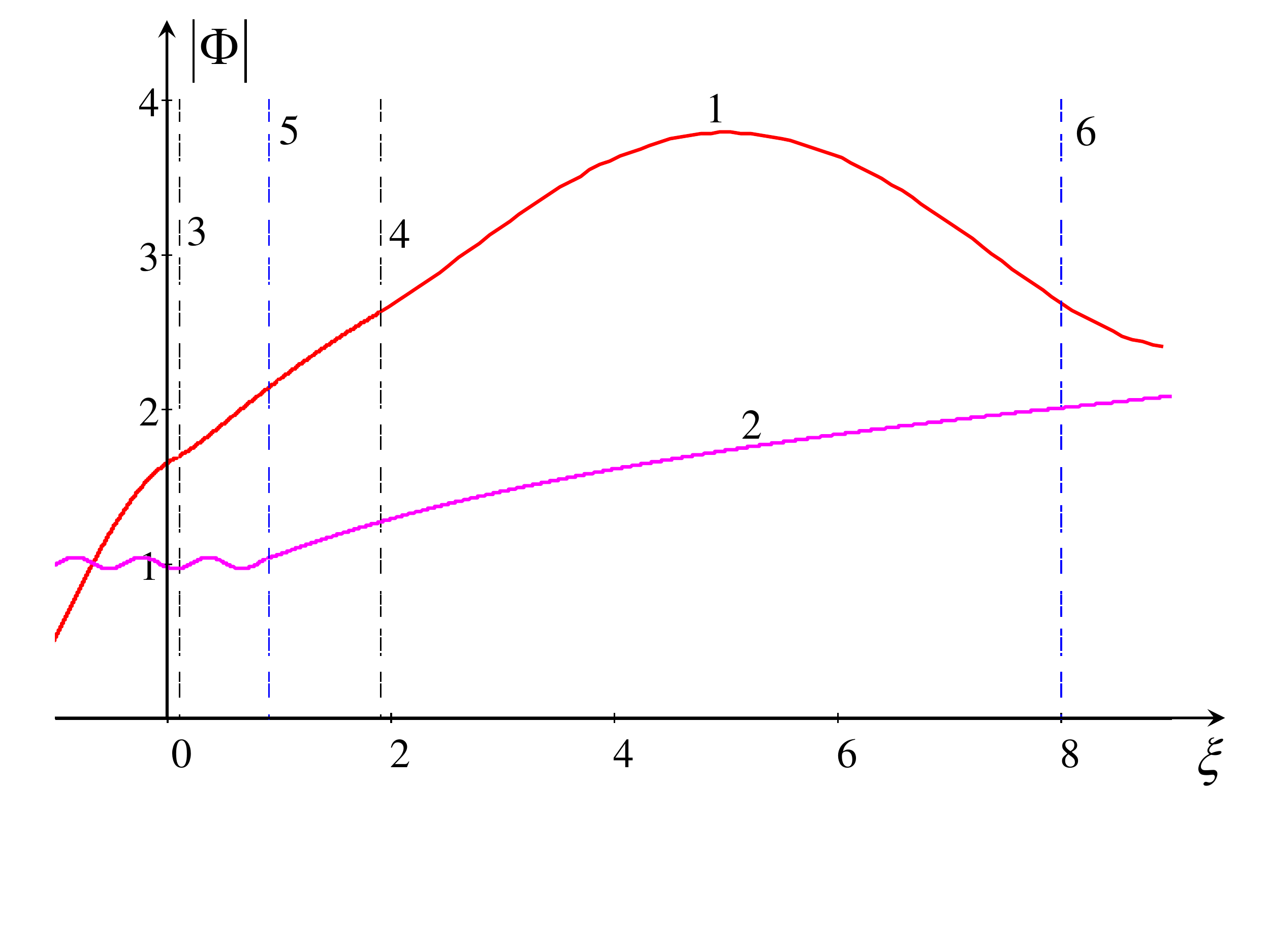}
\vspace*{-15mm}%
\caption{(color online) Modules of function $\Phi(\xi)$ for wave
scattering in accelerating trans-critical current with $V_1 = 0.1$
and $V_2 = 1.9$ (line 1) and $V_1 = 0.9$ and $V_2 = 8.0$ (line 2).
Dashed vertical lines 3 and 4 show the transition zone where the
current accelerates from $V_1 = 0.1$ to $V_2 = 1.9$, and dashed
vertical lines 5 and 6 show the transition zone where the current
accelerates from $V_1 = 0.9$ to $V_2 = 8.0$. The plot was
generated for $\hat\omega = 1$.}
\label{f11}%
\end{figure}

In Fig.~\ref{f11} we present graphics of $|\Phi(\xi)|$ as per
Eqs.~(\ref{SD1})--(\ref{SD4}) for $A_1 = 1$,  $A_2 = R$ as per
Eq.~(\ref{TrCoef1}), $D_1 = T_n$ as per Eq.~(\ref{TrCoef3}), and
$D_2 = T_p$  as per Eq.~(\ref{TrCoef4}). Coefficients $B_1 =
\breve B_1 = 0$ as per Eq.~(\ref{ConstantsB}), and $B_2 = \breve
B_2$ are given by Eq.~(\ref{TrCoef2}). Line 1 in this figure
pertains to the case when $V_1 = 0.1$, $V_2 = 1.9$, and line 2 --
to the case when $V_1 = 0.9$, $V_2 = 8.0$.

\subsection{\label{sec:level44}Wave transformation in trans-critical
decelerating currents  $V_1 > 1 > V_2 > 0$}

In this subsection we consider the wave transformation in
gradually decelerating background current assuming that the
current is super-critical in the left domain and sub-critical in
the right domain. For the sake of simplification of hypergeometric
functions used below we chose again the coordinate frame such as
shown in Fig.~\ref{f01}b). In such a current the transition
through the critical point, where $V(\xi) = 1$, occurs at $\xi =
-1$.

In the left domain, where the current is super-critical, only
downstream propagating waves can exist, with the positive or
negative energy. In contrast to that, in the right domain, where
the background current is sub-critical, two waves of positive
energy can coexist; one of them is co-current propagating and
another one is counter-current propagating.

The general solution of Eq.~(\ref{NormEq}) in the different
domains can be formally presented with the help of functions
$\tilde w$ as per Eq.~(\ref{DecelSub1}) and $\hat w$ as per
Eq.~(\ref{Gauss11De}):
\begin{widetext}%
\begin{eqnarray}
\Phi(\xi) &=& A_1e^{{\rm i}\,\kappa_1(\xi - \xi_1)} + A_2e^{{\rm
i}\,\kappa_2(\xi - \xi_1)}, \quad \phantom{Wwwwwwwww}\xi < \xi_1, \label{Solut1} \\
\Phi(\xi) &=& (-\xi)^{-{\rm i}\,\hat\omega}\left[\hat B_1\,\hat
w_1\left(\xi^{-2}\right) + \hat B_2\,\hat
w_3\left(\xi^{-2}\right)\right], \quad \xi_1 < \xi < -1, \label{Solut2} \\
\Phi(\xi) &=& B_1\,\tilde w_2\left(\xi^2\right) + B_2\,\tilde w_3
\left(\xi^2\right), \quad  \phantom{wwwwwww}-1 < \xi < \xi_2 < 0, \label{Solut3} \\
\Phi(\xi) &=& C_1\,e^{{\rm i}\,\kappa_3(\xi - \xi_2)} +
C_2\,e^{-{\rm i}\,\kappa_4(\xi - \xi_2)}, \quad
\phantom{wwwwwwww}\xi > \xi_2, \label{Solut4}
\end{eqnarray}
\end{widetext}%
where $\kappa_1 = \hat\omega/(V_1 + 1),\ \ \kappa_2 =
\hat\omega/(V_1 - 1),\ \ \kappa_3 = \hat\omega/(1 + V_2),\ \
\kappa_4 = \hat\omega/(1 - V_2)$.

The matching conditions at $\xi = \xi_1$ provide (cf.
Eqs.~(\ref{Match11})and (\ref{Match12})):
\begin{widetext}%
\begin{eqnarray}
A_1 + A_2 &=& V_1^{-{\rm i}\,\hat\omega}\left[\hat B_1\,\hat
w_1\left(V_1^{-2}\right) + \hat B_2\,\hat
w_3\left(V_1^{-2}\right)\right]\!, \label{MatCon1} \\
(V_1 - 1)A_1 - (V_1 + 1)A_2 &=& \frac{2{\rm
i}}{\hat\omega}\,V_1^{-{\rm i}\,\hat\omega - 2}\left(V_1^2 -
1\right)\left[\hat B_1\,\hat w'_1\left(V_1^{-2}\right) + \hat
B_2\,\hat w'_3\left(V_1^{-2}\right)\right]\!,\phantom{www}
\label{MatCon2}
\end{eqnarray}
\end{widetext}%

And similarly the matching conditions at $\xi = \xi_2$ provide:
\begin{widetext}%
\begin{eqnarray}
C_1 + C_2 &=& B_1\,\tilde w_2\left(V_2^2\right) + B_2\,\tilde w_3\left(V_2^2\right), \label{MatCon3} \\
(1 - V_2)C_1 - (1 + V_2)C_2 &=& \frac{2{\rm
i}}{\hat\omega}\,V_2\left(1 - V_2^2\right)\left[B_1\,\tilde
w'_2\left(V_2^2\right) + B_2\,\tilde
w'_3\left(V_2^2\right)\right]. \phantom{www} \label{MatCon4}
\end{eqnarray}
\end{widetext}%

\bigskip

With the help of Eq.~(\ref{HyperFuncEq}) we find the asymptotic
expansions when $\xi \to -1_{\pm 0}$
\begin{widetext}%
\begin{eqnarray}
\hspace{-8mm} \Phi(\xi) &=& \hat B_2 + \frac{\Gamma(1-{\rm
i}\,\hat\omega)\, \hat B_1}{2\Gamma^2(1-{\rm i}\,\hat\omega/2)} +
\frac{\Gamma(1+{\rm i}\,\hat\omega)\, \hat B_1}{2\Gamma^2(1+{\rm
i}\,\hat\omega/2)}\,\left(\xi^2-1\right)^{-{\rm i}\,\hat\omega} +
O\left(\xi^2-1\right), \quad \xi \to -1_{-0}, \label{Phi-m1} \\%
{} && \nonumber \\%
\hspace{-8mm} \Phi(\xi) &=& B_2 +
\frac{\Gamma(-{\rm i}\,\hat\omega)\,B_1}{\Gamma^2(1-{\rm
i}\,\hat\omega/2)} + \frac{\Gamma\left({\rm
i}\,\hat\omega\right)\,B_1}{\Gamma^2(1+{\rm
i}\,\hat\omega/2)}\,\left(1-\xi^2\right)^{-{\rm i}\,\hat\omega} +
O\left(1-\xi^2\right), \quad \xi \to  -1_{+0}. \label{Phi-p1}
\end{eqnarray}
\end{widetext}%
which are similar to Eqs.~(\ref{Phi-m}) and (\ref{Phi-p}), and
contain fast oscillating terms corresponding to counter-current
propagating B-waves as well.

To match solutions in the vicinity of critical point $\xi = -1$,
we again take into consideration a small viscosity. Bearing in
mind that $V(\xi) = -\xi$ (see Fig. \ref{f01}b)) and setting
$\zeta = \xi^2 = 1 + \varepsilon z$, $\nu = \varepsilon^2/2$, we
arrive at the equation similar to Eq.~(\ref{NormViscEq2}):
\begin{widetext}%
\begin{equation}
\label{NormViscEq3} %
\left(1 + \varepsilon z\right)^2\frac{d^3\Phi}{dz^3} + (1 +
\varepsilon z)\left(z + \frac{3 + {\rm
i}\,\hat\omega}{2}\varepsilon\right)\frac{d^2\Phi}{dz^2} +\left[(1
+ {\rm i}\,\hat\omega)(1 + \varepsilon z) + \frac{{\rm
i}\,\hat\omega}{4}\varepsilon^2\right]\frac{d\Phi}{dz} -
\frac{\varepsilon\hat\omega^2}{4}\Phi = 0.
\end{equation}
\end{widetext}%

This equation in the leading order on the small parameter
$\varepsilon \ll 1$ reduces to (cf. Eq.~(\ref{LeadOrderEq1})):
\begin{equation}
\label{LeadOrderEq2} %
\frac{d}{dz}\left(\frac{d^2\Phi_0}{dz^2} + z\frac{d\Phi_0}{dz} +
{\rm i}\,\hat\omega\Phi_0\right) = 0.
\end{equation}

Integrating this equation and substituting $\Phi_0(z) = D_0 +
e^{-z^2/4}G(z)$, we obtain again the equation of a parabolic
cylinder in the form (cf. Eq.~(\ref{ParabCyl2})):
\begin{equation}
\label{ParabCyl2}%
\frac{d^2G}{dz^2} + \left({\rm i}\,\hat\omega - \frac 12 -
\frac{z^2}{4}\right)G = 0.
\end{equation}

Thus, the general solution to Eq.~(\ref{LeadOrderEq2}) in the
vicinity of critical point $\xi = -1$ can be presented as: %
$$
\Phi_0(z) = D_0 + e^{-z^2/4}\left[D_1{\cal D}_{{\rm i}\,\hat\omega
- 1}(z) + D_2{\cal D}_{{\rm i}\,\hat\omega-1}(-z)\right],
$$
where $D_0$, $D_1$ and $D_2$ are arbitrary constants.

The asymptotic expansions (\ref{AssParCyl1})--(\ref{AssParCyl3})
show that this solution remains limited for any arbitrary
constants. Moreover, the oscillatory terms in Eqs.~(\ref{Phi-m1})
and (\ref{Phi-p1}) become exponentially small after transition
through the critical point $\xi = -1$. As was explained in
Sec.~\ref{sec:level3}, this means that the B-waves running toward
the critical point both from the left (negative-energy waves) and
from the right (counter-current propagating positive-energy waves)
dissipate in the vicinity of the critical point. For this reason
the wave energy flux does not conserve in the decelerating
trans-critical currents (see Eqs.~(\ref{44}) and (\ref{441})
below). Taking this fact into account, one can match solutions
(\ref{Phi-m1}) and (\ref{Phi-p1}): %
\begin{equation}
\label{38} %
D_0 = \frac{\Gamma(-{\rm i}\,\hat\omega)\,}{\Gamma^2(1-{\rm
i}\,\hat\omega/2)}B_1 + B_2 = \frac{\Gamma(1-{\rm
i}\,\hat\omega)\,}{2\Gamma^2(1-{\rm i}\,\hat\omega/2)}\hat B_1 +
\hat B_2.
\end{equation}

After that using the identity $\Gamma(x)\Gamma(1 - x) =
\pi/\sin{\pi x}$, we find for the constants $D_1$ and $D_2$ the
following expressions:
\begin{equation}
\label{TildeC}%
D_1 = \frac{-{\rm i}\sqrt{\pi/2}}{\Gamma^2(1 + {\rm
i}\hat\omega/2)}\frac{e^{-{\rm
i}\hat\omega\ln{\varepsilon}}}{\sinh{\pi\hat\omega}}B_1, \quad D_2
= \frac{\hat\omega\sqrt{\pi/2}}{2\Gamma^2(1 + {\rm
i}\hat\omega/2)}\frac{e^{-{\rm
i}\hat\omega\ln{\varepsilon}}}{\sinh{\pi\hat\omega}}\hat B_1.
\end{equation}

Using the prepared formulae we can now calculate the
transformation coefficients for incident waves of either positive
or negative energy travelling in the duct from the minus to plus
infinity.

\subsubsection{\label{sec:level441}Transformation of downstream
propagating positive-energy wave}

Assume first that the incident wave of unit amplitude has positive
energy and let us set in Eqs.~(\ref{Solut1}) and (\ref{Solut4})
$A_1 = 1$, $A_2 = 0$, $C_1 \equiv T_1$, and $C_2 = 0$. Then from
Eqs.~(\ref{MatCon1}) and (\ref{MatCon2}) we obtain (cf.
Eqs.~(\ref{Match11}) and (\ref{Match12})):
\begin{eqnarray}
\hat B_1\,\hat w_1\left(V_1^{-2}\right) + \hat B_2\,\hat
w_3\left(V_1^{-2}\right) &=& V_1^{{\rm i}\,\hat\omega}, \label{C1}
\\
\hat B_1\,\hat w'_1\left(V_1^{-2}\right) + \hat B_2\,\hat
w'_3\left(V_1^{-2}\right) &=& -\frac{{\rm
i}\,\hat\omega}{2}\frac{V_1^{{\rm i}\,\hat\omega+2}}{V_1+1}.
\label{C2}
\end{eqnarray}

From this set of equations using the Wronskian (\ref{Wronsk2De}),
one can find
\begin{widetext}%
\begin{eqnarray}
\hat B_1 &=& -\frac{\Gamma({\rm i}\,\hat\omega/2)\, \Gamma(1+{\rm
i}\,\hat\omega/2)}{\Gamma(1+{\rm i}\,\hat\omega)}\, V_1^{{-\rm
i}\,\hat\omega-2}\left(V_1^2 - 1\right)^{{\rm i}\,\hat\omega + 1}
\left[\frac{\hat w'_3\left(V_1^{-2}\right)}{V_1^2} + \frac{{\rm
i}\,\hat\omega}{2}\frac{\hat w_3\left(V_1^{-2}\right)}
{V_1+1}\right]\!, \phantom{www} \label{C11}\\
\hat B_2 &=& \frac{\Gamma({\rm i}\,\hat\omega/2)\, \Gamma(1+{\rm
i}\,\hat\omega/2)}{\Gamma(1+{\rm i}\,\hat\omega)}\, V_1^{-{\rm
i}\,\hat\omega - 2}\left(V_1^2 - 1\right)^{{\rm i}\,\hat\omega +
1} \left[\frac{\hat w'_1\left(V_1^{-2}\right)}{V_1^2} + \frac{{\rm
i}\,\hat\omega}{2}\frac{\hat w_1\left(V_1^{-2}\right)}{V_1+1}\right]\!. \label{C21}%
\end{eqnarray}
\end{widetext}%

\bigskip

Similarly from the matching conditions (\ref{MatCon3}) and
(\ref{MatCon4}) we obtain
\begin{eqnarray}
B_1\,\tilde w_2\left(V_2^2\right) + B_2\,\tilde w_3\left(V_2^2\right) &=& T_1, \label{MatCon31} \\
B_1\,\tilde w'_2\left(V_2^2\right) + B_2\,\tilde
w'_3\left(V_2^2\right) &=& \frac{-{\rm i}\,\hat\omega T_1}{2V_2(1
+ V_2)}. \label{MatCon41}
\end{eqnarray}

Using the Wronskian (\ref{Wronsk2}), we derive from these
equations
\begin{widetext}%
\begin{eqnarray}
B_1 &=& -\frac{\Gamma^2(1+{\rm i}\,\hat\omega/2)}{\Gamma(1+{\rm
i}\,\hat\omega)}\, \left(1 - V^2_2\right)^{{\rm i}\,\hat\omega +
1}\left[\tilde w'_3\left(V_2^2\right) + \frac{{\rm
i}\,\hat\omega}{2} \frac{\tilde w_3\left(V_2^2\right)}{V_2(1 +
V_2)}\right]T_1\!, \label{B1}\\
B_2 &=& \frac{\Gamma^2(1+{\rm i}\,\hat\omega/2)}{\Gamma(1+{\rm
i}\,\hat\omega)}\, \left(1 - V_2^2\right)^{{\rm i}\,\hat\omega +
1}\left[\tilde w'_2\left(V_2^2\right) + \frac{{\rm
i}\,\hat\omega}{2}\frac{\tilde w_2\left(V_2^2\right)}{V_2(1 + V_2)}\right]T_1\!. \label{B2}%
\end{eqnarray}
\end{widetext}%

Substituting $B_1$ and $B_2$, as well as $\hat B_1$ and $\hat
B_2$, in Eq.~(\ref{38}), we obtain the transmission coefficient %
\begin{widetext}%
$$
T_1 = -\frac{2{\rm i}}{\hat\omega}\,V_1^{-{\rm i}\,\hat\omega -
2}\left(\frac{V_1^2-1}{1-V_2^2}\right)^{{\rm
i}\,\hat\omega+1}\times {}
$$
\begin{equation}
\label{TransCoef} %
\frac{\displaystyle\frac{\hat w'_1\left(V_1^{-2}\right)}{V_1^2} +
\frac{{\rm i}\,\hat\omega}{2} \frac{\hat
w_1\left(V_1^{-2}\right)}{V_1 + 1} - \frac{\Gamma(1-{\rm
i}\,\hat\omega)}{2\Gamma^2(1-{\rm i}\,\hat\omega/2)}
\left[\frac{\hat w'_3(V_1^{-2})}{V_1^2} + \frac{{\rm
i}\,\hat\omega}{2} \frac{\hat w_3\left(V_1^{-2}\right)} {V_1+1}\right]}%
{\displaystyle\tilde w'_2\left(V_2^2\right) + \frac{{\rm
i}\,\hat\omega}{2} \frac{\tilde w_2\left(V_2^2\right)}{V_2(1+V_2)}
- \frac{\Gamma(-{\rm i}\,\hat\omega)}{\Gamma^2(1-{\rm
i}\,\hat\omega/2)}\left[\tilde w'_3\left(V_2^2\right) + \frac{{\rm
i}\,\hat\omega}{2} \frac{\tilde w_3\left(V_2^2\right)}{V_2(1+V_2)}\right]}.%
\end{equation}
\end{widetext}%

Calculations of the energy fluxes on each side of the transient
domain show that they are both positive, but generally different,
i.e., the energy flux does not conserve,%
\begin{equation}
\label{44}%
J_1 = J(\xi < -1) = \frac{2\hat\omega}{V_1}\ \ne\ J_2 =
J(\xi > -1) = \frac{2\hat\omega}{V_2}\,|T_1|^2.
\end{equation}
\vspace*{-8mm}%
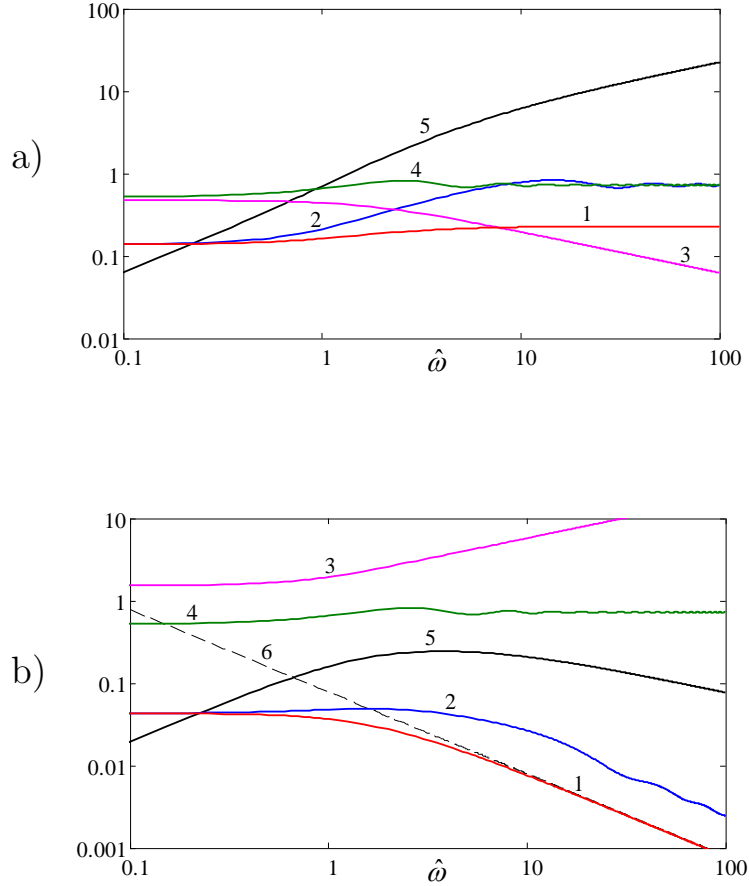
\begin{figure}[h!]
\centering
\includegraphics[width=90mm]{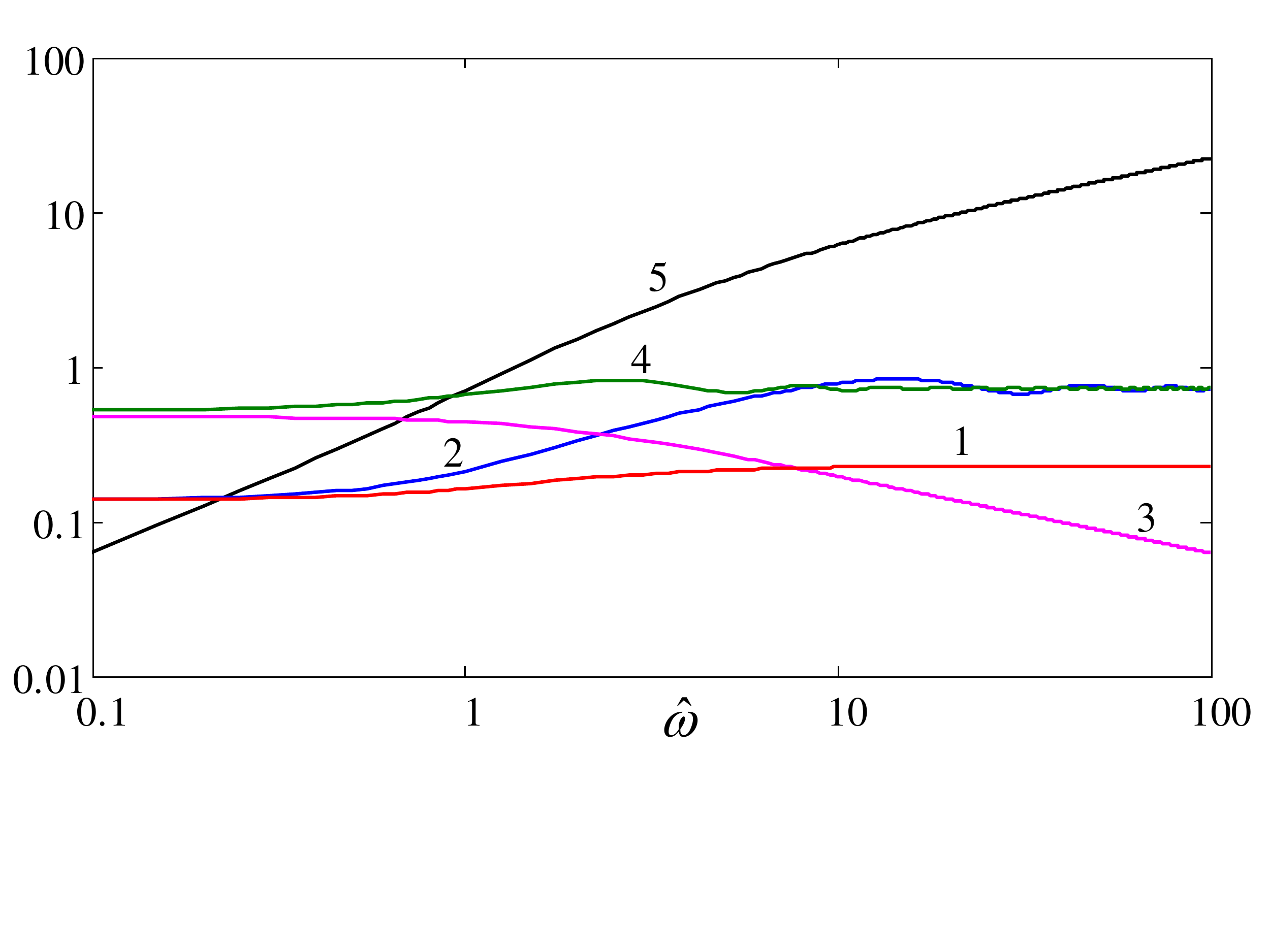}
\includegraphics[width=90mm]{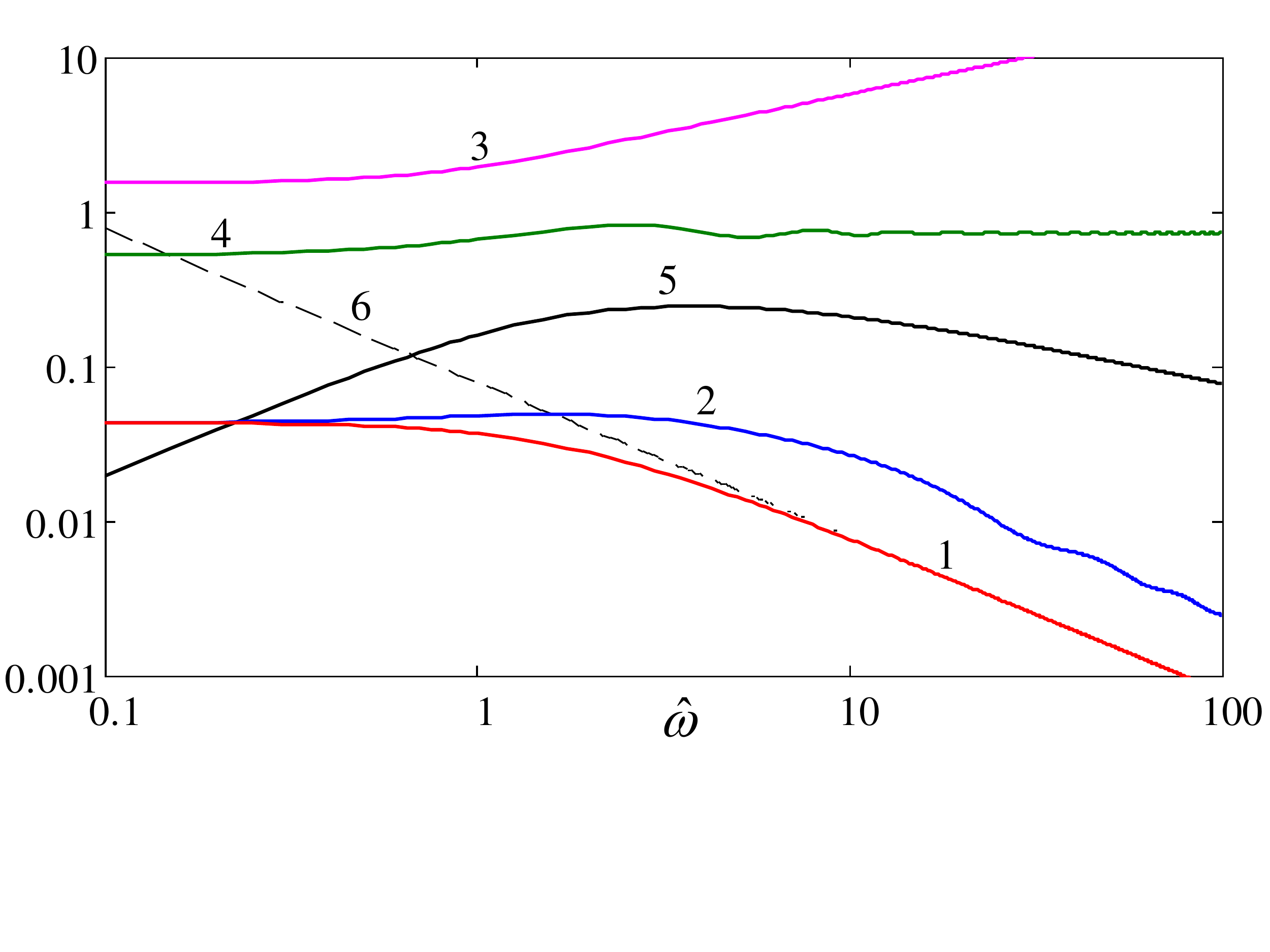}
\begin{picture}(300,6)%
\put(0,335){{\large a)}}%
\put(0,140){{\large b)}}%
\end{picture}
\vspace*{-20mm}%
\caption{(color online) Modules of the transmission coefficients
$|T_1|$ (line 1 in panel a) and $|T_2|$ (line 1 in panel b), as
well as coefficients of wave excitation in the transient domain,
$|B_1|$ (line 2), $|B_2|$ (line 3), $|\hat B_1|$ (line 4), and
$|\hat B_2|$ (line 5), for the scattering of positive energy wave
(panel a) and negative energy wave (panel b) as functions of
dimensionless frequency $\hat\omega$ for $V_1 = 1.9$, $V_2 = 0.1$.
Dashed line 6 in panel (b) represents the high-frequency
asymptotic for $|T_2| \sim \hat\omega^{-1}$.}
\label{f12}%
\end{figure}

This interesting fact can be explained by the partial wave
absorption in the critical point due to viscosity. The detailed
explanation of this is given in Section \ref{sec:level5}. The
difference in the energy flux in the incident and transmitted
waves is independent of the viscosity, when $\nu \to 0$: %
\begin{equation}
\label{FluxJump}%
\Delta J \equiv J_1 - J_2 = 2\hat\omega\left(1/V_1 -
|T_1|^2/V_2\right) \stackrel{\hat\omega \; \to \;
0}{\longrightarrow} \frac{1 - V_1V_2}{V_1(1 + V_2)^2}(V_1 -
V_2)J_1,
\end{equation}
and it is easily seen that it can be both positive and negative.

In Fig. \ref{f12}a) we present the transmission coefficient
$|T_1|$ together with the intermediate coefficients of wave
excitation in the transient domain, $|B_1|$, $|B_2|$, $|\hat
B_1|$, and $|\hat B_2|$, as functions of dimensionless frequency
$\hat\omega$ for the particular values of current speed $V_1 =
1.9$ and $V_2 = 0.1$. As one can see from this figure, the
transmission coefficient gradually increases with the frequency.

\begin{figure}[h]
\centering
\includegraphics[width=90mm]{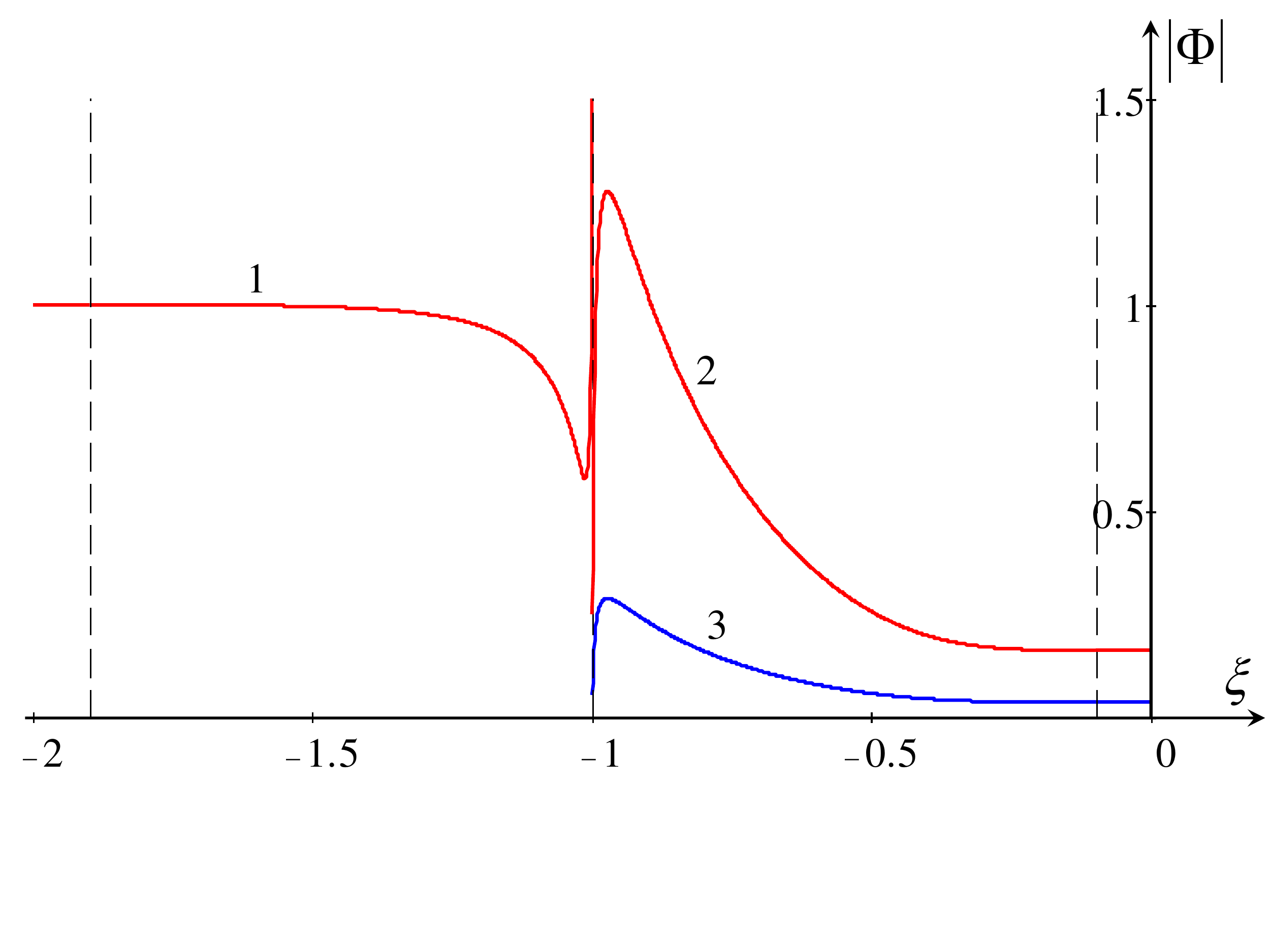}
\vspace*{-15mm}%
\caption{(color online) Modules of function $\Phi(\xi)$ for wave
scattering in decelerating trans-critical current with $V_1 = 1.9$
and $V_2 = 0.1$ for $\hat\omega = 1$. Lines 1 and 2 pertain to the
scattering of a positive-energy incident wave, and lines 1 and 3
pertain to the scattering of a negative-energy incident wave (line
1 is the same both for positive- and negative-energy waves).}
\label{f13}%
\end{figure}

The graphic of $|\Phi(\xi)|$ is shown in Fig.~\ref{f13} by lines 1
and 2. The plot was generated for $\hat\omega = 1$ on the basis of
solution Eqs.~(\ref{Solut1})--(\ref{Solut4}) with $A_1 = 1$, $A_2
= 0$, $D_1 = T_1$ as per Eq.~(\ref{TransCoef}), and $D_2 = 0$.
Coefficients $B_1$ and $B_2$ are given by Eqs.~(\ref{B1}) and
(\ref{B2}), and coefficients $\hat B_1$ and $\hat B_2$ are given
by Eqs.~(\ref{C11}) and (\ref{C21}). The module of function
$\Phi(\xi)$ is discontinuous only in the critical point $\xi =
-1$, and the phase of function $\Phi(\xi)$ quickly changes in the
small vicinity of this point.

\vspace*{-8mm}%

\subsubsection{\label{sec:level442}Transformation of downstream
propagating negative-energy wave}

Assume now that the incident wave is a unit amplitude wave of
negative energy and correspondingly set $A_1 = 0$, $A_2 = 1$, $C_1
= 0$, and $C_2 \equiv T_2$. Then from Eqs. (\ref{MatCon1}) and
(\ref{MatCon2}) we obtain:
\begin{eqnarray}
\hat B_1\,\hat w_1\left(V_1^{-2}\right) + \hat B_2\,\hat
w_3\left(V_1^{-2}\right) &=& V_1^{{\rm i}\,\hat\omega},
\label{C1N}
\\
\hat B_1\,\hat w'_1\left(V_1^{-2}\right) + \hat B_2\,\hat
w'_3\left(V_1^{-2}\right) &=& \frac{{\rm
i}\,\hat\omega}{2}\frac{V_1^{{\rm i}\,\hat\omega+2}}{V_1-1}.
\label{C2N}
\end{eqnarray}

From this set we find
\begin{widetext}%
\begin{eqnarray}
\hat B_1 &=& -\frac{\Gamma({\rm i}\,\hat\omega/2)\, \Gamma(1+{\rm
i}\,\hat\omega/2)}{\Gamma(1+{\rm i}\,\hat\omega)}\, V_1^{{-\rm
i}\,\hat\omega-2}\left(V_1^2 - 1\right)^{{\rm i}\,\hat\omega + 1}
\left[\frac{\hat w'_3\left(V_1^{-2}\right)}{V_1^2} - \frac{{\rm
i}\,\hat\omega}{2} \frac{\hat w_3\left(V_1^{-2}\right)}{V_1 -
1)}\right]\!, \phantom{www} \label{C11N}\\
\hat B_2 &=& \frac{\Gamma({\rm i}\,\hat\omega/2)\, \Gamma(1+{\rm
i}\,\hat\omega/2)}{\Gamma(1+{\rm i}\,\hat\omega)}\, V_1^{-{\rm
i}\,\hat\omega - 2}\left(V_1^2 - 1\right)^{{\rm i}\,\hat\omega +
1} \left[\frac{\hat w'_1\left(V_1^{-2}\right)}{V_1^2} - \frac{{\rm
i}\,\hat\omega}{2} \frac{\hat w_1\left(V_1^{-2}\right)}{V_1 -
1)}\right]\!. \label{C21N}%
\end{eqnarray}
\end{widetext}%

\bigskip

From the matching conditions at $\xi = \xi_2$ (see
Eqs.~(\ref{MatCon3}) and (\ref{MatCon4})) we obtain the similar
expressions for the coefficients $B_1$ and $B_2$ as in
Eqs.~(\ref{B1}) and (\ref{B2}) with the only replacement of $T_1$
by $T_2$. Substituting then all four coefficients $B_1$, $B_2$,
$\hat B_1$, and $\hat B_2$ in Eq.~(\ref{38}), we obtain the
transmission coefficient $T_2$:
\begin{widetext}%
$$
T_2 = -\frac{2{\rm i}}{\hat\omega}\,V_1^{-{\rm i}\,\hat\omega -
2}\left(\frac{V_1^2-1}{1-V_2^2}\right)^{{\rm i}\,\hat\omega+1}
\times {}%
$$
\begin{equation}
\label{TransCoef2} %
\frac{\displaystyle\frac{\hat w'_1\left(V_1^{-2}\right)}{V_1^2} -
\frac{{\rm i}\,\hat\omega}{2} \frac{\hat
w_1\left(V_1^{-2}\right)}{V_1 - 1} - \frac{\Gamma(1-{\rm
i}\,\hat\omega)}{2\Gamma^2(1-{\rm i}\,\hat\omega/2)}
\left[\frac{\hat w'_3(V_1^{-2})}{V_1^2} - \frac{{\rm
i}\,\hat\omega\,\hat w_3\left(V_1^{-2}\right)}
{2(V_1-1)}\right]}%
{\displaystyle\tilde w'_2\left(V_2^2\right) + \frac{{\rm
i}\,\hat\omega}{2} \frac{\tilde w_2\left(V_2^2\right)}{V_2(1+V_2)}
- \frac{\Gamma(-{\rm i}\,\hat\omega)}{\Gamma^2(1-{\rm
i}\,\hat\omega/2)}\left[\tilde w'_3\left(V_2^2\right) + \frac{{\rm
i}\,\hat\omega\,\tilde w_3\left(V_2^2\right)}{2V_2(1+V_2)}\right]}.%
\end{equation}
\end{widetext}%

Calculations of the energy fluxes on each side of the transient
domain show that they are not equal again, moreover, they have
opposite signs in the left and right domains:
\begin{equation}
\label{441}%
J_1 = J(\xi < -1) = -\frac{2\hat\omega}{V_1} < 0, \quad J_2 =
J(\xi > -1) = \frac{2\hat\omega}{V_2}\,|T_2|^2 > 0.
\end{equation}

The wave of negative energy in the left domain propagates to the
right, its group velocity $V_g$ is positive, but because it has a
negative energy $E$, its energy flux, $J = EV_g$ is negative.

In the long-wave approximation, $\hat\omega \to 0$ we obtain (see
Appendix \ref{appC}):
\begin{equation}
\label{Reduct6}%
T_1 = \frac{V_2(V_1 + 1)}{V_1(V_2 + 1)}, \quad
 T_2 = \frac{V_2(V_1 - 1)}{V_1(V_2 + 1)}, \quad K_{T1, T2} =
\frac{V_2}{V_1}\left(\frac{V_1 \pm 1}{V_2 + 1}\right)^2,
\end{equation}
where in the last formula sign plus pertains to the positive- and
sign minus -- to the negative-energy wave. As one can see, the
transmission coefficients are purely real and positive, $T_{1,2}
> 0$, in both cases.

The problem of surface wave transformation in a duct with the
stepwise change of cross-section and velocity profile is
undetermined for such current too; however from the results
obtained it follows that in terms of free surface perturbation the
transformation coefficients are
\begin{equation}
\label{Reduct7}%
T_{1\eta} = \frac{V_2}{V_1}\left(\frac{V_1 + 1}{V_2 + 1}\right)^2,
\quad T_{2\eta} = \frac{V_2}{V_1}\frac{V_1^2 - 1}{\left(V_2 +
1\right)^2}
\end{equation}
(for the relationships between the transformation coefficients in
terms of velocity potential and free surface perturbation see
Appendix \ref{appA}).

In Fig. \ref{f12}b) we present the transmission coefficient
$|T_2|$ together with the coefficients of wave excitation in the
intermediate domain, $|B_1|$, $|B_2|$, $|\hat B_1|$, and $|\hat
B_2|$, as functions of dimensionless frequency $\hat\omega$ for
the particular values of current speed $V_1 = 1.9$ and $V_2 =
0.1$. As one can see from this figure, the transmission
coefficient remains almost constant for small frequencies when
$\hat\omega < 1$, then it decreases with the frequency and
asymptotically vanishes as $|T_2| \sim \hat\omega^{-1}$ when
$\hat\omega \to \infty$.

The graphic of $|\Phi(\xi)|$ is shown in Fig.~\ref{f13} by lines 1
and 3 (the left branch of function $|\Phi(\xi)|$ for the incident
negative- and positive-energy waves are the same). The plot was
generated for $\hat\omega = 1$ on the basis of solution
Eqs.~(\ref{Solut1})--(\ref{Solut4}) with $A_1 = 0$, $A_2 = 1$,
$D_1 = 0$, and $D_2 = T_2$  as per Eq.~(\ref{TransCoef2}).
Coefficients $B_1$ and $B_2$ are given by Eqs.~(\ref{B1}) and
(\ref{B2}), and coefficients $\hat B_1$ and $\hat B_2$ are given
by Eqs.~(\ref{C11N}) and (\ref{C21N}). The module of function
$\Phi(\xi)$ is discontinuous only in the critical point $\xi =
-1$, but the phase of function $\Phi(\xi)$ quickly changes in the
small vicinity of this point.

\subsubsection{\label{sec:level443}Transformation of a counter-current
propagating wave}

Consider now the case when the incident wave propagates against
the mean current in the spatially variable current from the right
domain where the background current is sub-critical. There are no
waves capable to propagate against in the $\xi < -1$ domain where
$V > 1$, therefore there is no transmitted wave in this case.
However, the incident wave can propagate against the current and
even penetrate into the transient zone $\xi_1 < \xi < \xi_2$ up to
the critical point $\xi = -1$ until the current remains
subcritical.

Because there are no waves in the domain $\xi < -1$, we should set
in Eqs.~(\ref{Solut1})--(\ref{Solut4}) $A_1 = A_2 = \hat B_1 =
\hat B_2 = 0$, $C_1 \equiv R$, and $C_2 = 1$. Then the matching
condition (\ref{38}) yields
\begin{equation}
\label{SingleCoef}%
B_2 = -\frac{\Gamma(-{\rm i}\,\hat\omega)}{\Gamma^2(1-{\rm i}\,\hat\omega/2)}B_1,%
\end{equation}
and from Eqs. (\ref{MatCon3}) and (\ref{MatCon4}) we obtain for
the reflection coefficient
\begin{widetext}%
\begin{equation}
\label{ReflCoef} %
R = -\frac{\displaystyle\tilde w'_2\left(V_2^{2}\right) -
\frac{{\rm i}\,\hat\omega}{2}\frac{\tilde
w_2\left(V_2^{2}\right)}{V_2(1-V_2)} - \frac{\Gamma(-{\rm
i}\,\hat\omega)}{\Gamma^2(1-{\rm i}\,\hat\omega/2)} \left[\tilde
w'_3(V_2^{2}) - \frac{{\rm i}\,\hat\omega}{2}\frac{\tilde w_3\left(V_2^{2}\right)}{V_2(1-V_2)}\right]}%
{\displaystyle\tilde w'_2\left(V_2^2\right) + \frac{{\rm
i}\,\hat\omega}{2}\frac{\tilde w_2\left(V_2^2\right)}{V_2(1+V_2)}
- \frac{\Gamma(-{\rm i}\,\hat\omega)}{\Gamma^2(1-{\rm
i}\,\hat\omega/2)}\left[\tilde w'_3\left(V_2^2\right) + \frac{{\rm
i}\,\hat\omega}{2}\frac{\tilde w_3\left(V_2^2\right)}{V_2(1+V_2)}\right]}.%
\end{equation}
\end{widetext}%

\bigskip

Then, from Eqs.~(\ref{MatCon3}) and (\ref{SingleCoef}) we find
$B_1$ and $B_2$; in particular for $B_1$ we obtain:
\begin{widetext}%
\begin{equation}
\label{CoefB1} %
B_1 = \frac{1 + R}{\displaystyle\tilde w_2\left(V_2^2\right) -
\frac{\Gamma(-{\rm i}\,\hat\omega)}{\Gamma^2(1-{\rm
i}\,\hat\omega/2)}\tilde w_3\left(V_2^2\right)}.%
\end{equation}
\end{widetext}%

\begin{figure}[h]
\centering
\includegraphics[width=90mm]{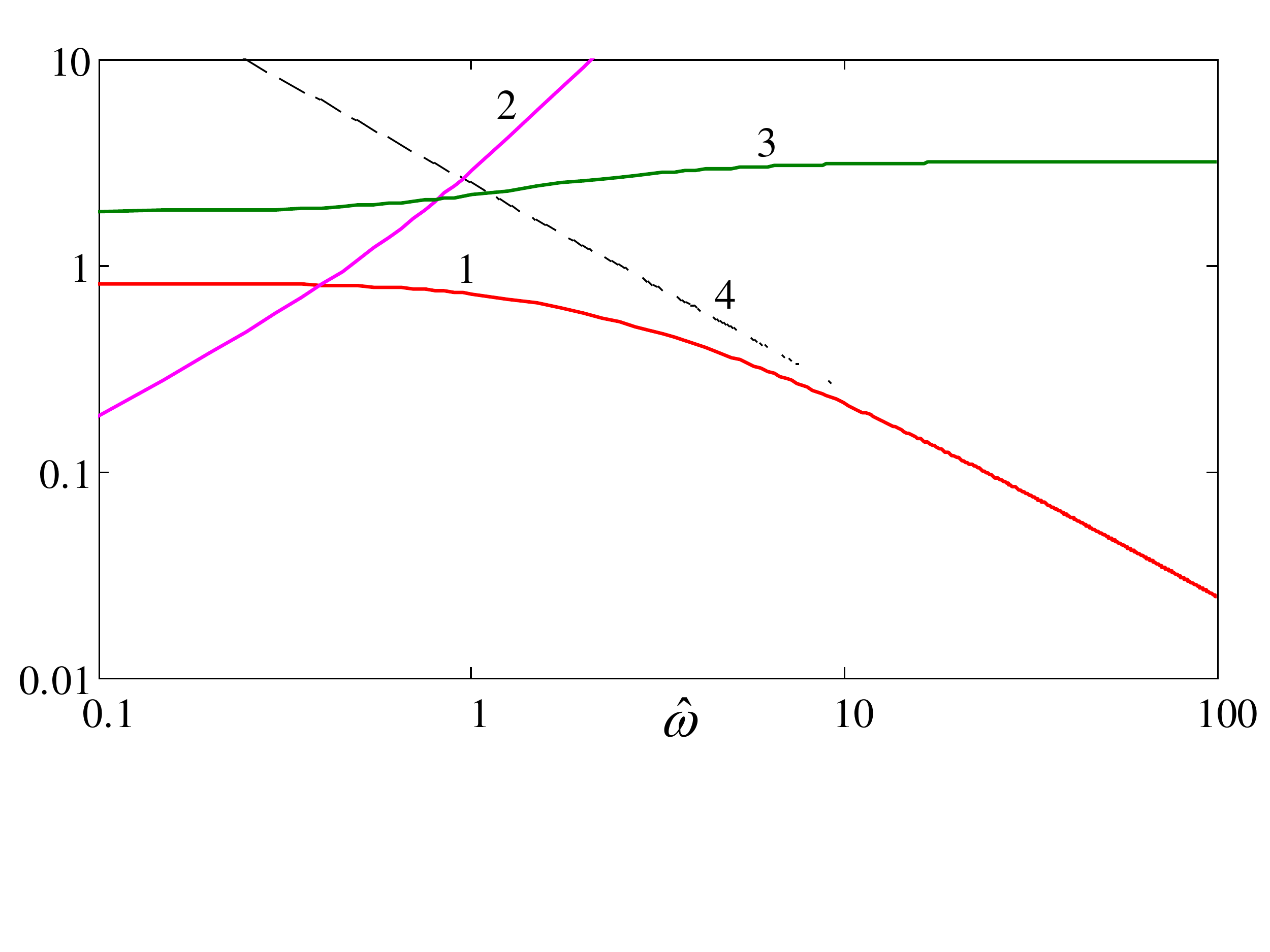}
\vspace*{-18mm}%
\caption{(color online) Modulus of the reflection coefficients
$|R|$ (line 1) and coefficients of wave excitation in the
transient domain, $|B_1|$ (line 2) and $|B_2|$ (line 3), as
functions of dimensionless frequency $\hat\omega$ for $V_1 = 1.9$,
$V_2 = 0.1$. Dashed line 4 represents the high-frequency
asymptotic for $|R| \sim \hat\omega^{-1}$.}
\label{f14}%
\end{figure}

Graphics of modulus of reflection coefficient $|R|$ as well as
coefficients $|B_1|$ and $|B_2|$ are shown in Fig.~\ref{f14} as
functions of dimensionless frequency $\hat\omega$ for the
particular values of $V_1 = 1.9$ and $V_2 = 0.1$.

In the long-wave approximation, $\hat\omega \to 0$, using the
asymptotics of hypergeometric function $_2F_1(a,b;c;d)$ (see
Appendix \ref{appC}), we obtain the limiting value of the
reflection coefficient
\begin{equation}
\label{Reduct8}%
R = \frac{1 - V_2}{1 + V_2}.
\end{equation}

In terms of free surface perturbation this value corresponds to
$R_\eta = 1$ (for the relationships between the transformation
coefficients in terms of velocity potential and free surface
perturbation see Appendix \ref{appA}). This formally agrees with
the solution found in Ref.~\cite{ChurErRousStep-2017}.

\begin{figure}[h]
\centering
\includegraphics[width=90mm]{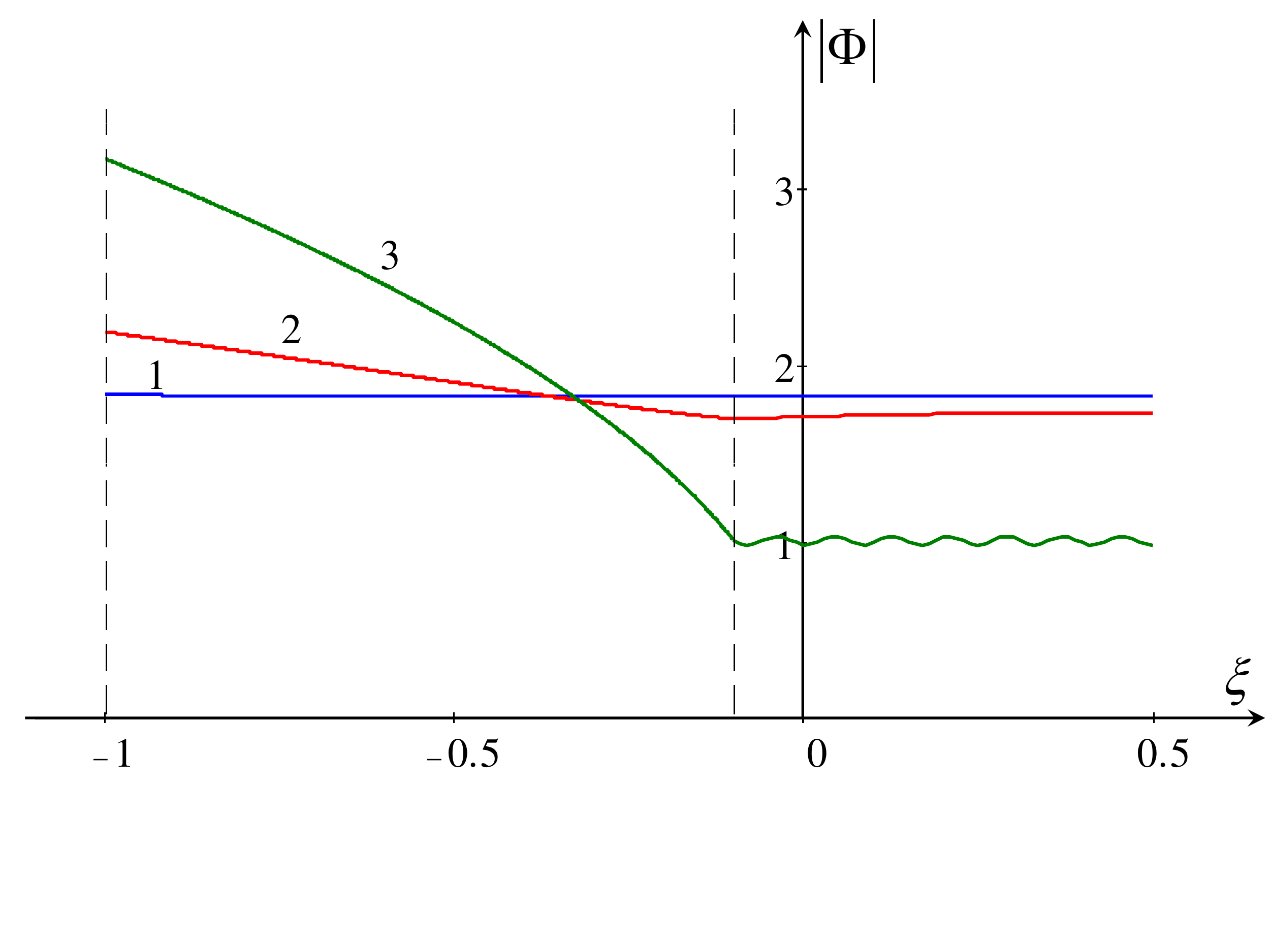}
\vspace*{-15mm}%
\caption{(color online) Module of function $\Phi(\xi)$ for a
counter-current propagating incident wave which scatters in the
decelerating trans-critical current with $V_1 = 1.9$ and $V_2 =
0.1$ for the particular values of $\hat\omega$: line 1 --
$\hat\omega = 0.1$, line 2 --
$\hat\omega = 1$, and line 3 -- $\hat\omega = 100$.}%
\label{f15}%
\end{figure}

In Fig.~\ref{f15} we present graphics of $|\Phi(\xi)|$ as per
Eqs.~(\ref{Solut1})--(\ref{Solut4}) for $A_1 = A_2 = 0$, $D_1 =
1$, and $D_2 = R$  as per Eq.~(\ref{ReflCoef}). Coefficients $B_1$
and $B_2$ are given by Eqs.~(\ref{CoefB1}) and (\ref{SingleCoef}),
and coefficients $\hat B_1 = \hat B_2 = 0$. A plot was generated
for three dimensionless frequencies: line 1 -- for $\hat\omega =
0.1$, line 2 -- for $\hat\omega = 1$, and line 3 -- for
$\hat\omega = 100$. The phase of function $\Phi(\xi)$ infinitely
increases when the incident wave approaches the critical point
$\xi = -1$.

The energy fluxes in the incident $J_i$ and reflected $J_r$ waves
in the right domain ($\xi > \xi_2$) are
\begin{equation}
J_i = -\frac{2\hat\omega}{V_2} < 0, \quad J_r =
\frac{2\hat\omega}{V_2}|R|^2 > 0.
\end{equation}

Thus, the total energy flux in the right domain $\Delta J \equiv
J_i - J_r = -\left(2\hat\omega/V_2\right)\left(1 - |R|^2\right) <
0$ is negative; it transfers towards the critical point, where it
is absorbed by the viscosity.

\section{\label{sec:level5}Discussion and Conclusion}

In this paper we have calculated the transformation coefficients
of shallow water gravity waves propagating on a longitudinally
varying quasi-one-dimensional current. Owing to the choice of a
piece-linear velocity profile $U(x)$ (or, in the dimensionless
variables, $V(\xi)$, see Fig.~\ref{f01}) we were able to calculate
analytically the scattering coefficients as functions of incident
wave frequency $\hat\omega$ for accelerating and decelerating
sub-, super-, and trans-critical currents, as well as for all
possible types of incident wave.

Presented analysis pertains to the dispersionless case when the
wavelengths of all waves participating in the scattering process
are much greater than the water depth in the canal. However, the
wavelengths $\lambda$ can be comparable with or even less than the
characteristic length of current inhomogeneity $L$. In the
long-wave limit $\lambda \gg L$, the scattering coefficients are
expressed through the simple algebraic formulae which are in
agreement with the formulae derived in \cite{ChurErRousStep-2017}
for the case of abrupt change of canal cross-section.

The most important property of scattering processes in sub-,
super-, and accelerating trans-critical currents is that the wave
energy flux conserves, $J =$ const, (see Eq.~(\ref{Chur57}) and
the text below Eq.~(\ref{NormEq})). This law provides a highly
convenient and physically transparent basis for the analysis of
wave scattering.

In the simplest case of sub-critical currents ($U(x) < c_0$, or
$V(\xi) < 1$), both accelerating and decelerating, all
participating waves possess a positive energy, and the energy flux
of the unit amplitude incident wave (no matter whether running
from the left or from the right) is divided between reflected and
transmitted waves in a such manner that $|R|^2 + K_T = 1$ (see
Eqs.~(\ref{FluxCons}) and (\ref{Reduct1}) for the wave running
from the left, and Eqs.~(\ref{FluxCons2}) and (\ref{Reduct2}) for
the wave running from the right).

In super-critical currents ($V(\xi) > 1$) there are positive- and
negative-energy waves both propagating downstream but carrying
energy fluxes of opposite signs. Propagating through the
inhomogeneity domain $\xi_1 < \xi < \xi_2$, they transform into
each other in a such way that the energy flux of each wave grows
in absolute value to the greater extent the greater the velocity
ratio is. As a result, at $\xi > \xi_2$ the energy flux of each
transmitted wave can become greater (in absolute value) than that
of the incident wave (see Fig. \ref{f08}). Quantitatively the
increase of wave-energy fluxes can be easily estimated in the
low-frequency limit using Eqs.~(\ref{Reduct3}) and
(\ref{Reduct4}).

The scattering process in accelerating trans-critical currents
($V_1 < 1 < V_2$) looks like a hybrid with respect to those in
sub- and super-critical currents. The incident wave can be the
only co-current propagating wave of positive energy. Initially, at
$\xi < \xi_1$, its energy flux (at unit amplitude) $J_0 =
2\hat\omega/V_1$, but in the domain $\xi_1 < \xi < 1$ it partially
transforms into the counter-current propagating reflected wave,
and at $\xi = 1$ its energy flux is only $J_0(1-|R|^2)$ (see
Eq.~(\ref{FluxCons023})). Further, in the super-critical domain $1
< \xi < \xi_2$, it generates a negative-energy wave, and the
energy fluxes of both waves grow in absolute value to the greater
extent the greater $V_2$ is. And again this process can be better
understand in the low-frequency limit by means of
Eqs.~(\ref{Reduct5}).

The most interesting scattering processes takes place in
decelerating trans-critical currents ($V_1 > 1 > V_2$) where
B-waves (which are either counter-current propagating
positive-energy waves or downstream propagating negative energy
waves, see Sec.~\ref{sec:level3}) run to the critical point (where
$V(\xi) = 1$) and become highly-oscillating in its vicinity. For
this reason we are forced to give up the model of ideal fluid and
to take into account an infinitesimal viscosity in the
neighborhood of the critical point. As a result, the energy flux
continues to conserve on the left and right of the critical point,
but changes in its small vicinity. Let us illuminate the details
of this phenomenon.

Consider first an incident positive-energy F-wave arriving from
the left. In the transient domain $\xi_1 < \xi < -1$ it partially
transforms into the negative-energy B-wave. The total wave flux
conserves, whereas the energy flux of each individual wave
increases in absolute value up to the critical point (to the more
greater extent the greater $V_1$ is). As follows from the
qualitative consideration on the basis of JWKB method (see Section
\ref{sec:level3}) and from exact analytical solutions (see
Subsection \ref{sec:level44}), near the critical point the B-wave
becomes highly oscillating in space. This causes its absorbtion
due to viscosity; as will be shown below, the absorbtion is
proportional to $\nu/\lambda^2$. In contrast to that, the
wavelength of the co-current propagating F-wave does not change
significantly in the process of transition through the critical
point (see Section \ref{sec:level3}), therefore the effect of
viscosity onto this wave is negligible. After transition this wave
runs through the non-unform subcritical domain $-1 < \xi < \xi_2$
and partially transforms into another B-wave -- a counter-current
propagating wave of positive energy. This wave approaching the
critical point also becomes highly oscillating and therefore
absorbs in the vicinity of that point. The energy flux of
transmitted F-wave decreases proportional to $V_2$. The total
change of energy flux in transition from the incident to the
transmitted wave is described by Eqs.~(\ref{FluxJump}), and in the
limit $\hat\omega \to 0$ is determined by the product $V_1V_2$.

If the incident wave arriving from minus infinity is the B-wave of
negative energy carrying a negative energy flux (see $J_1$ in
Eq.~(\ref{441})), then in the transient zone, $\xi_1 < \xi < -1$,
it generates due to scattering on inhomogeneous current the F-wave
of positive energy, so that the wave fluxes of both waves grow in
absolute value. The B-wave absorbs due to viscosity in the
vicinity of the critical point $\xi = -1$, whereas the F-wave
freely passes through this point with an insignificant change of
its wavelength. After passing through the critical point the
F-wave generates in the domain $-1 < \xi < \xi_2$ a new B-wave of
positive energy, which propagates a counter-current towards the
critical point and absorbs in its vicinity due to viscosity.
Therefore, the energy flux of the F-wave increases first from zero
at $\xi = \xi_1$ up to some maximal value at $\xi = -1$, then it
decreases due to transformation of wave energy into the B-wave to
some value at $\xi = \xi_2$ (see $J_2$ in Eq.~(\ref{441})), and
then remains constant. In the long-wave approximation, $\hat\omega
\to 0$, we obtain
$$
\Delta J \equiv |J_2| - |J_1| = \left(\dfrac {V_1}{V_2}|T_2|^2 -
1\right)|J_1| = \dfrac{V_2\,|J_1|}{V_1(1 + V_2)^2}\left[V_1^2 -
V_1\left(4 + V_2 + \frac{1}{V_2}\right) + 1\right].
$$

Analysis of this expression shows that because $V_2 + 1/V_2 \ge
2$, then $\Delta J$ can be positive (i.e., the energy flux of
transmitted wave can be greater than the energy flux of incident
wave by absolute value), if $V_1 > 3 + 2\sqrt{2} \approx 5.83$.

If there are two incident waves arriving simultaneously from minus
infinity so that one of them has positive energy and another one
-- negative energy, then at some relationships between their
amplitudes and phases it may happen that in the transient zone in
front of the critical point the superposition of these waves and
scattered waves generated by them can annihilate either the
positive-energy F-wave or negative-energy B-wave. In the former
case it will not be a transmitted wave behind the critical point,
because the B-wave in its vicinity completely absorbs (the
``opacity'' phenomena occurs). In the latter case there is no
negative-energy B-wave on the left of the critical point and
therefore there is nothing to absorb, and the F-wave passes
through this point without loss of energy (we assume that the
viscosity is negligible). Further, the F-wave spends some portion
of its energy transforming into the counter-current propagating
B-wave of positive energy which ultimately dissipates in the
vicinity of the critical point. Nevertheless, the residual energy
flux of transmitted wave at $\xi
> \xi_2$ turns to be equal to the total energy flux of two
incident waves at $\xi < \xi_1$, and in such a very particular
case the energy flux conserves.

Finally, if an incident B-wave of positive energy arrives from
plus infinity, then in the inhomogeneous zone, $-1 < \xi < \xi_2$,
it generates a co-current propagating F-wave of positive energy.
The energy fluxes of both these waves have opposite signs and
decrease in absolute value as one approaches the critical point.
In the critical point the energy flux of F-wave vanishes, and the
remainder of the B-wave absorbs. In this case the less the $V_2$
the higher the reflection coefficient $|R|$ is, and this is
especially clear  in the low-frequency approximation, see
Eq.~(\ref{Reduct8}).

The analysis presented above is based on the fact that the
wavelengths of scattered waves drastically decrease in the
vicinity of a critical point, where $V(\xi) = 1$. In such case
either the dispersion, or dissipation, or both these effects may
enter into play. We will show here that at certain situations the
viscosity can predominate over the dispersion. Considering the
harmonic solution $\sim e^{{\rm i}\kappa\xi}$ of Eq.
(\ref{NormViscEq}) in the vicinity of a critical point and
neglecting the term $\sim V'$, we obtain the dispersion relation
extending (\ref{DispRel}). In the dimensional form it is:
\begin{equation}
\label{ViscDispRel} %
\left(\omega - {\bf kU}\right)^2 = c_0^2k^2 - {\rm i}\nu k^2\left(\omega - {\bf kU}\right). %
\end{equation}

The solution to this equation for small a viscosity $\nu k \ll
c_0$ is
\begin{equation}
\label{SolViscDispRel} %
\omega = |c_0 \pm U||{\bf k}| - {\rm i}\nu k^2/2. %
\end{equation}

The viscosity effect becomes significant when the imaginary and
real parts of frequency become of the same order of magnitude.
This gives $|{\bf k}| \sim 2|c_0 \pm U|/\nu$. Multiplying both
sides of this relationship by $h$, we obtain $|{\bf k}|h \sim
2h|c_0 \pm U|/\nu$. For the counter-current propagating B-wave
$|c_0 - U| \to 0$, therefore the product $|{\bf k}|h$ can be small
despite of smallness on $\nu$. So, the condition $|{\bf k}|h \sim
2h|c_0 - U|/\nu \ll 1$ allows us to consider the influence of
viscosity in the vicinity of a critical point, whereas the
dispersion remains negligibly small. In the meantime, the
wavelength of co-current propagating F-wave does not change
dramatically in the process of transition through the critical
point (see Section \ref{sec:level3}). For such wave the viscosity
is significant when $|{\bf k}|h \sim 2h(c_0 + U)/\nu \gg 1$ which
corresponds to the deep-water approximation.

Notice in the conclusion that the wave-current interaction in
recent years became a very hot topic due to applications both to
the natural processes occurring in the oceans and as a model of
physical phenomena closely related with the Hawking radiation in
astrophysics \citep{Unruh-1981, Jacobson-1991, Unruh-1995,
Faccio-2013, CoutPar-2014, RobMichPar-2016, Coutant-2016,
Philbin-2016}. The influence of high-momentum dissipation on the
Hawking radiation was considered in astrophysical application
\citep{Robertson-2015} (see also \citep{Adamek-2013} where the
dissipative fields in de Sitter and black hole spacetimes metrics
were studied with application to the quantum entanglement due to
pair production and dissipation). The peculiarity of our paper is
in the finding of exactly solvable model which enabled us to
construct analytical solutions and calculate the scattering
coefficients in the dispersionless limit. We have shown, in
particular, that in the case of accelerating trans-critical
current both the reflection coefficient of positive-energy wave
and transmission coefficient of negative-energy wave decrease
asymptotically with the frequency as $|R| \sim T_n \sim
\hat\omega^{-1}$. This can be presented in terms of the Hawking
temperature $T_H = (1/2\pi)(dU/dx)$ (see, e.g., \citep{Unruh-1995,
RobMichPar-2016}) and dimensional frequency $\omega$ as $|R| \sim
T_n \sim 2\pi T_H/\omega$.

\begin{acknowledgments}
This work was initiated when one of the authors (Y.S.) was the
invited Visiting Professor at the Institut Pprime, Universit{\'e}
de Poitiers in August--October, 2016. Y.S. is very grateful to the
University and Region Poitou-Charentes for the invitation and
financial support during his visit. Y.S. acknowledges also the
funding of this study from the State task programme in the sphere
of scientific activity of the Ministry of Education and Science of
the Russian Federation (Project No. 5.1246.2017/4.6). The research
of A.E. was supported by the Australian Government Research
Training Program Scholarship.
\end{acknowledgments}

\appendix
\section{Energy Flux Conservation}%
\label{appA}%

Let us multiply equation (\ref{ansatz}) by the complex-conjugate
function $\overline{\varphi}$ and subtract from the result
complex-conjugate equation:
\begin{widetext}%
\begin{eqnarray}
\label{001} %
\ & \ & \overline{\varphi}\left(\frac{\partial }{\partial t} +
U\frac{\partial }{\partial x} \right)\left(\frac{\partial
\varphi}{\partial t} + U\frac{\partial \varphi}{\partial x}
\right) - \varphi\left(\frac{\partial }{\partial t} +
U\frac{\partial }{\partial x} \right)\left(\frac{\partial
\overline{\varphi}}{\partial t} + U\frac{\partial
\overline{\varphi}}{\partial x} \right) = \nonumber \\
\ & \ &
c_0^2U\left[\overline{\varphi}\frac{\partial}{\partial
x}\left(\frac{1}{U}\frac{\partial \varphi}{\partial x}\right) - %
\varphi \frac{\partial}{\partial x}\left(\frac{1}{U}\frac{\partial
\overline{\varphi}}{\partial x}\right)\right].
\end{eqnarray}
\end{widetext}%

Dividing this equation by $U$ and rearranging the terms we present
this equation in the form:
\begin{widetext}%
\begin{eqnarray}
\label{003} %
\ & \ & \frac{\partial}{\partial
t}\left[\frac{\overline{\varphi}}{U}\left(\frac{\partial
\varphi}{\partial t} + U\frac{\partial \varphi}{\partial x}
\right) - \frac{\varphi}{U}\left(\frac{\partial
\overline{\varphi}}{\partial t} + U\frac{\partial
\overline{\varphi}}{\partial x} \right)\right] + \nonumber \\
\ & \ &
\frac{\partial}{\partial x}\left[\overline{\varphi} \left(\frac{\partial
\varphi}{\partial t} + U\frac{\partial \varphi}{\partial x}
\right) - \varphi\left(\frac{\partial \overline{\varphi}}{\partial
t} + U\frac{\partial \overline{\varphi}}{\partial x} \right) -
\frac{c_0^2}{U}\left(\overline{\varphi}\frac{\partial
\varphi}{\partial x} - \varphi\frac{\partial
\overline{\varphi}}{\partial x}\right)\right] = 0.
\end{eqnarray}
\end{widetext}%

If we denote
\begin{eqnarray}
{\cal E} &=& \frac{\rm
i}{U}\left[\bar\varphi\left(\frac{\partial\varphi} {\partial t} +
U\frac{\partial\varphi}{\partial x}\right) -
\varphi\left(\frac{\partial\bar\varphi}{\partial t} +
U\frac{\partial\bar\varphi}{\partial x}\right)\right],
\phantom{ww} \label{006} \\%
&& \nonumber \\
J &=& {\cal E}U - {\rm
i}\,\frac{c_0^2}{U}\left(\bar\varphi\frac{\partial\varphi}
{\partial x} - \varphi\frac{\partial\bar\varphi}{\partial
x}\right), \label{007}
\end{eqnarray}
then Eq.~(\ref{003}) can be presented in the form of the
conservation law
\begin{equation}
\label{005} %
\frac{\partial {\cal E}}{\partial t} + \frac{\partial J}{\partial
x}=0,
\end{equation}

For the waves harmonic in time, $\varphi = \Phi(x)e^{-i\omega t}$,
both ${\cal E}$ and $J$ do not depend on time, and Eq.~(\ref{005})
reduces to $J = $ const. Substituting in Eq.~(\ref{007}) written
in the dimensionless form solution (\ref{SolA1}) for $\xi < \xi_1$
and solution (\ref{SolA3}) for $\xi > \xi_2$, after simple
manipulations we obtain
\begin{eqnarray}
J &=& \frac{2\hat\omega}{V_1}\left(1 - |R|^2\right), \quad \xi <
\xi_1; \label{008} \\%
&& \nonumber \\
J &=& \frac{2\hat\omega}{V_2}\,|T|^2, \quad\quad\qquad \xi >
\xi_2. \label{009}
\end{eqnarray}

Equating $J$ calculated in Eqs.~(\ref{008}) and (\ref{009}), we
obtain the relationship between the transformation coefficients
presented in Eq.~(\ref{FluxCons}).

Using then solution (\ref{SolA2}) for $\xi_1 < \xi < \xi_2$, we
obtain
\begin{widetext}%
\[J = 2\hat\omega\left|B_1w_2(\xi^2) + B_2w_3(\xi^2)\right|^2
 - {}\]
\[2i(1 - \xi^2)\left\{|B_1|^2\left[w'_2(\xi^2)\overline{w_2}(\xi^2)
 - \overline{w'_2}(\xi^2)w_2(\xi^2)\right] + |B_2|^2\left[w'_3(\xi^2)\overline{w_3}(\xi^2) -
\overline{w'_3}(\xi^2)w_3(\xi^2)\right] + {} \right.
\]
\begin{equation}
B_1\overline{B_2}\left[w'_2(\xi^2)\overline{w_3}(\xi^2) -
w_2(\xi^2)\overline{w'_3}(\xi^2)\right] - \left.
\overline{B_1}B_2\left[\overline{w'_2}(\xi^2)w_3(\xi^2) -
\overline{w_2}(\xi^2)w'_3(\xi^2)\right]\right\} = {\rm const}.
\label{010}
\end{equation}
\end{widetext}%

It was confirmed by direct calculations with the solutions
(\ref{SolA1})--(\ref{SolA3}) that $J$ is indeed independent of
$\xi$ for given other parameters.

In a similar way, for the super-critical accelerating current one
can obtain in the intermediate interval $\xi_1 < \xi < \xi_2$
\begin{widetext}%
\begin{eqnarray}
J &=& \frac{2\hat\omega}{\xi^2}\left|B_1\breve w_1(\xi^{-2}) +
B_2\breve w_3(\xi^{-2})\right|^2 - \frac{2{\rm i}\,(\xi^2 -
1)}{\xi^4}\left\{|B_1|^2\left[\breve
w'_1(\xi^{-2})\overline{\breve w_1}(\xi^{-2})
 - \overline{\breve w'_1}(\xi^{-2})\breve w_1(\xi^{-2})\right] + {} \right. \nonumber \\
{} && |B_2|^2\left[\breve w'_3(\xi^{-2})\overline{\breve
w_3}(\xi^{-2}) - \overline{\breve w'_3}(\xi^{-2})\breve
w_3(\xi^{-2})\right] + B_1\overline{B_2}\left[\breve
w'_1(\xi^{-2})\overline{\breve w_3}(\xi^{-2}) -
\breve w_1(\xi^{-2})\overline{\breve w'_3}(\xi^{-2})\right] - {} \nonumber \\
{} && \left. \overline{B_1}B_2\left[\overline{\breve
w'_1}(\xi^{-2})\breve w_3(\xi^{-2}) - \overline{\breve
w_1}(\xi^{-2})\breve w'_3(\xi^{-2})\right]\right\} = {\rm const}.
\label{0010}
\end{eqnarray}
\end{widetext}%
Here the coefficients $B_1$ and $B_2$ should be taken either from
Eqs.~(\ref{CoefB1p}) and (\ref{CoefB2p}) for the scattering of
positive energy wave or from Eqs.~(\ref{CoefB1n}) and
(\ref{CoefB2n}) for the scattering of negative energy wave.

The transformation coefficients $R$ and $T$ were derived in this
paper in terms of the velocity potential $\varphi$. But they can
be also presented in terms of elevation of a free surface $\eta$.
Using Eq.~(\ref{RedMassCons1}) for $x < x_1$ and definition of
$\varphi$ just after that equation we obtain for a wave sinusoidal
in space
\begin{equation}
\label{011} %
(\omega - {\bf k}\cdot{\bf U_1})\eta = khu = {\rm i}hk^2\varphi.
\end{equation}

Bearing in mind that according to the dispersion relation $\omega
- {\bf k}\cdot{\bf U_1} = c_0|k|$, we find from Eq.~(\ref{011})
\begin{equation}
\label{012} %
\varphi = -{\rm i}\frac{c_0}{h|k|}\eta = -{\rm i}\frac{c_0(c_0 \pm
U_1)}{h\omega}\eta,
\end{equation}
where sign plus pertains to co-current propagating incident wave
and sign minus -- to counter-current propagating reflected wave.

Similarly for the transmitted wave for $x > x_2$ we derive
\begin{equation}
\label{013} %
\varphi = -{\rm i}\frac{c_0(c_0 + U_2)}{h\omega}\eta.
\end{equation}

Substitute expressions (\ref{012}) and (\ref{013}) for incident,
reflected and transmitted waves into Eq.~(\ref{FluxCons}) and bear
in mind that $R \equiv \varphi_r/\varphi_i$, $T \equiv
\varphi_t/\varphi_i$, and $\omega$ and $c_0$ are constant
parameters:
\begin{equation}
\label{014} %
V_2\left[\left(1 + V_1\right)^2 - \left(1 -
V_1\right)^2|R_\eta|^2\right] = V_1\left(1 +
V_2\right)^2|T_\eta|^2,
\end{equation}
where
\begin{equation}
\label{015} %
R_\eta \equiv \frac{\eta_r}{\eta_i} = \frac{1 + V_1}{1 - V_1}R,
\quad \mbox{and} \quad T_\eta \equiv \frac{\eta_t}{\eta_i} =
\frac{1 + V_1}{1 + V_2}T.
\end{equation}

In such form Eq.~(\ref{FluxCons}) represents exactly the
conservation of energy flux (see \citep{MaiRusStep-2016,
ChurErRousStep-2017}).

\section{Derivation of Matching Conditions for Equation (\ref{NormEq})}%
\label{appB} %

To derive the matching conditions in the point $\xi_1$, let us
present Eq.~(\ref{NormEq}) in two equivalent forms:
\begin{equation}
\label{NormEq2} %
\left[\left(V - \frac{1}{V}\right)\Phi\right]'' - \left[\left(1 +
\frac{1}{V^2}\right)V'\Phi\right]' - 2\,{\rm i}\,\hat\omega\,\Phi'
- \frac{\hat\omega^2}{V}\,\Phi = 0.
\end{equation}
\begin{equation}
\label{NormEq1} %
\left[\left(V - \frac{1}{V}\right)\Phi'\right]' - 2\,{\rm
i}\,\hat\omega\,\Phi' - \frac{\hat\omega^2}{V}\,\Phi = 0,
\end{equation}

Let us multiply now Eq.~(\ref{NormEq1}) by $\zeta = \xi - \xi_1$
and integrate it by parts with respect to $\zeta$ from
$-\varepsilon$ to $\varepsilon$:
\begin{widetext}%
\begin{equation}
\label{IntNormEq2} %
\left.\left\{\zeta\left[\left(V-\frac{1}{V}\right)\Phi' - 2\,{\rm
i}\,\hat\omega\, \Phi\right] - \left(V - \frac{1}{V}\right)
\Phi\right\}\right|_{-\varepsilon}^{\varepsilon} +
\int\limits_{-\varepsilon}^{\varepsilon}\left[
\left(1+\frac{1}{V^2}\right)V' + 2\,{\rm i}\,\hat\omega -
\frac{\hat\omega^2}{V}\,\zeta\right]\Phi\,d\zeta = 0.
\end{equation}
\end{widetext}%

In accordance with our assumption about the velocity, function
$V(\zeta)$ is piece-linear, and its derivative is piece-constant.
Assuming that function $\Phi(\zeta)$ is limited on the entire
$\zeta$-axis, $|\Phi| \le M$, where $M  < \infty$ is a constant,
we see that the integral term vanishes when $\varepsilon \to 0$.
The very first term, which contains $\zeta$ in front of the curly
brackets $\{\ldots\}$, also vanishes when $\varepsilon \to 0$, and
we have
\begin{equation}
\label{Cond2} %
\left. \left[\left(V -
\frac{1}{V}\right)\Phi\right]\right|_{-\varepsilon}^{\varepsilon}
 = 0.
\end{equation}
This implies that $\Phi(\zeta)$ is a continuous function in the
point $\zeta = \zeta_1$.

If we integrate then Eq.~(\ref{NormEq2}) with respect to $\zeta$
in the same limits as above, we obtain:
\begin{equation}
\label{IntNormEq1} %
\left.\left[\left(V - \frac{1}{V}\right)\Phi' - 2\,{\rm
i}\,\hat\omega\,\Phi\right]\right|_{-\varepsilon}^{\varepsilon} -
\hat\omega^2\int\limits_{-\varepsilon}^{\varepsilon}
\frac{\Phi(\zeta)}{V(\zeta)}\,d\zeta = 0.
\end{equation}

Under the same assumptions about functions $V(\zeta)$ and
$\Phi(\zeta)$, the integral term here vanishes when $\varepsilon
\to 0$ and we obtain:
\begin{equation}
\label{Cond1} %
\left[\left(V - \frac{1}{V}\right)\Phi' - 2\,{\rm
i}\,\hat\omega\,\Phi\right]_{-\varepsilon}
 = \left[\left(V - \frac{1}{V}\right)\Phi' - 2\,{\rm
i}\,\hat\omega\,\Phi\right]_{\varepsilon}.
\end{equation}

Due to continuity of functions $V(\zeta)$ and $\Phi(\zeta)$ in the
point $\zeta = \zeta_1$, we conclude that the derivative
$\Phi'(\zeta)$ is a continuous function in this point too. The
same matching conditions can be derived for the point $\xi_2$ as
well.

\section{The Transformation Coefficients in the Long-Wave Limit}%
\label{appC} %

The long-wave approximation in the dispresionless case considered
here corresponds to the limit $\omega \to 0$. In such a case the
wavelength of each wave is much greater than the length of the
transient domain, $\lambda \gg L$, so that the current speed
transition from the left domain $\xi < \xi_1$ to the right domain
$\xi
> \xi_2$ can be considered as sharp and stepwise. Then using the
relationships (see \cite{Luke-1975})
$$
{_2F_1}(a,b;b;s) = (1 - s)^{-a} \quad \mbox{and} \quad s\,
{_2F_1}(1,1;2;s) = -\ln{(1 - s)},
$$
we can calculate functions (\ref{AccelSub1}) and
(\ref{DecelSub1}), as well as (\ref{Gauss11Ac}) and
(\ref{Gauss11De}), and their derivatives and obtain in the leading
order in $\hat \omega$ the following asymptotic
expressions (bearing in mind that $\zeta = \xi^2$ and $\eta = \xi^{-2}$): %
$$
\begin{array}{rcl c rcl c rcl c rcl} %
w_2(\zeta) &=& -\ln{(1 - \zeta)}, &\quad& w'_2(\zeta) &=&
\displaystyle\frac{1}{1 - \zeta}, &\quad& w_3(\zeta) &=& 1,
&\quad&
w'_3(\zeta) &=& O(\hat\omega^2), \label{Approx1} \\%
\tilde w_2(\zeta) &=& -\ln{(1 - \zeta)}, &\quad& \tilde
w'_2(\zeta) &=& \displaystyle\frac{1}{1 - \zeta}, &\quad& \tilde
w_3(\zeta) &=& 1, &\quad& \tilde w'_3(\zeta) &=& O(\hat\omega^2),
\label{Approx2}
\\%
\breve w_1(\eta) &=& 1, &\quad& \breve w'_1(\eta) &=&
\displaystyle -\frac{{\rm i}\,\hat\omega}{2}\frac{1}{1 - \eta},
&\quad& \breve w_3(\eta) &=& 1, &\quad& \breve w'_3(\eta) &=&
\displaystyle \frac{{\rm i}\hat\omega}{2\eta},
\label{Approx3} \\%
\hat w_1(\eta) &=& 1, &\quad& \hat w'_1(\eta) &=&
\displaystyle\frac{{\rm i}\,\hat\omega}{2}\frac{1}{1 - \eta},
&\quad& \hat w_3(\eta) &=& 1, &\quad& \hat w'_3(\eta) &=&
\displaystyle -\frac{{\rm i}\hat\omega}{2\eta}. \label{Approx4}%
\end{array}
$$

Using these formulae, one can readily calculate the limiting
values of transformation coefficients in the long-wave
approximation when $\hat\omega \to 0$. Their values are presented
in the corresponding subsections.

\bibliography{ChErSt}

\end{document}